\documentclass[11pt,a4paper]{article}
\usepackage{jheppub}
\pdfoutput=1
\usepackage[english]{babel}
\usepackage[latin1]{inputenc} 

\usepackage{color}
\usepackage{float}
\usepackage{amsxtra}
\usepackage{amstext}
\usepackage{amssymb}
\usepackage{latexsym}
\usepackage{dsfont}
\usepackage{hyperref}
\usepackage{multirow}
\usepackage{booktabs}
\usepackage{ifthen}

\usepackage{rotating}
\usepackage{longtable}
\usepackage{pdflscape}

\usepackage{subfig} %added by Remco

\usepackage{appendix}

\newcommand{\forloop}[5][1]{%
\setcounter{#2}{#3}%
\ifthenelse{#4}{#5\addtocounter{#2}{#1}%
\forloop[#1]{#2}{\value{#2}}{#4}{#5}}%
{}}

\newcounter{crcounter}

\newcommand{\etmis}{E$_{\text{T}}^{\text{miss}}$}

\newcommand{\atl}{ATLAS}

\title{Higgs, di-Higgs and tri-Higgs production via SUSY processes at the LHC with 14 TeV}
\author[a]{Melissa van Beekveld,} %order?
\author[a,b]{Wim Beenakker,}
\author[a]{Sascha Caron,}
\author[a]{Remco Castelijn,}
\author[a]{Marie Lanfermann,}
\author[a]{Antonia Struebig}

\affiliation[a]{Institute for Mathematics, Astrophysics and Particle Physics,
                Faculty of Science, Mailbox 79,\\
                Radboud University Nijmegen, P.O. Box 9010, NL-6500 GL Nijmegen,
                The Netherlands\\ 
                and Nikhef, Science Park, Amsterdam, The Netherlands}
\affiliation[b]{Institute of Physics, University of Amsterdam, Science Park 904,
                1018 XE Amsterdam, The Netherlands}

\emailAdd{mcbeekveld@gmail.com}
\emailAdd{W.Beenakker@science.ru.nl}
\emailAdd{scaron@cern.ch}
\emailAdd{r.castelijn@nikhef.nl}
\emailAdd{marie.lanfermann@gmail.com}
\emailAdd{antonia.struebig@cern.ch}

\abstract{
We have systematically investigated the production of a Higgs boson with a mass
of about $125$~GeV in the decays of supersymmetric particles within the 
phenomenological MSSM (pMSSM). We find regions of parameter space that are
consistent with all world data and that predict a sizeable rate of anomalous 
Higgs, di-Higgs and even tri-Higgs events at the 14 TeV LHC. All relevant SUSY 
production processes are investigated. We find that Higgs bosons can be produced
in a large variety of SUSY processes, resulting in a large range of different 
detector signatures containing missing transverse momentum. Such Higgs events 
are outstanding signatures for new physics already for the early 14 TeV LHC 
data. SUSY processes are also important to interprete deviations
found in upcoming Standard Model Higgs and di-Higgs production measurements.
}


\begin{document}
\maketitle

\section{Introduction}
The Higgs-boson discovery at the Large Hadron Collider (LHC) 
\cite{Aad:2012tfa,Chatrchyan:2012ufa} marks the beginning of a new era in 
particle physics. It gives us exciting new possibilities to study the physics 
of the Standard Model (SM) of particle physics. 
In this paper we investigate 
the next level of Higgs-boson searches, namely the possibility 
that Higgs bosons with a mass of about $125$~GeV 
are produced by processes involving physics beyond the SM.
\\[3mm]
Supersymmetry (SUSY) \cite{Miyazawa:1966,Ramond:1971gb,Golfand:1971iw,Neveu:1971rx,Neveu:1971iv,Gervais:1971ji,Volkov:1973ix,Wess:1973kz,Wess:1974tw,Fayet:1976et,Fayet:1977yc,Farrar:1978xj,Fayet:1979sa,Dimopoulos:1981zb} 
is one of the conceivable extensions of the SM. It could provide a natural 
candidate for cold dark matter if R-parity is conserved 
\cite{Goldberg:1983nd,Ellis:1983ew} and it allows for a stabilization
of the electroweak scale by reducing the fine tuning of higher-order 
corrections to the Higgs mass \cite{Dimopoulos:1981zb,Witten:1981nf,Dine:1981za,Dimopoulos:1981au,Sakai:1981gr,Kaul:1981hi}. 
In its minimal version, i.e.~the Minimal Supersymmetric Standard Model (MSSM), 
SUSY predicts superpartners for the existing SM particles and two Higgs 
doublets instead of one. On top of that, R-parity is assumed to be conserved in
the MSSM, which results in the existence of a lightest supersymmetric particle 
(LSP). If the LSP is a neutralino, i.e.~a Majorana-fermion superpartner 
associated with the neutral SM bosons in the electroweak sector, it is only 
weakly interacting and stable. It escapes detection, which results in missing 
transverse momentum in the detector. 
%Squarks and gluinos, the superpartners of the SM quarks and gluons, 
%are heavy coloured particles, which can decay into jets and this LSP. 
%Channels with jets and missing transverse momentum have a large discovery 
%potential at the LHC~\cite{Aad:2009wy}, since the coupling strength of the 
%strong force would cause an abundance of squarks and gluinos to be produced if 
%these particles are not too heavy. 
\\[3mm]
In the present study we investigate systematically the possibilities to produce
Higgs bosons with a mass \,$m_{h^0}\approx 125$~GeV \footnote{This study assumes
       that the discovered Higgs boson at $m_{h^0}\approx 125$~GeV is the
       lightest neutral CP-even Higgs boson of the MSSM.}
in the decay of SUSY particles. This analysis is based on the phenomenological 
MSSM (pMSSM)~\cite{Djouadi:2002ze,Berger:2008cq}. The pMSSM is 
scanned for parameter regions 
where the SUSY particles have a viable branching ratio to Higgs bosons. 
Only those models are selected that fulfil the current constraints on SUSY.
The relevance for the upcoming LHC runs at 14 TeV is discussed in detail and the
most relevant Higgs production processes are identified. Higgs production via 
particular SUSY processes has been studied 
e.g.~in~\cite{Ghosh:2012mc,Howe:2012xe,Gori:2011hj,Ghosh:2013qga}. We calculate the 
allowed production rates for anomalous Higgs, di-Higgs and tri-Higgs events. 
Subsequently, LHC events are simulated for each interesting model. These events 
are classified into topologies according to the SM particles produced in 
association with the Higgs boson(s) and the Higgs kinematics is studied.
We identify topologies that are interesting for extending the current SUSY 
searches. Experimentally the events might be best detectable by explicitly 
``tagging'' the Higgs boson(s) in SUSY searches. Since the invariant mass of 
the (lightest) Higgs boson is known and well reconstructable in many decay 
modes, and since we know that the SM rate to produce Higgs events with large 
missing transverse momentum (and maybe other SM particles) is small, 
a ``Higgs-tag'' can provide a unique signature for new physics.\\[3mm]
A few analyses have already searched for such events in ATLAS and CMS data.
Higgs production via $\widetilde{\chi}_2^0 \widetilde{\chi}_1^{\pm}$  
neutralino-chargino production has been 
investigated in ATLAS~\cite{ATLAS-Higgs} and CMS~\cite{Khachatryan:2014mma,Khachatryan:2014qwa}. 
In addition, searches have been pursued by CMS for a simplified model with a 
Higgs produced in top squark decays~\cite{Khachatryan:2014doa,Chatrchyan:2013mya}.
The present study aims to systematically investigate the possibility to produce
Higgs bosons within the current constraints on SUSY by considering all relevant
SUSY processes and decays.\\[3mm]
This paper is organized as follows. In section~2 the most important 
supersymmetric decay mechanisms for producing light Higgs bosons are discussed.
In section~3 the pMSSM parameter space is scanned for models that are
consistent with all current experimental constraints on SUSY and that have the
potential to produce sizeable Higgs-boson event rates. Finally, in section~4 
the surviving pMSSM models are studied with regard to the expected 
Higgs-boson event rates at the early stages of the upcoming LHC run and with 
regard to special kinematical features, such as boosts and missing transverse 
momentum. 
%\\{\bf More?}

\section{Supersymmetric decays into the lightest Higgs}

In view of its important role in producing Higgs bosons, we start with a 
detailed discussion of the neutralino/chargino sector.
In the MSSM some of the superfields mix as a result of SUSY breaking 
to form new mass eigenstates. Let's first consider the neutral SM bosons in the
electroweak sector, i.e.~the hypercharge $B$ boson, neutral weak $W^3$ boson and
neutral components of the two Higgs doublets. The associated Majorana-fermion 
superpartners, i.e.~the Bino $\widetilde{B}$, neutral Wino $\widetilde{W}^3$ and
neutral Higgsinos $\widetilde{H}_d^0$ and $\widetilde{H}_u^0$, mix to form 
neutral mass eigenstates called neutralinos ($\widetilde{\chi}_{1,2,3,4}^0$\,,
numbered in increasing mass order). This mixing is caused by off-diagonal terms
in the neutralino mass matrix, which acts on the Bino, Wino and Higgsino fields
\cite{Martin:1997ns}:
\begin{equation*}
{\cal L}_{\footnotesize\mbox{neutralino mass}} 
= -\,\frac{1}{2}\,(\psi^0)^{T} M_{\widetilde{\chi}^0}\,\psi^0 + \,\mbox{h.c.}~,
\end{equation*}
where
\begin{equation*}
\psi^0 = {\begin{pmatrix} \widetilde{B} \\ \widetilde{W}^3 \\ \widetilde{H}_d^0
                          \\ \widetilde{H}_u^0 \end{pmatrix}}
\end{equation*}
and
\begin{equation}
M_{\widetilde{\chi}^0} =  {\begin{pmatrix} 
M_1 & 0 & -c_\beta s_{\theta_W} m_Z & s_\beta s_{\theta_W} m_Z\\
0 & M_2 & c_\beta c_{\theta_W} m_Z & -s_\beta c_{\theta_W} m_Z\\
-c_\beta s_{\theta_W} m_Z & c_\beta c_{\theta_W} m_Z & 0 & -\mu \\
s_\beta s_{\theta_W} m_Z & -s_\beta c_{\theta_W} m_Z & -\mu & 0  \label{nmm}
\end{pmatrix}}~.
\vspace*{3mm}
\end{equation}
Here $s_\alpha \equiv \sin\alpha\,$ and $\,c_\alpha \equiv \cos\alpha$.
The parameters $M_1$ and $M_2$ are the SUSY-breaking mass parameters for the 
Bino and Winos, $\mu$ is the SUSY version of the SM Higgs-mass parameter, 
$\cos\theta_W=m_W/m_Z$ is the ratio of the SM $W$-boson and $Z$-boson masses,
and $\tan\beta$ is the ratio of the two Higgs vacuum expectation values.\\[3mm]
A similar mixing phenomenon occurs in the charged sector, belonging to the 
charged weak bosons $W^{\pm}$ and the charged components of the Higgs doublets. 
The associated Dirac-fermion superpartners, i.e.~the charged Winos 
$\,\widetilde{W}^{\pm}$ and Higgsinos $\widetilde{H}_{u/d}^\pm$\,, mix to form 
charged mass eigenstates called charginos ($\widetilde{\chi}_{1,2}^\pm$\,, 
numbered in increasing mass order) as a result of the mixing in the chargino 
mass matrix \cite{Martin:1997ns}:

\begin{equation}
M_{\widetilde{\chi}^\pm} =  {\begin{pmatrix} 
M_2 & \sqrt{2} c_\beta c_{\theta_W} m_Z\\
\sqrt{2} s_\beta c_{\theta_W} m_Z & \mu  \label{cmm} 
\end{pmatrix}}~.
\vspace*{3mm}
\end{equation}
The mixing in the neutralino and chargino mass matrices stems from 
terms that go with the $Z$-boson mass. However, in the case that $M_1$, $M_2$ 
and $|\mu|$ largely exceed the mass of the $Z$-boson, the mixing terms are 
relatively small. If we neglect the mixing terms, the neutralinos are either a 
Bino, a Wino or a symmetric/antisymmetric mix of both Higgsino states, 
$\widetilde{H}_{S/A}^0 \equiv \frac{1}{\sqrt{2}}\,\bigl(\widetilde{H}_u^0
\,{}^{\displaystyle +}\!/\!{}_{\displaystyle -}\,\widetilde{H}_d^0\bigr)$. The charginos
are in that case either a charged Wino or a charged Higgsino. The composition 
for all possible regimes is shown in table~\ref{tab:regimes}. In this 
simplified case the mass of the Bino neutralino is $M_1$, the masses of the 
Wino neutralino and charginos are $M_2$, and the masses of the Higgsino 
neutralinos and charginos are $|\mu|$.\\[3mm]
In fact some of the eigenvalues of the mass matrices will turn out to be 
negative. For instance, $\widetilde{H}_{S}^0$ corresponds to the eigenvalue 
$-\mu$, whereas $\widetilde{H}_{A}^0$ corresponds to the opposite-sign 
eigenvalue $+\mu$. In order to arrive at a proper (non-negative) definition of 
the mass of all particles, an extra factor $\gamma^5$ will have to be absorbed
into the definition of the negative-mass eigenstates, which flips the sign of 
the corresponding mass eigenvalue. As we will see, this extra factor $\gamma^5$
has important consequences for the decay properties of the neutralinos. \\

\begin{table}[!h]
\centering
{\renewcommand{\arraystretch}{1.5}
\renewcommand{\tabcolsep}{0.2cm}
\begin{tabular}{ccccc}
\hline
Regime & Composition neutralinos & Composition charginos \\
\hline
$M_1 < M_2 < |\mu|$ & $(\widetilde{B}, \widetilde{W}, \widetilde{H}, 
\widetilde{H})$ & $(\widetilde{W}, \widetilde{H})$ \\
$M_1 < |\mu| < M_2$ & $(\widetilde{B}, \widetilde{H}, \widetilde{H}, 
\widetilde{W})$ & $(\widetilde{H}, \widetilde{W})$ \\
$|\mu| < M_1 < M_2$ & $(\widetilde{H}, \widetilde{H}, \widetilde{B}, 
\widetilde{W})$ & $(\widetilde{H}, \widetilde{W})$ \\
$|\mu| < M_2 < M_1$ & $(\widetilde{H}, \widetilde{H}, \widetilde{W}, 
\widetilde{B})$ & $(\widetilde{H}, \widetilde{W})$ \\
$M_2 < |\mu| < M_1$ & $(\widetilde{W}, \widetilde{H}, \widetilde{H}, 
\widetilde{B})$ & $(\widetilde{W}, \widetilde{H})$ \\
$M_2 < M_1 < |\mu|$ & $(\widetilde{W}, \widetilde{B}, \widetilde{H}, 
\widetilde{H})$ & $(\widetilde{W}, \widetilde{H})$ \\
\hline
\end{tabular}}
 \caption{Composition of the neutralinos ($\widetilde{\chi}_1^0\,,
          \widetilde{\chi}_2^0\,,\widetilde{\chi}_3^0\,,\widetilde{\chi}_4^0$\,) 
          and charginos ($\widetilde{\chi}_1^\pm,\widetilde{\chi}_2^\pm$).}\label{tab:regimes}
\end{table}

When we switch on the mixing again, mixed neutralino states consisting of Binos,
Winos and Higgsinos will exist. However, since the mixing is small, there will 
always be a part that dominates the state, which we then refer to as Binolike, 
Winolike or Higgsinolike. The true masses of all the neutralinos and charginos 
behave as in the previously discussed simplified case, which is governed by 
the three mass parameters $M_1$, $M_2$ and $\mu$.

\subsection{Neutralino and chargino decays into the lightest Higgs}

\begin{figure}[H]
  \centering
  \subfloat[BR$(\widetilde{\chi}^0_2 \rightarrow \widetilde{\chi}^0_1+h^0)$]
  {\label{fig:br21}\includegraphics[width=0.50\textwidth]{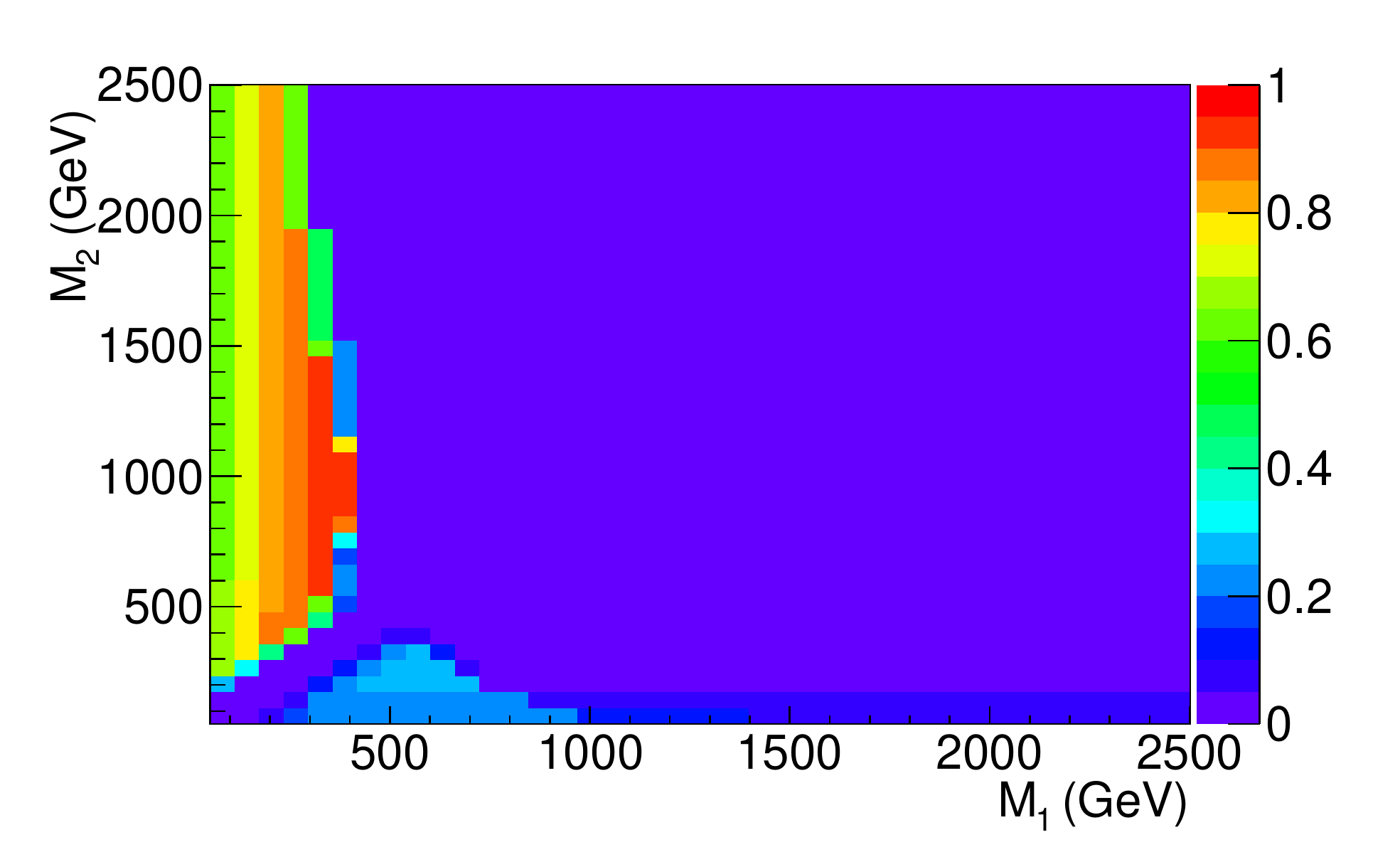}}
  \subfloat[BR$(\widetilde{\chi}^0_3 \rightarrow \widetilde{\chi}^0_1+h^0)$]
  {\label{fig:br31}\includegraphics[width=0.50\textwidth]{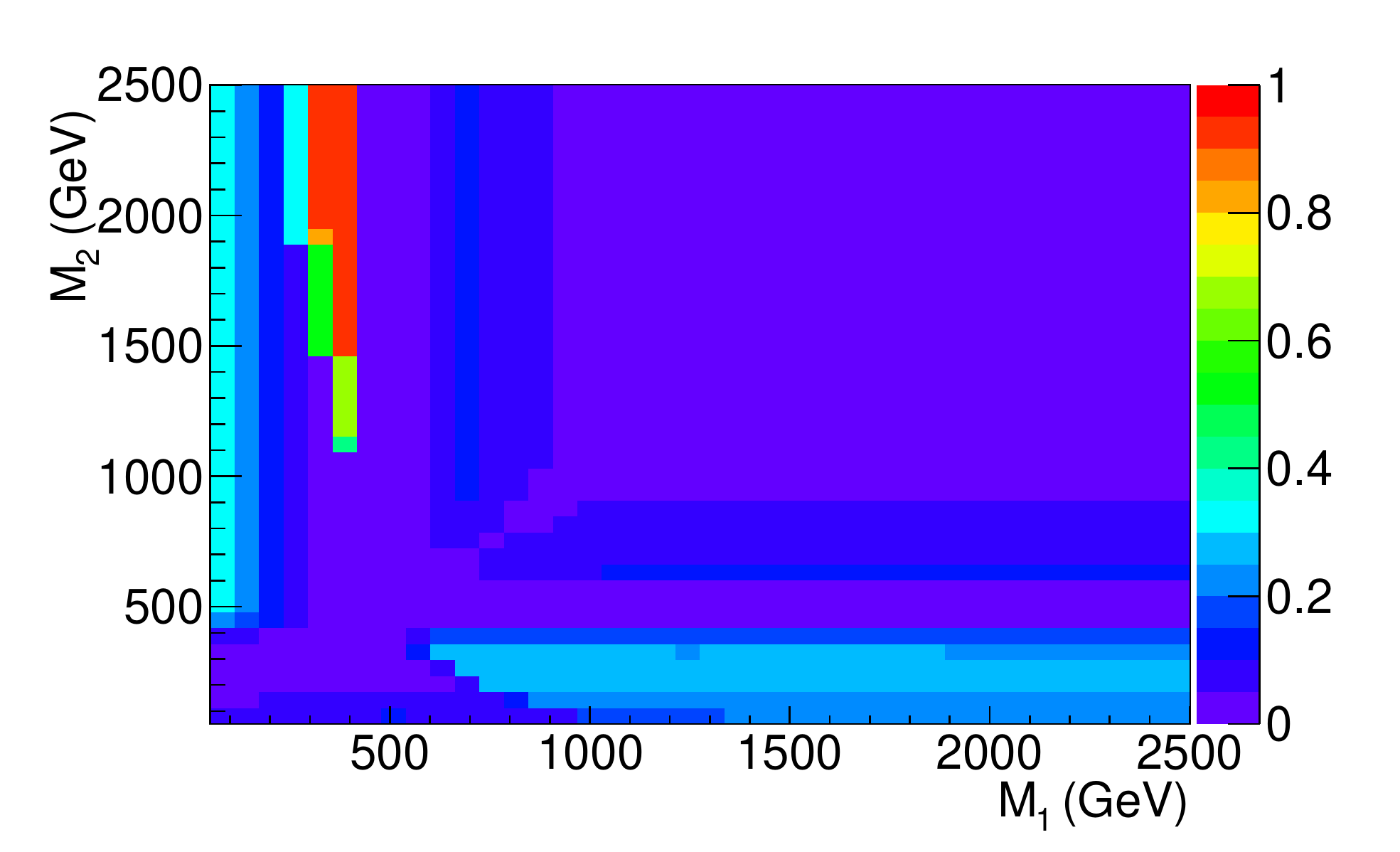}}\\
  \subfloat[BR$(\widetilde{\chi}^0_4 \rightarrow \widetilde{\chi}^0_1+h^0)$]
  {\label{fig:br41}\includegraphics[width=0.50\textwidth]{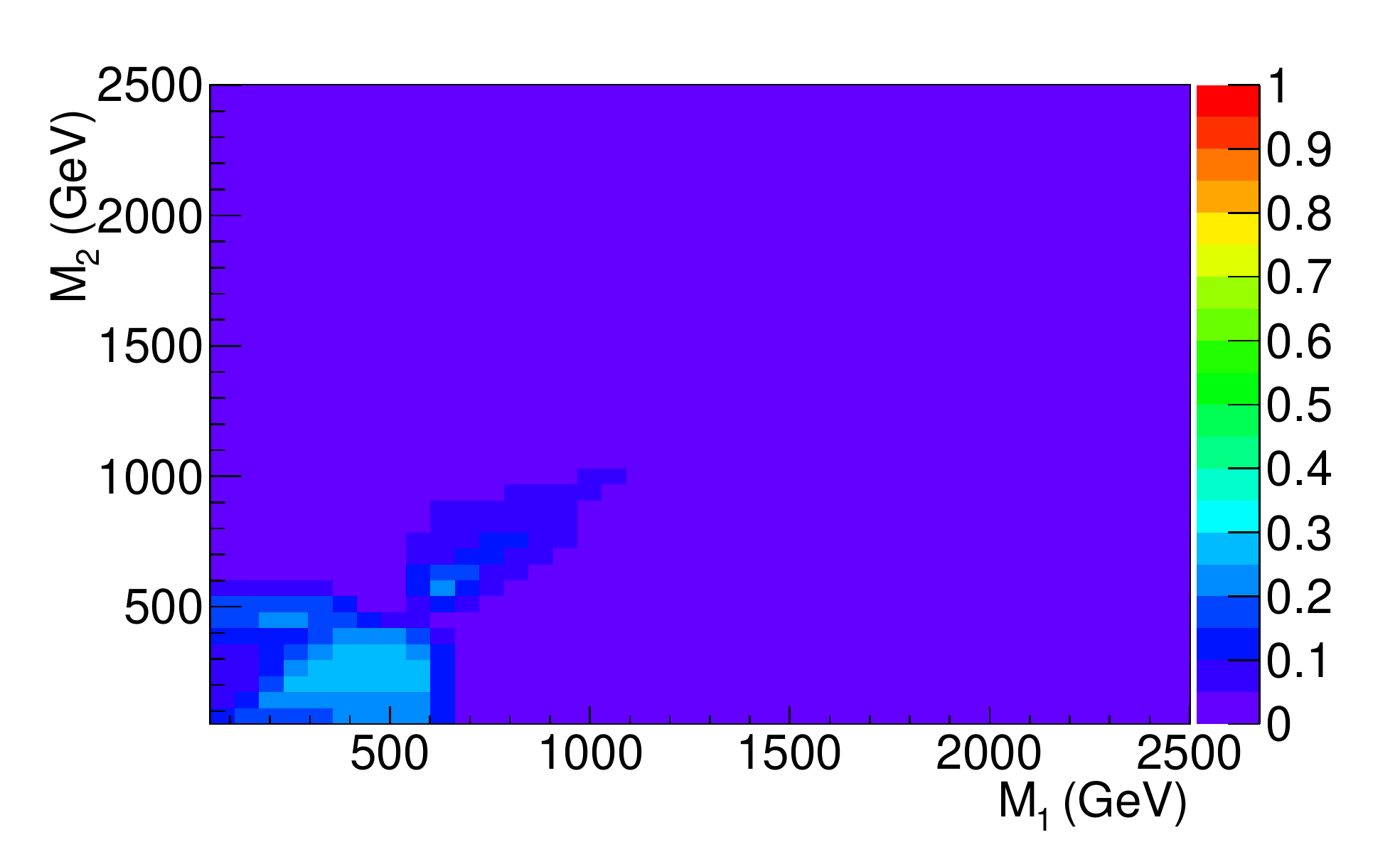}} 
  \subfloat[BR$(\widetilde{\chi}^\pm_2 \rightarrow \widetilde{\chi}^\pm_1+h^0)$]
  {\label{fig:brcc}\includegraphics[width=0.50\textwidth]{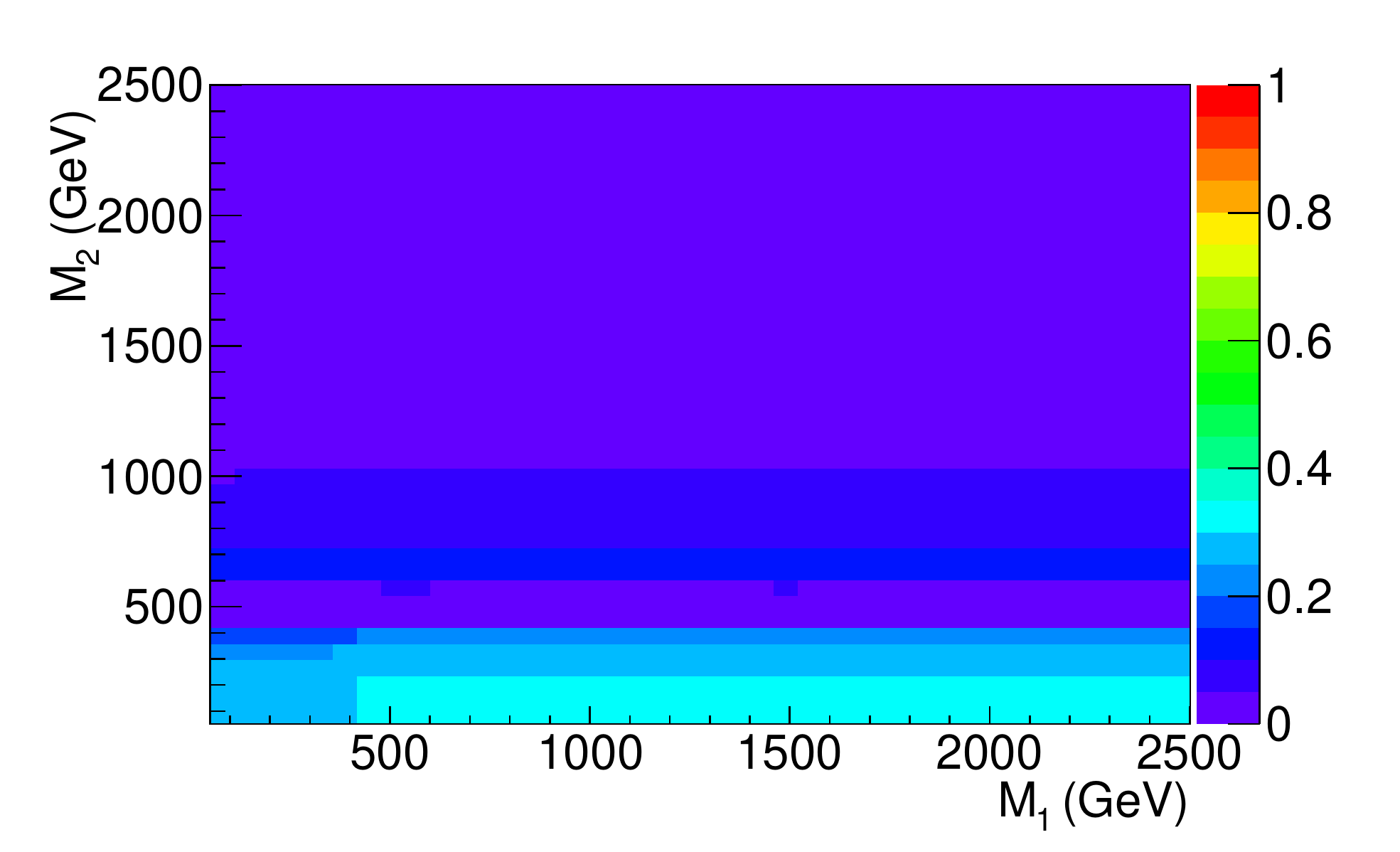}}
  \\[4mm]
  \caption{Prominent direct branching ratios for the decay of 
           neutralinos/charginos into the lightest Higgs boson and the lightest
           neutralino/chargino in the case that $\mu$ = 500 GeV, 
           $\tan\beta$ = 50 and all other parameters scaled up to very high 
           values.}
  \label{fig:br}
\end{figure}

If we choose the lightest neutralino to be the LSP, all supersymmetric 
particles will eventually decay into a lightest neutralino. The branching 
ratios of the most important direct decay channels of neutralinos into the 
lightest Higgs boson $h^0$ accompanied by a LSP are shown in 
figure~\ref{fig:br}. The lightest chargino plays an important role if it is of 
almost the same mass as the lightest neutralino. Therefore the branching ratio 
of figure~\ref{fig:brcc} is also included.\\[3mm]
Some of the features of these decay processes can be explained very well 
kinematically with the previously discussed simplified case. For example, the 
decay $\widetilde{\chi}^0_2 \rightarrow \widetilde{\chi}^0_1+h^0$ is very 
unlikely in the case that $M_1, M_2 > |\mu|$\, or when the smallest two 
parameters of the set $M_1$, $M_2$, $|\mu|$\, are relatively close (i.e.~less 
than $m_{h^0}$ apart), as can be seen in figure~\ref{fig:br21}. This is because 
both neutralinos have more or less the same mass in that case, which means that
the decay $\widetilde{\chi}^0_2 \rightarrow \widetilde{\chi}^0_1+h^0$ is 
kinematically not allowed. For the same reason the decay  
$\widetilde{\chi}^\pm_2 \rightarrow \widetilde{\chi}^\pm_1+h^0$ is greatly 
suppressed in the region around $M_2 \approx |\mu|$, as can be seen in 
figure~\ref{fig:brcc}. In figure~\ref{fig:br31} we see that a similar thing
holds for the decay $\widetilde{\chi}^0_3 \rightarrow \widetilde{\chi}^0_1+h^0$ 
for $M_2\approx|\mu|<M_1$\, or \,$M_1\approx|\mu|<M_2$, since in that case the 
lightest three neutralinos have more or less the same mass.

For some of the features of these decay processes, such as the apparent 
complementarity of 
BR$(\widetilde{\chi}^0_2 \rightarrow \widetilde{\chi}^0_1+h^0)$\, and 
\,BR$(\widetilde{\chi}^0_3 \rightarrow \widetilde{\chi}^0_1+h^0)$\, for 
$M_2>|\mu|>M_1$, we have to dig a little bit deeper. In order to facilitate the 
discussion we first list in table~\ref{gaugino_int} the possible interactions 
between the Binos, Winos and Higgsinos, from which the neutralinos and 
charginos inherit their decay properties. In order to identify the interactions
that involve the light Higgs boson, the two Higgs doublets are represented
by the associated five Higgs mass eigenstates $h^0,\,H^0,\,A^0$ and $H^\pm$.
The $h^0$ field corresponds to the light CP-even Higgs boson, whereas the other 
four fields correspond to the heavy CP-even, CP-odd and charged Higgs bosons,
respectively.\\[3mm]
\begin{table}[!h]
\centering
{\renewcommand{\arraystretch}{1.5}
\renewcommand{\tabcolsep}{0.2cm}
\begin{tabular}{c||c|c|c|c|c|c}
\hline
& $\widetilde{B}$ & $\widetilde{W}^3$ & $\widetilde{H}_S^0$ & $\widetilde{H}_A^0$
& $\widetilde{W}^\pm$ & $\widetilde{H}_{u/d}^\pm$\\
\hline\hline
$\widetilde{B}$ & & & $h^0,H^0,A^0$ & $h^0,H^0,A^0$ & & $H^\mp$\\
$\widetilde{W}^3$ & & & $h^0,H^0,A^0$ & $h^0,H^0,A^0$ & $W^\mp$ & $H^\mp$\\
$\widetilde{H}_S^0$ & $h^0,H^0,A^0$ & $h^0,H^0,A^0$ & & $Z$ & $H^\mp$ & $W^\mp$\\
$\widetilde{H}_A^0$ & $h^0,H^0,A^0$ & $h^0,H^0,A^0$ & $Z$ & & $H^\mp$ & $W^\mp$\\
$\widetilde{W}^\pm$ & & $W^\mp$ & $H^\mp$ & $H^\mp$ & $Z$ & $h^0,H^0,A^0$\\
$\widetilde{H}_{u/d}^\pm$ & $H^\mp$ & $H^\mp$ & $W^\mp$ & $W^\mp$ & $h^0,H^0,A^0$ & 
$Z$\\
\hline
\end{tabular}}
 \caption{Interactions between the Binos, Winos and Higgsinos. The entries 
          indicate which fields are involved in the interaction.}
 \label{gaugino_int}
\end{table}
Besides the kinematical observations mentioned before, we observe the following
features in the various mass domains. These features mainly involve the 
competition between the decay modes into the \,$h^0$ and alternative decay 
modes involving $W$ or $Z$ bosons.
\\[3mm]
{\it The Binolike-Winolike-Higgsinolike mass domain $M_1 < M_2 < |\mu|$\,:} 
\begin{itemize}
\item The decay $\widetilde{\chi}^0_2 \rightarrow \widetilde{\chi}^0_1+h^0$ 
      tends to dominate the branching ratio of $\widetilde{\chi}_2^0$ if it is 
      kinematically allowed, resulting in values for the branching ratio
      BR$(\widetilde{\chi}^0_2 \rightarrow \widetilde{\chi}^0_1+h^0)$\,
      that can get close to unity (see figure~\ref{fig:br21}). As can be read 
      off from table~\ref{gaugino_int}, this is caused by the fact that the 
      decay into the $h^0$ involves the (suppressed) Higgsino component of 
      either the Binolike $\widetilde{\chi}^0_1$ or Winolike 
      $\widetilde{\chi}^0_2$\,, whereas the decay into a $Z$ boson involves the 
      (double suppressed) Higgsino components of both these neutralinos. 
      At the same time the decay 
      $\widetilde{\chi}^0_2 \rightarrow \widetilde{\chi}^{\pm}_1+W^{\mp}$ is 
      not allowed  kinematically since 
      $m_{\widetilde{\chi}_1^\pm}\approx m_{\widetilde{\chi}_2^0}$\,.
\item The branching ratios
      BR$(\widetilde{\chi}^0_{3,4} \rightarrow \widetilde{\chi}^0_1+h^0)$ can only
      reach values that are substantially smaller (see figures~\ref{fig:br31} 
      and~\ref{fig:br41}), since the alternative decay modes
      \,$\widetilde{\chi}^0_{3,4} \rightarrow \widetilde{\chi}^{\pm}_1+W^{\mp}$\, 
      and \,$\widetilde{\chi}^0_{3,4} \rightarrow \widetilde{\chi}^0_2+h^0/Z$\, 
      seriously reduce the maximum branching ratio for direct decays into the 
      LSP. Note, however, that the $\widetilde{\chi}_{3,4}^0$ neutralinos might 
      be of interest for di-Higgs decay modes in view of the possible two-step
      decays $\widetilde{\chi}^0_{3,4} \rightarrow \widetilde{\chi}^0_2+h^0$\, 
      followed by \,$\widetilde{\chi}^0_2 \rightarrow \widetilde{\chi}^0_1+h^0$. 
\end{itemize}
{\it The Winolike-Binolike-Higgsinolike mass domain $M_2 < M_1 < |\mu|$\,:}
\\[2mm]
the same arguments in principle apply to this mass domain. However, in this case
the (single suppressed) decay mode
$\widetilde{\chi}^0_2 \rightarrow \widetilde{\chi}^{\pm}_1+W^{\mp}$ cannot be
avoided since \,$m_{\widetilde{\chi}_1^\pm}\approx m_{\widetilde{\chi}_1^0}$\,.
As a result BR$(\widetilde{\chi}^0_2 \rightarrow \widetilde{\chi}^0_1+h^0)$
will at most reach 0.3 in this mass regime.\\[3mm]
{\it The Binolike-Higgsinolike-Winolike mass domain $M_1<|\mu|<M_2$\,:}
\begin{itemize}
\item  We observe both very large (almost unity) and very small branching 
       ratios for the decay into the $h^0$, with as additional striking 
       feature the apparent complementarity of 
       BR$(\widetilde{\chi}^0_2 \rightarrow \widetilde{\chi}^0_1+h^0)$\, and 
       \,BR$(\widetilde{\chi}^0_3 \rightarrow \widetilde{\chi}^0_1+h^0)$
       (see figures~\ref{fig:br21} and~\ref{fig:br31}). 
       This has to do with the occurrence of negative-mass eigenstates in the 
       Higgsino sector and the associated factor $\gamma^5$ that is introduced 
       in order to flip the sign of the mass eigenvalue. 
       If $\widetilde{\chi}_2^0$ corresponds to a genuine positive-mass 
       eigentstate, then 
       $\widetilde{\chi}^0_2 \rightarrow \widetilde{\chi}^0_1+h^0$ is an 
       unsuppressed scalar decay mode (see table~\ref{gaugino_int}) that tends 
       to dominate the single suppressed decay into a $Z$ boson. 
       If $\widetilde{\chi}_2^0$ corresponds to a negative-mass eigenstate, then
       \,$\widetilde{\chi}^0_2 \rightarrow \widetilde{\chi}^0_1+h^0$\, is a
       velocity-suppressed pseudoscalar decay mode and this time the decay mode
       into a $Z$ boson dominates. The observed complementarity follows from 
       the fact that $\widetilde{\chi}_2^0$ and $\widetilde{\chi}_3^0$ 
       correspond to opposite-sign mass eigenvalues, while at the same time the
       suppression factors are such that the role of $h^0$ and $Z$ are 
       effectively interchanged in the two cases~\cite{Arbey:2012fa}. 
\item The branching ratio
      BR$(\widetilde{\chi}^0_4 \rightarrow \widetilde{\chi}^0_1+h^0)$ is even
      more reduced than in the previously discussed mass domains 
      (see figure~\ref{fig:br41}). This is caused by the larger number of 
      competing alternative decay modes, i.e.
      $\widetilde{\chi}^0_4 \rightarrow \widetilde{\chi}^{\pm}_1+W^{\mp}$ and
      \,$\widetilde{\chi}^0_4 \rightarrow \widetilde{\chi}^0_{2,3}+h^0/Z$.
      Note, though, that the $\widetilde{\chi}_4^0$ neutralino might be of
      interest for di-Higgs decay modes in view of the possible two-step decays 
      $\widetilde{\chi}^0_4 \rightarrow \widetilde{\chi}^0_{2,3}+h^0$\, followed 
      by \,$\widetilde{\chi}^0_{2,3} \rightarrow \widetilde{\chi}^0_1+h^0$. 
\end{itemize}
{\it The Winolike-Higgsinolike-Binolike mass domain $M_2<|\mu|<M_1$\,:}\\[2mm]
the previous arguments in principle apply to this mass domain as well. However,
in this case the decay modes
\,$\widetilde{\chi}^0_{2,3} \rightarrow \widetilde{\chi}^{\pm}_1+W^{\mp}$ cannot be
avoided. This reduces the maximum combined branching ratio for the decays
into $h^0$ and $Z$ to roughly 0.3. For this maximum combined 
branching ratio again a complementarity phenomenon is observed 
(see figures~\ref{fig:br21} and~\ref{fig:br31}).\\[3mm]
{\it The Higgsinolike LSP mass domain $|\mu|<M_{1,2}$\,:}
\begin{itemize}
\item As mentioned before, the decay
      $\widetilde{\chi}^0_2 \rightarrow \widetilde{\chi}^0_1+h^0$ is not allowed
      kinematically since both neutralinos have more or less the same mass in 
      the Higgsinolike LSP case.
\item The branching ratio 
      BR$(\widetilde{\chi}^0_3 \rightarrow \widetilde{\chi}^0_1+h^0)$ is
      strongly reduced, since the alternative decay modes 
      $\widetilde{\chi}^0_3 \rightarrow \widetilde{\chi}^{\pm}_1+W^{\mp}$ and
      \,$\widetilde{\chi}^0_3 \rightarrow \widetilde{\chi}^0_2+h^0/Z$ cannot be
      avoided as a result of the approximate mass-degeneracy of 
      $\widetilde{\chi}_1^0$\,, $\widetilde{\chi}_2^0$ and 
      $\widetilde{\chi}^{\pm}_1$. Moreover, based on the above-given 
      complementarity discussion we know that at least one of the 
      \mbox{$\widetilde{\chi}^0_3 \rightarrow \widetilde{\chi}^0_2+h^0/Z$} 
      decay modes will not be suppressed. Note, though, that
      $\widetilde{\chi}^0_3 \rightarrow \widetilde{\chi}^0_2+h^0$ in itself 
      already constitutes a direct SUSY decay channel into the lightest Higgs 
      boson. 
\end{itemize}
{\it The chargino decays\,:}
\begin{itemize}
\item The lightest chargino can never decay directly into the lightest Higgs
      boson. This is due to the fact that R-parity conservation forbids a decay
      to a neutral Higgs boson. For this a lighter charged supersymmetric 
      fermion is needed, which is not present in the case of the lightest 
      chargino. It is possible for a lightest chargino to decay 
      into three particles, but such three-particle chargino decay modes with 
      a Higgs boson featuring in the final state are rather rare.
\item For $M_2<|\mu|$ the branching ratio 
      BR$(\widetilde{\chi}^\pm_2 \rightarrow \widetilde{\chi}^\pm_1+h^0)$ can 
      reach maximum values of up to 0.35 as a result of the competition from 
      the unavoidable decay mode 
      $\widetilde{\chi}^\pm_2 \rightarrow \widetilde{\chi}^0_1+W^\pm$ as well as
      the decay mode 
      $\widetilde{\chi}^\pm_2 \rightarrow \widetilde{\chi}^\pm_1+Z$. The total
      branching ratio for multi-step decays into the lightest Higgs boson can 
      substantially exceed 0.35 in view of the possibility of two-step decays
      of the form
      $\widetilde{\chi}^\pm_2 \rightarrow \widetilde{\chi}^0_2+W^\pm$
      followed by 
      $\widetilde{\chi}^0_2 \rightarrow \widetilde{\chi}^0_1+h^0$\, if
      \,$M_{1,2}<|\mu|$. 
\item For $|\mu|<M_2$ the branching ratio 
      BR$(\widetilde{\chi}^\pm_2 \rightarrow \widetilde{\chi}^\pm_1+h^0)$ gets
      additionally reduced by the alternative decay mode to the second 
      Higgsinolike neutralino. Again multi-step decay modes can substantially 
      enhance the branching ratio for the decay into the $h^0$. 
\end{itemize}
In conclusion, the branching ratios for direct decays of neutralinos/charginos
into the LSP and the lightest Higgs boson can be pretty large, reaching maximum
values close to one for $\widetilde{\chi}_{2,3}^0$. For $\widetilde{\chi}^\pm_1$ 
there is effectively no decay into the lightest Higgs boson. For the heavier 
states $\widetilde{\chi}_4^0$ and $\widetilde{\chi}^\pm_2$ the direct-decay 
branching ratios can reach 0.35 at best. However, for these heavy SUSY particles
the total branching ratio for multi-step decays into the LSP and the lightest 
Higgs boson can be substantially larger if the non-Higgs decay step gives rise 
to $\widetilde{\chi}_{2,3}^0$\,, which can subsequently decay into the lightest 
Higgs boson with high probability.

\subsection{Sfermion decays into the lightest Higgs}

Next we give a brief summary of the other supersymmetric decay channels that
can produce a lightest Higgs boson, starting with the
sfermions (squarks and sleptons), the scalar superpartners of the SM fermions
(quarks and leptons). Such decay modes will play a role later on when the
masses of the sfermions are not artificially scaled up to very high values.
In this context it should also be noted that, apart from the interactions 
listed in table~\ref{gaugino_int}, the Binos, Winos and Higgsinos can also 
decay into fermion--sfermion pairs, involving the Yukawa interactions.\\[3mm]
Since the Wino couples only to left-handed sfermions, the decays of left-handed
($\widetilde{f}_L$) and right-handed ($\widetilde{f}_R$) sfermions are 
different. In addition, the couplings to Higgsinos are Yukawa suppressed. 
This results in a profound difference between the decays of 1st/2nd generation 
sfermions and 3rd generation sfermions, since only the latter may have a large 
coupling to the Higgsinos.\\[3mm] 
{\it Direct decays of sfermions into the lightest Higgs boson\,:}
\begin{itemize}
\item First of all there is the possibility to have a mass difference between 
      left- and right-handed sfermions. As a result, there is the possibility 
      for $\widetilde{f}_{L,R} \rightarrow \widetilde{f}_{R,L} + h^0$ decay modes
      if the mass difference between the left- and right-handed sfermions 
      exceeds 125~GeV. The couplings involved in this decay mode are Yukawa 
      suppressed in the pMSSM. Therefore, the direct decay is mostly relevant 
      for 3rd generation sfermions.
\item The sfermions of the 3rd generation are mixtures of left- and right-handed
      states, indicated by $\widetilde{f}_{1,2}$ (numbered in increasing mass 
      order). Therefore, there is an additional possibility for a direct decay
      via $\widetilde{f}_{2} \rightarrow \widetilde{f}_{1} + h^0$. This decay can
      involve a non-Yukawa-suppressed (gauge) coupling between two left- or two 
      right-handed components of the sfermion mass eigenstates. For 3rd 
      generation squarks the gauge coupling and the Yukawa coupling can be of 
      the same order of magnitude, which can lead to unexpected cancellations
      between both direct $h^0$ production mechanisms in that case. 
\end{itemize}
{\it Indirect decays of 1st/2nd generation sfermions into the lightest Higgs 
     boson\,:}\\[2mm]
sfermions can decay to heavy neutralinos or the heavy chargino, which can
subsequently decay into lighter neutralinos or charginos and the lightest Higgs
boson. The decay pattern differs for the left- and right-handed sfermions, 
depending on the composition of the LSP.
\begin{itemize}
\item {\it Winolike LSP:}\, the direct decay of the right-handed sfermions to 
      the LSP is suppressed. If kinematically allowed the right-handed sfermions
      will decay to the Binolike neutralino, with the decay to the Higgsinolike
      states being Yukawa suppressed. As explained above, this Binolike 
      neutralino can decay with a moderately large branching ratio to the $h^0$,
      since the decay to the $Z$ boson is double suppressed. The left-handed 
      sfermions predominantly decay to the LSP, which strongly reduces indirect
      decays into the lightest Higgs boson.
\item {\it Higgsinolike LSP:}\, the decay of the right and left-handed 1st/2nd
      generation sfermions to the LSP is Yukawa suppressed. If possible, these 
      sfermions will decay to the heavier Bino- or Winolike states. 
      As explained above, these states can decay with reduced branching 
      fraction to the $h^0$.
\item {\it Binolike LSP:}\, if the $\widetilde{\chi}^0_1$ is Binolike, the
      right-handed sfermions predominantly decay to the LSP. However, the
      left-handed sfermions still prefer to decay (if kinematically allowed) to
      the heavier Winolike neutralino/chargino. This is caused by an intrinsic 
      $c_{\theta_W}/s_{\theta_W}\approx 1.9$ enhancement factor of the weak coupling 
      of sleptons compared to the hypercharge coupling, with an additional 
      factor 3 enhancement for squarks. As explained above, in these models the
      Winolike neutralino can have a large branching ratio to Higgs bosons. 
      If it is a $\widetilde{\chi}^0_4$\,, then decays to charginos are also 
      possible. If it is a $\widetilde{\chi}^0_2$\,, then its branching ratio
      to $h^0$ bosons is potentially very large and can be close to unity.
\end{itemize}
For 3rd generation sfermions the couplings to the Higgsinolike states are not
Yukawa suppressed anymore and can even become large for top squarks. This 
results in a richer structure of possible decay modes and a more prominent role
of Higgsinolike states as decay products. In that case indirect Higgs production
can also become important in scenarios where Higgsinolike states have a large 
branching ratio into the lightest Higgs boson, as described in the previous 
subsection. 

\subsection{Heavy Higgs-boson decays into the lightest Higgs}

Also the heavy Higgs particles can decay into the $h^0$. These particles are a 
consequence of SUSY, which requires more than one Higgs doublet, 
but as far as R-parity is concerned they qualify as ``SM''
particles. Consequently, these particles do not necessarily have to decay into 
the LSP and therefore do not necessarily give rise to large
missing transverse momentum in their decay chains. A comprehensive overview of 
the decays of the heavy Higgs bosons is given in~\cite{Djouadi:2005gj}.

\begin{itemize}
\item {\it Direct decays:}\, as will be discussed later, all surviving MSSM 
      models have $M_A$ values exceeding $300$~GeV. This is known as the 
      decoupling limit (large $M_A$) and consequently all heavy Higgs bosons
      have similar masses, which blocks decays among heavy Higgs bosons.
      The heavy CP-even Higgs boson $H^0$ can directly decay to two $h^0$ 
      bosons. The CP-odd Higgs boson $A^0$ can decay to $h^0 Z$. The charged 
      Higgs bosons $H^{\pm}$ can decay to $W^{\pm} h^0$. The corresponding 
      branching ratios tend to be rather small, because in most surviving MSSM 
      models $\tan\beta$ is relatively large (\,$>10$) and consequently the 
      decays of the heavy Higgs bosons to $b$-quarks are dominant. Later on we 
      will encounter a couple of exceptional models that have the lowest values
      for $M_A$ and at the same time a relatively small value for $\tan\beta$ in
      order to survive the experimental constraints. Such models have noticeable
      branching ratios for direct heavy Higgs-boson decays into the lightest 
      Higgs. More details can be found in Ref.~\cite{Djouadi:2005gj}.
\item {\it Indirect decays:}\, the heavy Higgs bosons also have the possibility
      to decay into heavy neutralinos or charginos (if kinematically allowed), 
      especially if one of these decay states is Higgsinolike and the other 
      Bino- or Winolike (see table 2). Those states can subsequently decay to
      $h^0$, sometimes with high branching ratios. This can, for instance, 
      result in di-Higgs production from an $A^0$ decay.
\end{itemize}
 %the chapter of Wim
\section{Finding candidate pMSSM models: simulation and constraints}

The MSSM has more than 100 free parameters. Most of those parameters
are not relevant for LHC physics. In the pMSSM the free parameters are
reduced to 19 by demanding CP-conservation, 
minimal flavour violation and degenerate mass spectra for the 1st and 2nd 
generations of sfermions.
The LSP is required to be the neutralino $\widetilde{\chi}_1^0$ in
order to have a viable dark-matter candidate. This reduced model should cover
a large fraction of the relevant SUSY phase space for $h^0$ production.
The 19 remaining parameters are 
10 sfermion masses\footnote{The corresponding sfermion labels are 
\ensuremath{\widetilde{Q}_{\mathrm{1}}}, 
\ensuremath{\widetilde{Q}_{\mathrm{3}}}, 
\ensuremath{\widetilde{L}_{\mathrm{1}}}, 
\ensuremath{\widetilde{L}_{\mathrm{3}}}, 
\ensuremath{\widetilde{u}_{\mathrm{1}}}, 
\ensuremath{\widetilde{d}_{\mathrm{1}}}, 
\ensuremath{\widetilde{u}_{\mathrm{3}}}, 
\ensuremath{\widetilde{d}_{\mathrm{3}}}, 
\ensuremath{\widetilde{e}_{\mathrm{1}}} and 
\ensuremath{\widetilde{e}_{\mathrm{3}}}. Here 1 indicates the light-flavoured 
mass-degenerate 1st and 2nd generation sfermions and 3 the heavy-flavoured 
3rd generation. The labels $\widetilde{Q}$ and $\widetilde{L}$ refer to the 
superpartners of the left-handed fermionic $SU(2)$ doublets, whereas the other 
labels refer to the superpartners of the right-handed fermionic $SU(2)$ 
singlets.}, 3~gaugino masses $M_{1,2,3}$\,, the ratio of the Higgs vacuum 
expectation values $\tan\beta$, the Higgsino mixing parameter $\mu$, 
the mass $m_A$ of the CP-odd Higgs-boson $A^0$ and 3~trilinear scalar couplings 
$A_{b,t,\tau}$. 

\subsection{Generation and pre-selection of pMSSM model-sets}
SUSY-HIT \cite{Skands:2003cj} is used to generate the particle spectra of the 
19-parameter pMSSM models. Only models are selected with a neutralino
as LSP. The Higgs mass has been precisely determined by ATLAS and CMS to be 
$125.4$ (ATLAS \cite{Aad:2014aba}) and $125.0$~GeV (CMS \cite{CMS:2014ega}) 
with uncertainties of $0.3-0.4$~GeV for each experiment. We select only models 
with a lightest Higgs boson $h^0$ within the range:

\begin{equation}
124.4 \text{ GeV} \le m_{h^0} \le 126.5 \text{ GeV}~.
\end{equation} 

In addition we produce two statistically independent sets of models:
\begin{itemize}

\item{Set A: Higgs production via direct decay of an arbitrary SUSY particle
or a heavy Higgs boson.} 
\\
As described in the previous section, Higgs production can occur via 
various different decays of SUSY particles.
In addition, $h^0$ bosons can be produced in
the decay of heavy Higgs bosons.
For this set we require in the preselection that at least one SUSY particle 
or heavy Higgs boson has a direct branching ratio to $h^0$ exceeding 20\%. 

\item{Set B: Higgs production via direct decays of charginos or neutralinos.}\\
Since Higgs production via neutralino or chargino decays 
is most important, a second set of models dedicated to 
these decays is produced. For this set we required that at
least one of the following direct branching ratios exceeds 20\%:
\begin{alignat}{2}
\begin{split}
&\text{BR}( \widetilde{\chi}^0_{2,3,4} \rightarrow \widetilde{\chi}^0_1+h^0 ) 
\,>\, 0.2 
\quad\text{ or} \\
&\text{BR}(\widetilde{\chi}^\pm_2\!\rightarrow\widetilde{\chi}^\pm_1\!+h^0 )
\,>\, 0.2~.
\end{split} \label{cons_2}
\end{alignat}
\end{itemize}

The advantage of set B is that less model points are needed to study the most
relevant Higgs production modes, since Higgs production predominantly 
originates from the decay of a heavy neutralino or chargino. Those neutralinos
and charginos can be directly produced or they are produced in cascade decays 
of predominantly squarks or gluinos, since these coloured SUSY particles can 
have a large cross section. The advantage of set A is its larger coverage of 
possible 3rd-generation and heavy-Higgs decay modes. 

\subsection{Parameter space coverage with a particle filter}
This study has not the objective to provide a statistical interpretation like 
a ``coverage'' or a ``likelihood'' for a given parameter region. The objective 
is to find regions in the parameter space that are consistent with the global 
constraints on SUSY and where in addition the production of $h^0$ bosons is 
large (or close to maximal) in order to determine possible rates and topologies
for SUSY Higgs production at the LHC. Each of our parameter sets represents
a viable model point that could be realized in nature. \\[3mm]
We use a simplified two-step particle filter algorithm \cite{particlefilter} to
find model points in the pMSSM parameter space. 
\begin{enumerate}
\item First the 19 parameters of the pMSSM (3 gaugino masses, 6 squark masses, 
4 slepton masses, 3 trilineair couplings, $M_A$, $\mu\,$ and $\,\tan\beta$)
are chosen randomly from a flat prior distribution. The squark and slepton 
masses and $M_A$ have ranges between 100~GeV and 3000~GeV. The Higgsino mixing 
term $\mu$, which in principle can be negative, ranges between -3000~GeV and 
3000 GeV. 
This is also the case for the trilineair couplings, although we choose the 
ranges in that case between -5000~GeV and 5000~GeV to be sure that the trilinear
couplings will not restrict the simulation too much. The lower bound on the 
gaugino masses is chosen to be 10~GeV to ensure that neutralinos, charginos and
gluinos with very low masses are also evaluated. Finally, the ratio $\tan\beta$
of the Higgs vacuum expectation values is chosen between 1 and 50. For each set
of pMSSM parameters SUSY-HIT \cite{Skands:2003cj} is used to generate the SUSY 
particle spectra and mixing matrices. Subsequently the preselection criteria 
of the previous subsection are checked. Model-sets are generated randomly 
within the given parameter range until we find 10000 model-sets fulfilling the 
preselection requirements.
\item These model-sets are then used as seeds (or particles) to build 
a posterior probability distribution from which further model-sets are 
generated. The posterior probability distribution is chosen as a sum of 
multi-dimensional Gaussian distributions centered around the parameter values 
$\vec{S}$ of each seed point. The multi-dimensional width of the Gaussian 
distributions is set to $10\%$, $25\%$ and $40\%$ of $\vec{L}_d$, where $L_d$ 
is the extent of the parameter space in dimension $d$ out of 19. Around each 
seed further models are generated. The three sets are compared in order to 
evaluate the dependence on the width of the sampling. A comparison of the width
dependence and a comparison of sets A and B is shown in 
figure~\ref{fig:width_comparison} in the appendix. Since no
significant difference is found all sets have been merged.
This simulation process continues until in
total at least 250000  models survive all preselection criteria.
\end{enumerate}

\subsection{Experimental constraints}

The code micrOMEGAs \cite{Belanger:2001fz} is used to calculate specific 
observables for each model-set in order to compare them with the experimental 
restrictions. The following constraints impact especially the neutralino and
chargino mixing and can also influence their decay to the lightest Higgs boson.

\begin{itemize}
\item From the WMAP and the Planck data we adopt the cold dark-matter (DM) 
relic density in the universe \cite{Komatsu:2008hk,Ade:2013zuv}.
We select a region corresponding to the last Planck central value
$0.1186 \pm 0.0031$ including an 10\% (upper) and 20\% (lower) theoretical 
uncertainty%
\footnote{A value of approximately 10\% is due to uncertainties entering the
          calculations of the relic density from SUSY parameters. The lower 
          uncertainty is slightly larger, in order to include the possibility 
          to have small additional sources of Dark Matter.}:
\begin{equation}
0.094 < \Omega_ch^2 < 0.131~. \vspace*{2mm}
\end{equation}
\item The limits from the 85.3 days Large Underground Xenon (LUX) data
\cite{Akerib:2013tjd} are taken into account. 
To compare the calculated proton/neutron cross sections 
$\sigma^{(p/n)}$ to the experimental limits, we use a normalized cross section 
for a point-like nucleus \cite{Belanger:2010cd}:
\begin{equation}
\sigma \,= \, \frac{\bigl(Z\sqrt{\sigma^{(p)}}+(A-Z)\sqrt{\sigma^{(n)}}\,\bigr)^2}
                   {A^2}~,
\end{equation}
with $A$ and $Z$ the mass number and atomic number of the target. 
In our case the target is xenon with $A = 131$ and $Z=54$.
\item We implement the LHCb and CMS measurements of the 
branching ratio of the strange B meson to two muons
~\cite{Aaij:2013aka,Chatrchyan:2013bka} by demanding
\begin{equation}
BR(B_s^0 \rightarrow \mu^+ \mu^-) = (\,3.0^{\,+\,1.0}_{\,-\,0.9}\,) \times 10^{-9}~.
\end{equation} 
\item 
We impose the LEP limits on the invisible width of the $Z$ boson and on the 
SUSY particle masses as implemented in micrOmegas. 
\end{itemize}

\begin{figure}
\centering
\includegraphics[width = \textwidth]{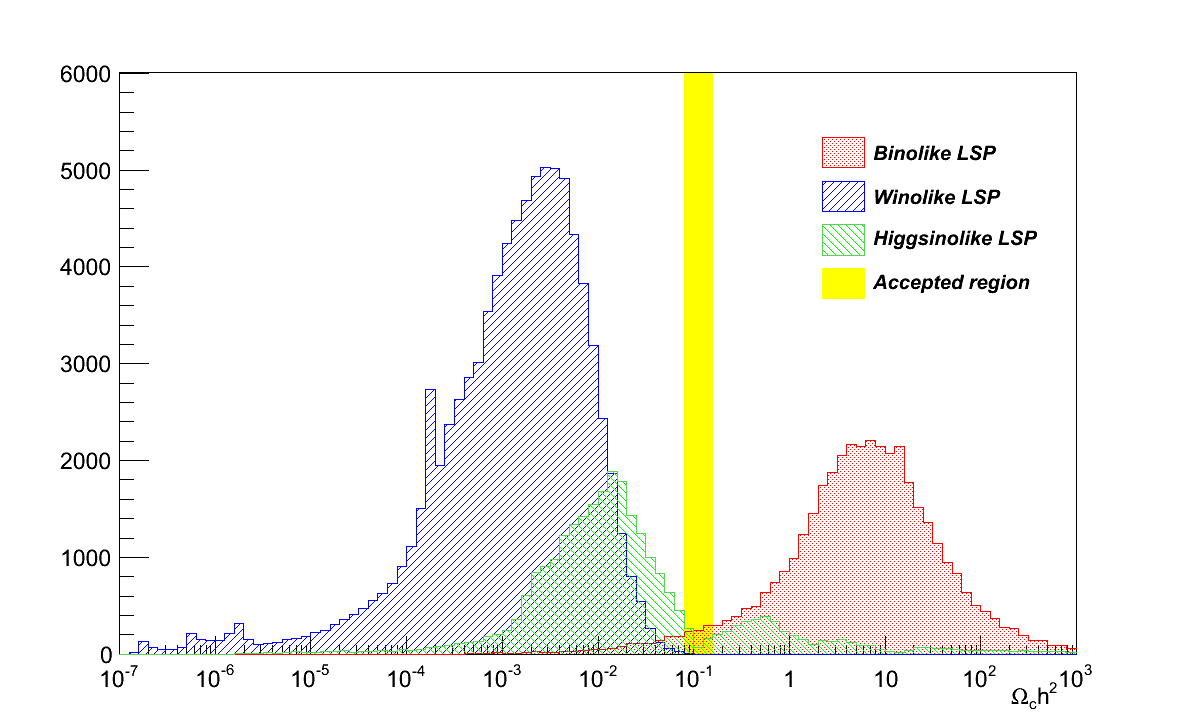}
\caption{Dark Matter relic density  $\,\Omega_ch^2$  obtained from the 19 parameter pMSSM models compared with the accepted region.
         The number of models is shown as a function of $\,\Omega_ch^2$.}
\label{fig:wmap_all}
\end{figure}

The WMAP/Planck results place severe constraints on the models as can be 
seen in figure~\ref{fig:wmap_all}. The LSP's of the surviving models turn out 
to be mostly Binolike, with a relatively low mass, and to a lesser extent 
Higgsinolike or Winolike, with a relatively high mass. 
This is caused by the possibility of coannihilation of the LSP 
together with the lightest chargino or next-to-lightest neutralino, which is 
mostly absent for Binolike LSP's. In order to reduce the efficiency of the 
coannihilation and have a DM relic density that is not too low, Higgsinolike 
and Winolike LSP's are substantially heavier than Binolike LSP's. 
The occurrence of Winolike LSP's is suppressed within the simulated parameter 
space, since in that case the coannihilation turns out to be very efficient. 
Among the useful models that survive the WMAP/Planck constraint we have found 
only a few with a Winolike LSP. \\[3mm]
\begin{figure}
\centering
\includegraphics[width = \textwidth]{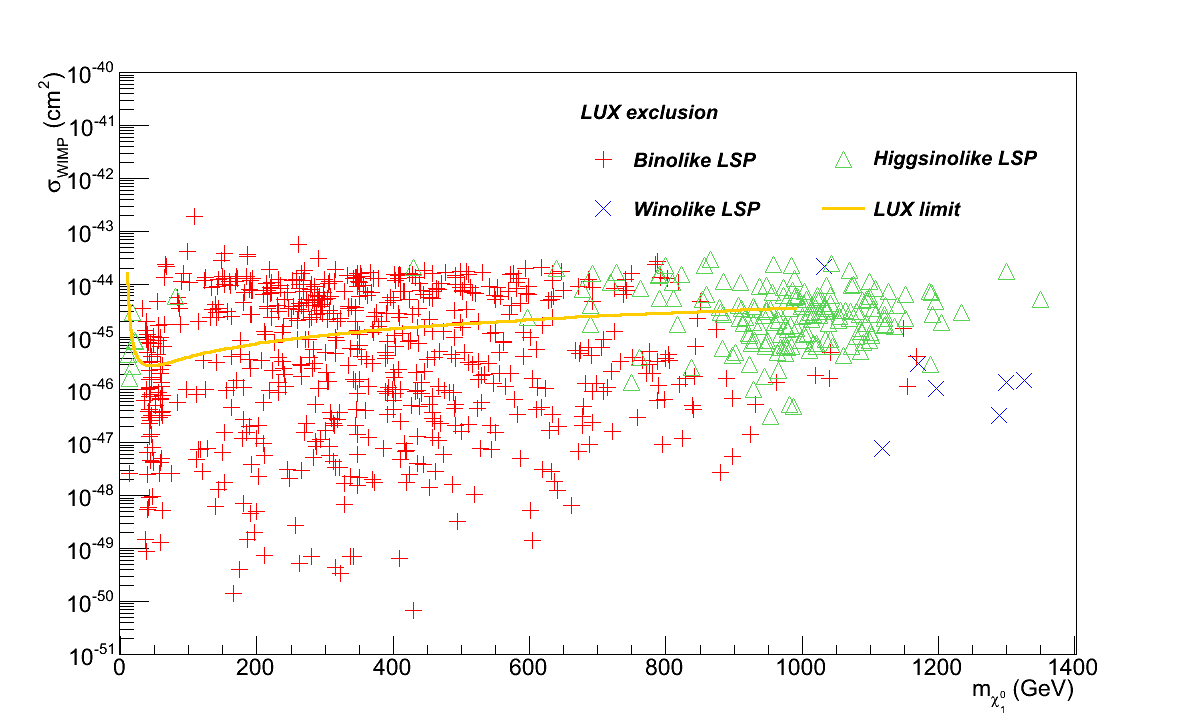}
\caption{Upper limits on the LSP\,--\,nucleon scattering cross section for the 
         LUX experiment, compared with the 19 parameter pMSSM after the 
         WMAP/Planck constraint. LEP constraints are not yet applied.}
\label{fig:luxx_all}
\end{figure}
Having checked the impact of the WMAP/Planck constraint, we now impose the LUX 
limits on the surviving models. The additional impact of the LUX limits is much 
smaller, as can be seen in figure~\ref{fig:luxx_all} where the LUX  
experimental limits are imposed. Given the 
WMAP/Planck and LUX constraints, the $B_s$ and LEP constraints have little 
additional impact on the number of viable models. Notable exceptions are the 
surviving models in figure~\ref{fig:luxx_all} with a very light Higgsinolike 
LSP, which are removed by the LEP constraints on the lightest chargino mass.
\\[3mm]
After a first iteration about $250$ models survived. These models were
again used to construct a posterior for a second particle-filter iteration, 
producing in total about 430 models that fulfil all constraints. 
As expected the success rate with this posterior is increased.

\subsection{ATLAS constraints: event generation, fast simulation and analysis}\label{sec:ATLASconstraints}

\begin{figure} [H]
\centering
  {\label{fig:sqsg_2}
  \includegraphics[width=\textwidth]{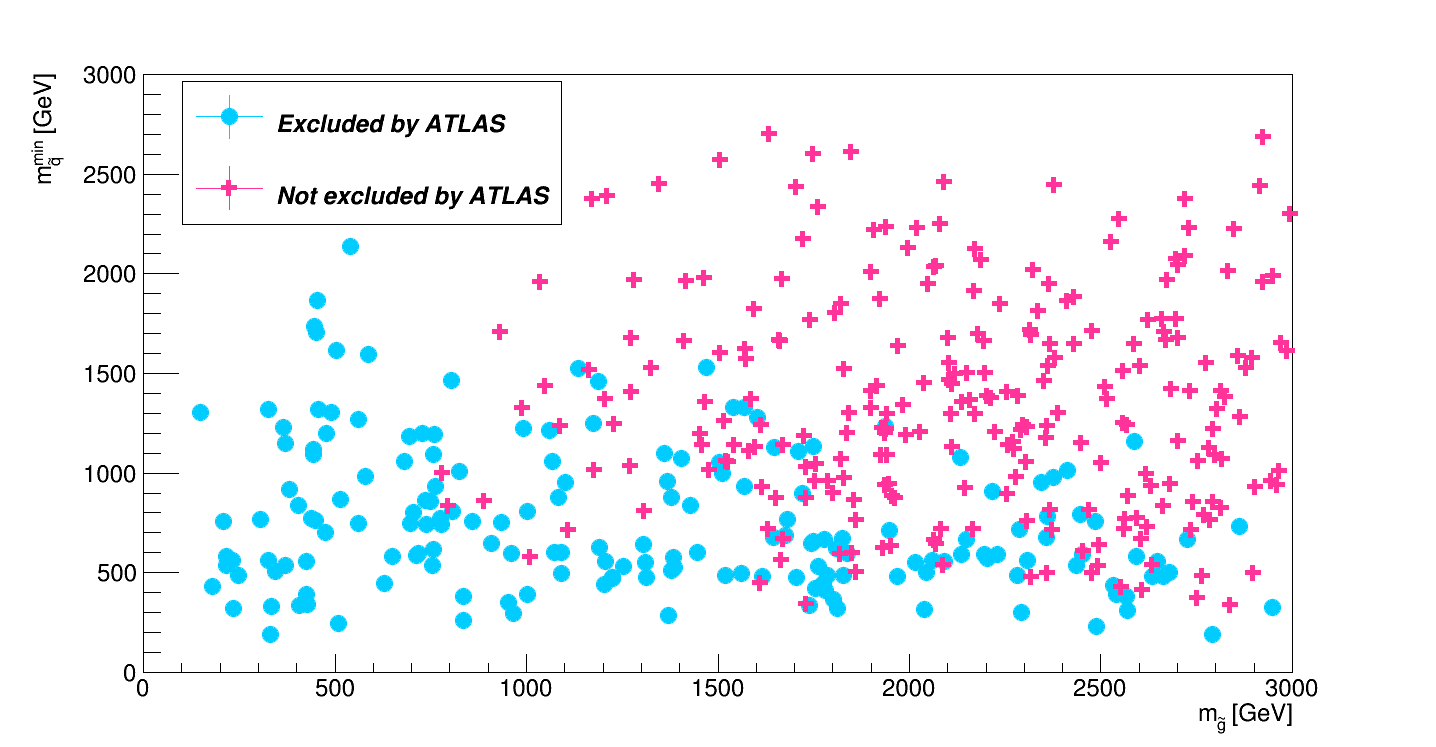}}    \\
\caption{Lightest light-flavoured squark mass against the gluino mass for 
         models that survive the ATLAS constraints (red crosses) 
         and models that are excluded by ATLAS (blue dots).}
\label{fig:sqsg}
\end{figure}

The remaining model-sets are compared with recent constraints from LHC SUSY 
searches at 7~TeV and 8~TeV centre-of-mass energies. Most important are 
constraints from searches for squarks and/or gluinos and chargino-neutralino 
production. Searches for heavy Higgs production had no influence on the 
remaining models. The mass constraint on the $\,h^0$ boson demands either large 
$M_A$ or very heavy top squarks. In fact, the model with the lightest $A^0$ 
boson in our sample has $M_A\approx 330$~GeV. Since $\tan\beta=6.9$ is 
relatively small for this model, it is not excluded by the ATLAS and CMS 
searches for heavy Higgs production~\cite{Aad:2014vgg,Khachatryan:2014wca}. 
\\[3mm]
The limits of the ATLAS experiment on light squarks, gluinos and 
chargino-neutralino production are implemented by emulating the ATLAS analysis 
chain. Events from LHC collisions are generated for each pMSSM model and
the detector response is simulated by a fast detector simulation.
The acceptance and efficiency is determined by applying the most important
ATLAS analysis cuts on the simulated events. Finally, these numbers are used to 
calculate the expected number of signal events for each signal region and 
analysis. Subsequently, these expected yields are compared to the 
model-independent $95\%$ C.L. limits provided by ATLAS.\\[3mm]
PYTHIA 6.4 \cite{Sjostrand:2006za} is used for the event simulation of 
proton--proton ($pp$) collisions at a 7~TeV and 8~TeV centre-of-mass energy. 
All SUSY production processes are enabled. For every model point and each 
centre-of-mass energy 10000 events are generated, which we found to be enough 
even for the models with the smallest selection efficiencies. To get as close 
as possible to the ATLAS analysis we use DELPHES~3.0 \cite{deFavereau:2013fsa} 
as a fast detector simulation with the default ATLAS detector card, modified by
setting the jet cone radius to $0.4$. 
The PYTHIA output is read in by DELPHES 
in HepMC format, which is produced by HepMC~2.04.02 \cite{Dobbs:2001ck}. 
The object reconstruction is done by DELPHES, which uses the same anti-k$_T$ 
jet algorithm \cite{Cacciari:2008gp} as ATLAS. Also included in the 
reconstruction are isolation criteria for electrons and muons. We do not 
emulate pile-up events.\\[3mm]
The 7 TeV analysis implementation is identical to Ref.~\cite{Strege:2014ija}.
The selection efficiencies of our own implementation were compared to ATLAS
in Ref.~\cite{Strege:2014ija} and were found to agree within uncertainties.
For this study the implementation used in Ref.~\cite{Strege:2014ija} was 
updated using the recent 8 TeV selection criteria. For the chargino-neutralino 
searches the SR$0\tau$a selection with all 20 bins was implemented as 
described in Ref.~\cite{Aad:2014nua}. For the squark and gluino searches all 
13 signal regions without explicit $W$ selection of Ref.~\cite{Aad:2014wea} are 
considered. In order to check constraints from multi-b-jet searches
we included also signal region SR-$0\ell$-A from \cite{Aad:2014lra}.\\[3mm]
Preliminary direct searches for decays into $h^0$ bosons from 
neutralinos do not influence the remaining models. The mass of the lightest 
neutralino with a sizeable branching ratio to Higgs bosons is about 185~GeV
and the mass of the LSP is at least 40~GeV. 
This is well beyond the exclusion reach of the  
ATLAS and CMS searches in these channels 
\cite{ATLAS-Higgs,Khachatryan:2014mma,Khachatryan:2014qwa}.\\[3mm]
After the event selection, the event counts are scaled to the luminosities 
considered in the analyses %, i.e.~35 \lumipb\ and 1.04 \lumifb, respectively. 
with leading order cross sections as provided by Pythia.
%The next-to-leading order (NLO) cross sections used for this are calculated by 
%LHC-Faser light \cite{LHCFaserweb,Dreiner:2010gv} 
%from PROSPINO2.1~\cite{Beenakker:1996ch,Beenakker:1997ut} cross-section grids.
The limits on the effective cross sections given by the \atl\ analyses are 
%given in table~\ref{tab:sigreg35} and \ref{tab:sigreg104}. They can be 
used to calculate a limit on the number of signal events passing the cuts.
No attempt was made to include theoretical uncertainties. In the studied SUSY 
mass range these uncertainties are small compared to the differences of the 
ATLAS and DELPHES setups and would not change drastically any conclusion of 
this work. \\[3mm]
In the end, $252$ of the models passed all selection criteria. 
Figure \ref{fig:sqsg} shows the excluded and non-excluded models as a function
of the gluino mass and the minimal mass of the first and second generation 
squarks $\,m^{min}_{\tilde{q}}$. Most excluded model points are due to limits on 
squarks and gluinos and have squark or gluino masses below about 1500 GeV,
in agreement with current LHC limits. All models with a gluino mass below 
750 GeV are excluded. Remarkably, a large fraction of models with low squark 
masses is still allowed. One well-known reason for this is that the lightest 
squark can be compressed with the $\widetilde{\chi}^0_1$ as shown in 
Figure~\ref{fig:MSUSY_A}. This leads to very soft jets from squark
decays. The squarks might only be visible via mono-jet signatures. 
\begin{figure} [H]
\centering
  {\label{fig:MSUSY_A}
  \includegraphics[width=\textwidth]{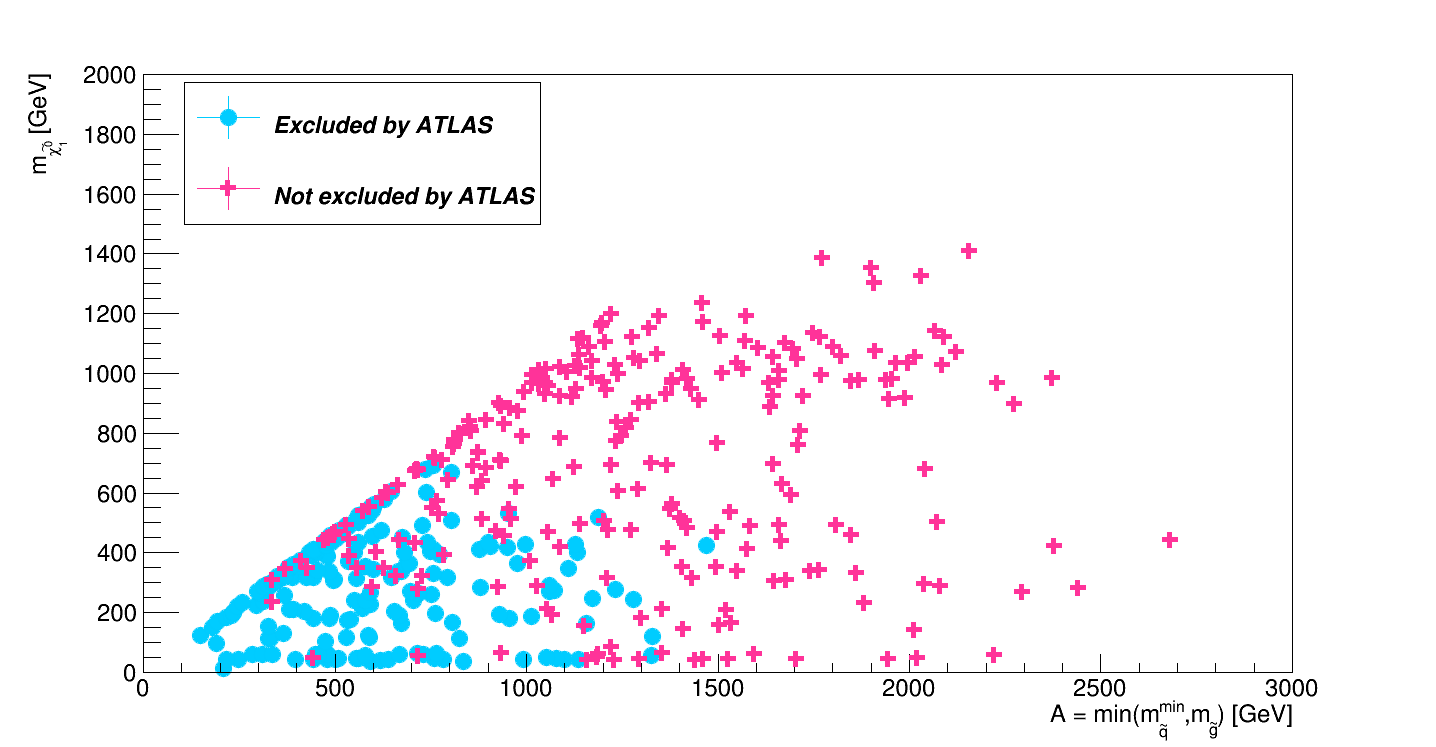}}    \\
\caption{The excluded and non-excluded models as a function of the LSP mass and 
the minimum of the gluino mass and the masses of the first and second 
generation squarks min$(m^{min}_{\tilde{q}},m_{\tilde{g}})$. }
\label{fig:MSUSY_A}
\end{figure}
The enhancement of Higgs production in the studied models leads to a second 
interesting feature that causes the fraction of non-excluded models in this 
study to be larger than previously found in other scans 
(e.g. in ~\cite{CahillRowley:2012kx}). In many non-excluded models the lightest
squarks are compressed with a heavy neutralino/chargino. To illustrate this we 
indicate the minimal mass of all first and second generation squarks and the 
gluino by $A$. Figure~\ref{fig:MSUSY_DeltaA} shows the smallest difference 
min$(\Delta A)$ between $A$ and the masses of all neutralinos and charginos as 
a function of $A$ (given that the neutralino or chargino mass is smaller than 
$A$). In contrast to Figure~\ref{fig:MSUSY_A}, all non-excluded models with 
$A<800$~GeV have a mass difference $\Delta A$ below $300$~GeV, which implies 
that many squarks are compressed with $\widetilde{\chi}^0_{2,3,4}$ or a chargino. 
\begin{figure} [H]
\centering
  {\label{fig:MSUSY_DeltaA}
  \includegraphics[width=\textwidth]{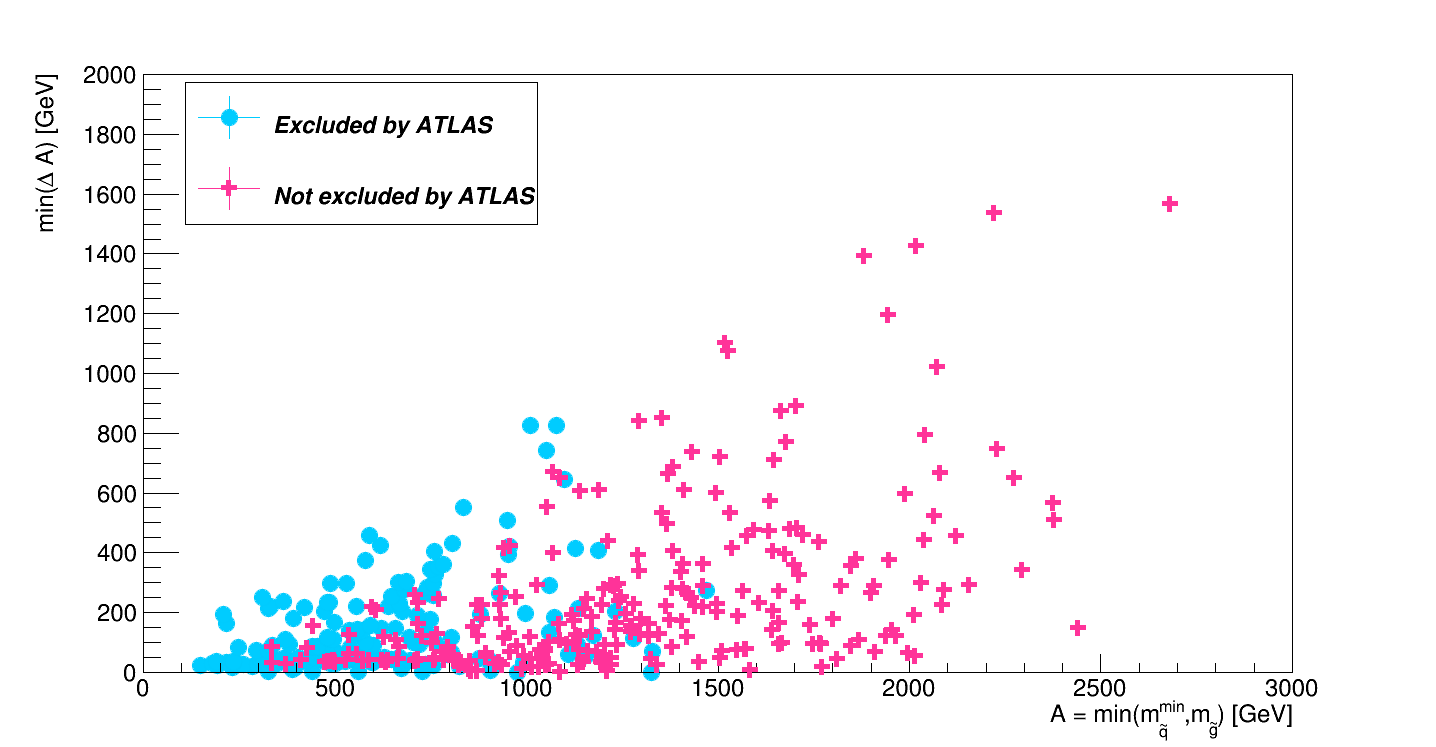}}    
\\
\caption{The same as in Figure~\ref{fig:MSUSY_A}, but this time the LSP mass 
         is replaced by the smallest difference min$(\Delta A)$ between 
         $A=$ min$(m^{min}_{\tilde{q}},m_{\tilde{g}})$ and the masses of all 
         neutralinos and charginos (given that the neutralino or chargino mass 
         is smaller than $A$).}
\label{fig:MSUSY_DeltaA}
\end{figure}
As discussed in section~2, in many cases the squarks do not directly decay
to the LSP, especially in the model points selected for this study.
If e.g.~the LSP is Binolike, the lightest $\widetilde{q}_L$ prefers to decay 
(if kinematically allowed) into the heavier Winolike neutralino or chargino. 
This is caused
by a $3c_w/s_w\approx 5.5$ enhancement factor of the weak coupling with respect
to the hypercharge coupling. If the squark happens to be compressed with the
Winolike neutralino, jets from the squark decays are also soft and the 
remaining signature is determined by the branching ratios of the heavier 
Winolike neutralino. In these models the Winolike neutralino can have a large 
branching ratio to Higgs bosons. If it is a $\widetilde{\chi}^0_4$\,, then 
decays to charginos are also possible. If the chargino decay is dominant, SUSY 
searches with leptons might be sensitive to these points. Searches asking for 
one lepton in the final state typically exclude simplified models with 
degenerate squarks decaying to charginos if $m_{\widetilde{q}} < 800$~GeV and 
$m_{\widetilde{\chi}^0_1} < 300$~GeV (see e.g.~\cite{ATLAS-CONF-2013-062}).  
After applying all other search constraints we find no model that fulfills 
these requirements. If the Winolike neutralino is a $\widetilde{\chi}^0_2$\,, 
then explicit searches for Higgs production from squarks might give a unique 
discovery possibility. Similar multi-step decays are possible in other cases as
outlined in~section~2.

\section{Analysis of the candidate models}

\subsection{Branching Ratios}

\begin{figure}[H]
  \centering
  \includegraphics[width=\textwidth]{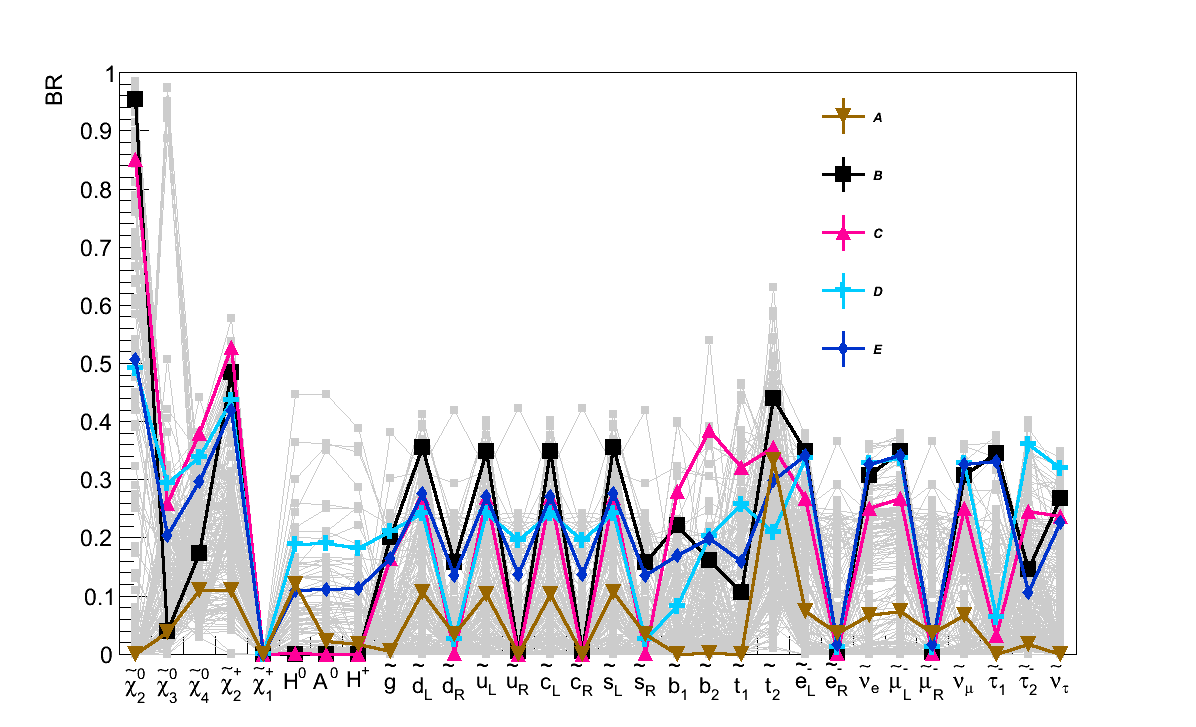}  
\caption{Branching ratio of all supersymmetric particles for decays into at 
         least one $h^0$ boson for the candidate models that survive all 
         constraints.}
\label{fig:br_fin_2}
\end{figure}

The branching ratios of all SUSY particles for decays into the lightest Higgs
boson $h^0$ have been determined for all surviving models. These branching 
ratios include also multi-step decays to the Higgs boson, i.e.~particles can 
have a non-zero branching ratio even if they do not couple to the Higgs boson 
directly. We show in Figure~\ref{fig:br_fin_2} the branching ratio for all MSSM
particles to the light Higgs boson $h^0$. All models are shown in grey in order 
to indicate the ranges of these branching ratios. The possible decay processes 
have been described in more detail in section~2.\\[3mm]
The sfermions can have decay branching ratios of up to $0.4$, with the values
for left- and right-handed sfermions strongly varying from model to model. 
The $\widetilde{b}_2$ and $\widetilde{t}_2$ squarks have a larger branching 
ratio due to the direct decay $\widetilde{f}_2 \rightarrow \widetilde{f}_1+h^0$.
As explained in section~2, the $\widetilde{\chi}^0_{2,3}$ neutralinos can have 
branching ratios close to unity. The $\widetilde{\chi}^{\pm}_2$ charginos can 
have a branching ratio that substantially exceeds $0.35$ due to multi-step 
decays. The branching ratios of the heavy Higgs bosons range up to
$\,\sim 0.4$ due to direct as well as multi-step decays.
Some models with interesting features are shown in colour. These models, which
are labeled A\,--\,E, are shown in Table~\ref{table:Parameters_MoI} and are
discussed in more detail below.

\begin{figure}[H]
  \centering
  \includegraphics[width=\textwidth]{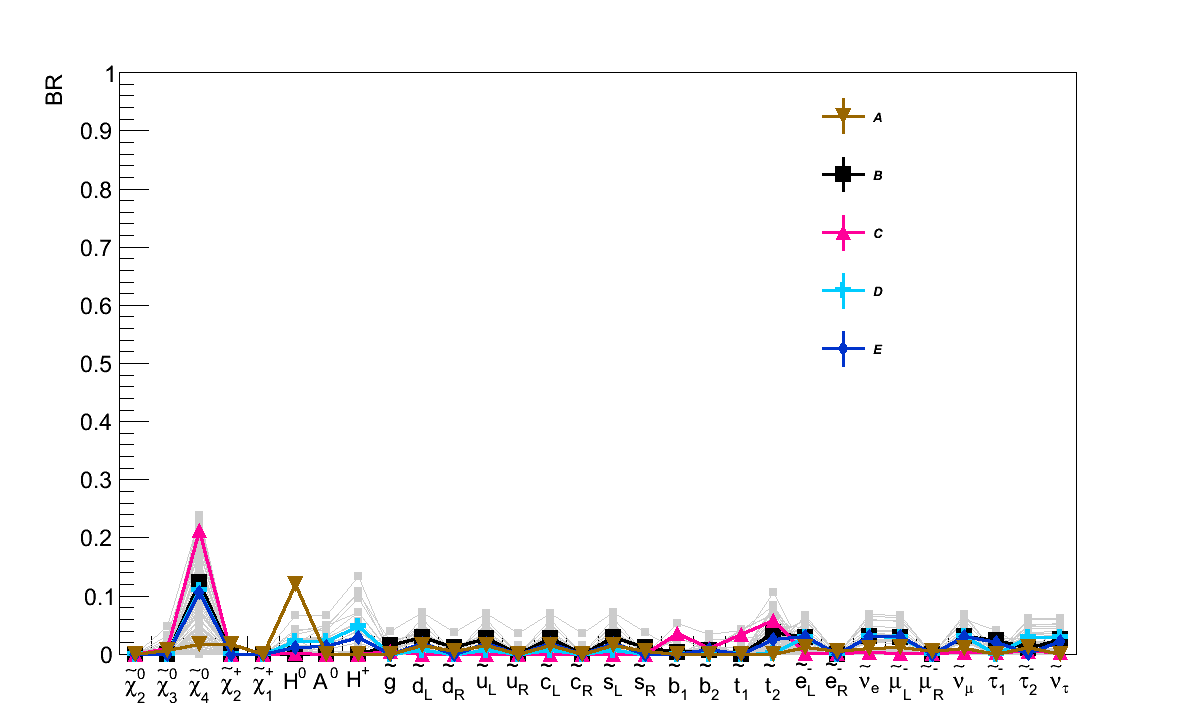}  
\caption{Branching ratio of all supersymmetric particles for decays into at 
         least two $h^0$ bosons for the candidate models that survive all 
         constraints.}
\label{fig:br_fin_5}
\end{figure}

If we go one step further and take a look at all supersymmetric particles that 
decay into at least two Higgs bosons, the heaviest neutralino has the highest 
branching ratio, as can be seen in figure~\ref{fig:br_fin_5}. 
Although it is not the preferred decay channel, it can decay into a 
$\widetilde{\chi}^0_{2,3}$\,, which can subsequently decay into a LSP. 
Both decays can produce one $h^0$ boson. The sfermions can decay into two $h^0$
bosons via the intermediate decay into a heavy neutralino. The $H^0$ boson can 
directly decay to two $h^0$ bosons, the $A^0$ and $H^{\pm}$ bosons decay via 
heavy neutralinos/charginos. The $\widetilde{t}_2$ top squark can decay to 
$\widetilde{t}_1+h^0$ and $\widetilde{t}_1$ can subsequently decay to one more
$h^0$.

%Gluinos will typically decay into left- and right-handed 
%squarks equally, where in most cases only the decay products of the 
%left-handed squarks will contain Higgs bosons, as discussed before. As can be
%seen in figure~\ref{fig:br_fin_2} the stops are additional candidates. 
%The expected LHC production rates are 
%studied in the next section.

\subsection{Event generation, fast simulation and analysis}

In order to determine the phenomenological relevance of $h^0$ production via 
SUSY processes, the LHC production rate needs to be determined. The generation 
of simulated events of $pp$ collisions at a centre-of-mass energy of 14~TeV for
each candidate model utilises PYTHIA6.4, HepMC~2.04.02 and DELPHES~3.0, just as 
described in section~\ref{sec:ATLASconstraints}.
In the simulation the branching ratio of the $h^{0}$ boson into two photons has
been set to unity manually. This is done in order to prevent interference of 
jets or leptons originating from $h^0$ decays when analysing the jet 
multiplicities in the final states later on in this section.
All other decay channels of the $h^0$
boson have been assigned a zero branching ratio. This does not affect the total
$h^0$ production rates.
A total number of 100000 events are generated for each of the approximately
$250$ candidate models that survive all previous constraints. 
All SUSY production processes are enabled. 
\begin{sidewaystable}[htb]
    \centering
    \caption{Parameters of the SUSY models with large Higgs production cross sections as discussed in the text.}
   \begin{tabular}{|l|c|c|c|c|c|} 
\cline{1-6}
%  Model point                             & sigma$10_{184}$ & sigma$10_{7}$ ($\sigma$20 dir.)    &  sigma$10_{33}$ &  sigma$25_{75}$ &  sigma$25_{127}$   \\ 
  Model point                             & model A & model B    &  model C &  model D &  model E   \\  \cline{1-6}
  Special feature 1                            & large $H^0\rightarrow h^0h^0$ rate & tri-Higgs    & $h^0$ via squarks & $h^0$ via gluinos & tri-Higgs   \\  \cline{1-6}\\[-5mm]
Special feature 2                            & medium $A^0\rightarrow h^0Z$ & large $\widetilde{\chi}^0_2\rightarrow \widetilde{\chi}^0_1 h^0$    & $h^0$ via $\widetilde{q}\,\widetilde{\chi}$ & $h^0$ via stops & di-higgs  \\  \cline{1-6}
 \cline{1-6}
$M_{1}$ (GeV)                                  &  2174 &  509    & 454  &  41   &  49 \\
$M_{2}$ (GeV)                                  &  2388 & 878    & 624  &  898  &  554     \\
$M_{\widetilde{g}}$ (GeV)                                  & 2755  &  2027   & 2765 & 1231   & 2020    \\ 
$\mu$  (GeV)                                  & -946   & 799   & -914 &  -260  &  -227     \\
$\tan\beta$                                   & 6.9 &   26   &  19  &  24    &   42      \\ 
$m_A$  (GeV)                                  &  330  & 980   &  820 & 1680    &  1370     \\
$m_{\widetilde{q}_L}$ (GeV)                & 2700 &   1290  &  760 & 2820   &  2500     \\
$m_{\widetilde{u}_R}$ (GeV)                & 1060     & 1200  &  480 & 1970   &  2530     \\
$m_{\widetilde{d}_R}$ (GeV)                & 2230 &    2200  &  1300 &  1250   &  2230     \\
m$_{\widetilde{\chi}_1^{\pm}}$ (GeV)              & 956 &   780   & 627  &  263    &  228          \\
m$_{\widetilde{\chi}_2^{\pm}}$ (GeV)              & 2410 &   926  & 931  & 944    &  595         \\
m$_{\widetilde{t}_1}$ (GeV)                       & 1670 &   980  &  1930 &  850  &   1030 \\
m$_{\widetilde{t}_2}$ (GeV)                       & 2530 &   2630  &  2190  & 1130   &   2740 \\[0.5mm]
\cline{1-6}\\[-5mm]
BR $\widetilde{\chi}_2^{\pm} $ $\rightarrow h^0$  & 11\% & 48\%     & 52\% & 43\%  & 42\% \\%REMCO
BR $\widetilde{\chi}_2^{0}$ $\rightarrow h^0$ , $\widetilde{\chi}_3^{0}$ $\rightarrow h^0$ , $\widetilde{\chi}_4^{0}$ $\rightarrow h^0$ & 0\%,4\%,11\% & 95\%,4\%,17\%   & 85\%,26\%,38\% &   49\%, 29\%, 34\% & 51\%,20\%,30\%  \\%REMCO
BR $\widetilde{q}_L$ $\rightarrow h^0$  & 11\% & 34\%     & 27\% & 24\%  & 27\%  \\%REMCO
BR $\widetilde{u}_R$ $\rightarrow h^0$  & 0\% & 1\%     & 0\% & 20\%  & 14\% \\%REMCO
BR $\widetilde{d}_R$ $\rightarrow h^0$  & 4\% & 15\%     & 0\% & 2\%  & 14\% \\%REMCO
BR $\widetilde{t}_1$ $\rightarrow h^0$  & 0\% & 11\%     & 32\% & 26\%  & 16\% \\%REMCO
BR $\widetilde{t}_2$ $\rightarrow h^0$  & 33\% & 44\%     & 36\% & 21\%  & 30\% \\%REMCO
Heavy Higgs with largest Higgs BR & $H^0$ (12\%) & too small & $H^0$ (0.05\%) & $A^0,H^0$ (19\%) & all (11\%) \\
MSSM particle with largest di-Higgs BR & $H^0$ (12\%) & $\widetilde{\chi}_4^{0}$ ($12\%$)  & $\widetilde{\chi}_4^{0}$ ($21\%$)  & $\widetilde{\chi}_4^{0}$ ($11\%$)   &  $\widetilde{\chi}_4^{0}$ ($11\%$)  \\
\cline{1-6}\\[-5mm]
Largest $h^0$ production process      &  $H^0$   & $\widetilde{q}\,\widetilde{q}+\widetilde{q}\,\widetilde{\bar{q}}$  &$\widetilde{q}\,\widetilde{q}+\widetilde{q}\,\widetilde{\bar{q}}$ & $\widetilde{\chi}^{0}_2 \widetilde{\chi}^{\pm}_1$ & $\widetilde{\chi}^{0}_2 \widetilde{\chi}^{\pm}_{1}$\\
Second largest $h^0$ production process & $A^0$ & $\widetilde{q}\,\widetilde{g}$   & $\widetilde{q}_L \widetilde{\chi}_2^0$ & $\widetilde{\chi}^{0}_3 \widetilde{\chi}^{\pm}_1$ &$\widetilde{\chi}^{0}_{2} \widetilde{\chi}^{0}_{3}$\\
\cline{1-6}
         \end{tabular}
\label{table:Parameters_MoI}
\end{sidewaystable}
\clearpage

\subsection{Determining the expected number of events with Higgs bosons}
The number of expected Higgs ($h^0$), di-Higgs and tri-Higgs events is 
calculated for an integrated luminosity of 10 fb$^{-1}$ for each SUSY production
process. For each model point the branching ratios to the $h^0$ as well as the 
SUSY production cross sections $\sigma_{ISUB_{i}}$ for each subprocess $ISUB_i$ are
considered. All cross sections are determined by PYTHIA. No attempt was made to
include NLO corrections. In general these NLO corrections would further 
increase the production rate. So, in that sense our estimates are conservative.
The number of events with at least one, two or three $h^0$ bosons is calculated
as \\[-2mm]
\begin{align}
N_{events}^{\geq 1 h^{0}} =& \mathcal{L}_{int} \cdot \sigma_{ISUB_{i}} \cdot 
[BR_{\widetilde{C}}(\widetilde{C} \rightarrow \geq 1 h^{0} + X) \nonumber\\
&+ BR_{\widetilde{C}}(\widetilde{C} \rightarrow 0 h^{0} + X) 
   \cdot BR_{\widetilde{D}}(\widetilde{D} \rightarrow \geq 1 h^{0} + Y)]~, 
\\[-5mm]\nonumber
\label{eqn:Nh}
\end{align}
\begin{align}
N_{events}^{\geq 2 h^{0}} =& \mathcal{L}_{int} \cdot \sigma_{ISUB_{i}} \cdot 
[BR_{\widetilde{C}}(\widetilde{C} \rightarrow \geq 1 h^{0} + X) 
 \cdot BR_{\widetilde{D}}(\widetilde{D} \rightarrow \geq 1 h^{0} + Y)\nonumber\\
&+ BR_{\widetilde{C}}(\widetilde{C} \rightarrow 0 h^{0} + X) 
   \cdot BR_{\widetilde{D}}(\widetilde{D} \rightarrow \geq 2 h^{0} + Y)\nonumber\\
&+ BR_{\widetilde{C}}(\widetilde{C} \rightarrow \geq 2 h^{0} + X) 
   \cdot BR_{\widetilde{D}}(\widetilde{D} \rightarrow 0 h^{0} + Y)]~,
\\[-5mm]\nonumber
\label{eqn:Nhh}
\end{align}
\begin{align}
N_{events}^{\geq 3 h^{0}} =& \mathcal{L}_{int} \cdot \sigma_{ISUB_{i}} \cdot 
[BR_{\widetilde{C}}(\widetilde{C} \rightarrow \geq 1 h^{0} + X) 
 \cdot BR_{\widetilde{D}}(\widetilde{D} \rightarrow \geq 2 h^{0} + Y)\nonumber\\
&+ BR_{\widetilde{C}}(\widetilde{C} \rightarrow \geq 2 h^{0} + X) 
   \cdot BR_{\widetilde{D}}(\widetilde{D} \rightarrow 1 h^{0} + Y)\nonumber\\
&+ BR_{\widetilde{C}}(\widetilde{C} \rightarrow 0 h^{0} + X) 
   \cdot BR_{\widetilde{D}}(\widetilde{D} \rightarrow \geq 3 h^{0} + Y)\nonumber\\
&+ BR_{\widetilde{C}}(\widetilde{C} \rightarrow \geq 3 h^{0} + X) 
   \cdot BR_{\widetilde{D}}(\widetilde{D} \rightarrow 0 h^{0} + Y)]~.
\\[-2mm]\nonumber
\label{eqn:Nhhh}
\end{align}

\subsubsection{Higgs production via SUSY processes}

\begin{figure}[H]
  \centering
  \includegraphics[angle=270,width=0.9\textwidth]{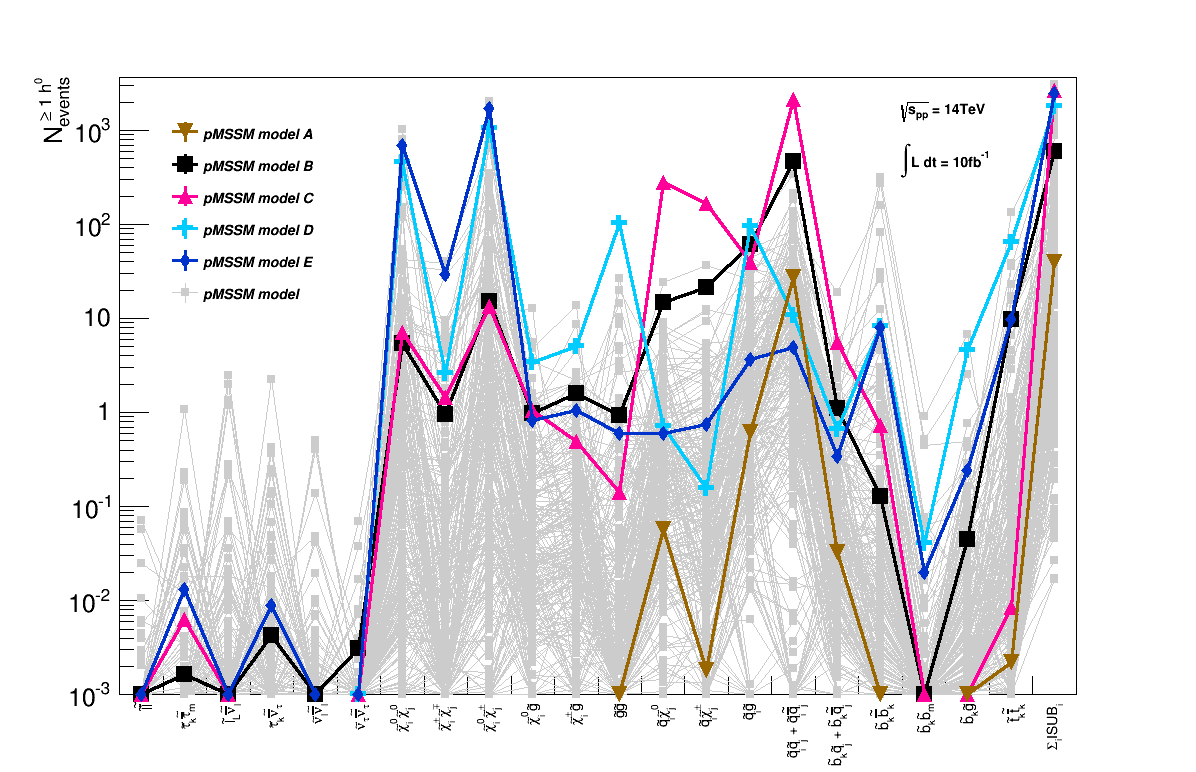}\vspace*{-8mm}
  \caption{Number of expected events containing at least one $h^0$ boson in 
           cascades of supersymmetric origin for all candidate models. 
           The vertical axis indicates the supersymmetric final states of the 
           primary interaction process.}
  \label{fig:N_events_1h_gray}
\end{figure}

Figure~\ref{fig:N_events_1h_gray} shows the rate of events with $\ge 1 h^0$ for 
all SUSY production processes normalized to an integrated luminosity of 
$10$~fb$^{-1}$. The most important classes of production processes are 
squark-(anti)squark production, in particular for left-handed squarks 
(see figure~\ref{fig:Summary_squarksquark} in the appendix), chargino-neutralino
production (see also figure~\ref{fig:Summary_charginos}  in the appendix) and 
neutralino pair production (see also figure~\ref{fig:Summary_neutralinos} in the
appendix). Next in line are the associated production of a neutralino/chargino 
and a light squark (see also figures~\ref{fig:Summary_squarkneutralino} and 
\ref{fig:Summary_squarkchargino}  in the appendix), and the production of pairs
of bottom or top squarks (see also figure~\ref{fig:Summary_heavysquarks} in the
appendix). \\[3mm]
Due to the nature of the mixing matrix, $h^0$ production via neutralino-pair and
chargino-neutralino processes are correlated. Large Higgs production rates are 
possible if the heavier neutralinos/chargino are relatively
light and decay to $h^0$. Examples are model $D$ and $E$ shown in 
Table~\ref{table:Parameters_MoI}. 
\clearpage
As can be seen in model $B$ or $C$, Higgs production from squarks can still be 
large when at the same time Higgs production via chargino/neutralino processes 
is supressed. This happens when the charginos and non-LSP neutralinos are too 
heavy (about 600-800 GeV in model $B$ and $C$) for direct production. These 
heavy neutralinos/charginos can, however, still be produced in the decay of a 
slightly heavier squark. As described in section~2 and~3 such squark decays 
might be dominant. Searching for squark production might then be the only 
possibility to detect these models. Similar neutralino/chargino decays are 
important to produce light Higgs bosons in bottom squark decays.\\[3mm] 
The associated production of a chargino/neutralino and a squark can also be 
interesting for $h^0$ production. For the same mass of the produced particles
the associated production cross section is in between the (electroweak) 
chargino/neutralino production and the (strong) squark production. This process
can be important if the mass of the chargino/neutralino is similar to the mass 
of the squark, i.e.~if one of the squarks is rather light. As explained before 
squarks can still be light in our models, e.g.~if the squark decays via a heavy
chargino/neutralino rather than directly to $\widetilde{\chi}_1^0$. It is 
then difficult to detect the squarks in the conventional way at the LHC. 
An example is model $C$ where the left-handed squark (mass 760 GeV) decays with
65\% branching ratio into the $\widetilde{\chi}^{\pm}_1$ and with 30\% branching 
ratio into $\widetilde{\chi}^{0}_2$ (both with a mass of 627 GeV). 
The $\widetilde{\chi}^{0}_2$ decays with 85\% branching ratio into a $h^0$ boson.
\\[3mm]
Higgs-boson production via top squarks can be enhanced for light stops. 
An example is model $D$, which has a $\widetilde{t}_1$ mass of 850 GeV and a 
$\widetilde{t}_2$ mass of 1130 GeV. Both stops decay to $h^0$ bosons 
predominantly via heavy neutralinos with branching ratios of $20-25$\%. This 
gives rise to special final-state topologies, involving top quarks, 
(possibly multiple) Higgs bosons and missing transverse momentum. 
Higgs-boson production via gluinos proceeds through the decay into light 
squarks. In the case of model $D$, these light squarks are top squarks, leading
to spectacular event topologies where the gluino (or even both gluinos) can 
decay into $h^0t\bar{t}\,\widetilde{\chi}_1^0$. \\[3mm]
The most important Higgs production processes are summarized in 
Table~\ref{table:processes}. In some models (e.g.~model $D$) $h^0$ production 
is significant for almost all important SUSY production processes. When all 
contributions to $h^0$ production from SUSY interactions are summed up, 
realistic models are found that lead to about 3000 events with at least one 
$h^0$ for 10~fb$^{-1}$ of data. In almost all models a significant amount of 
missing tranverse momentum due to the LSP's is expected. This makes the events
different from  $h^0$ production via SM processes.

\subsubsection{Di-Higgs production via SUSY processes}

\begin{figure}[H]
  \centering
  \includegraphics[angle=270,width=0.9\textwidth]{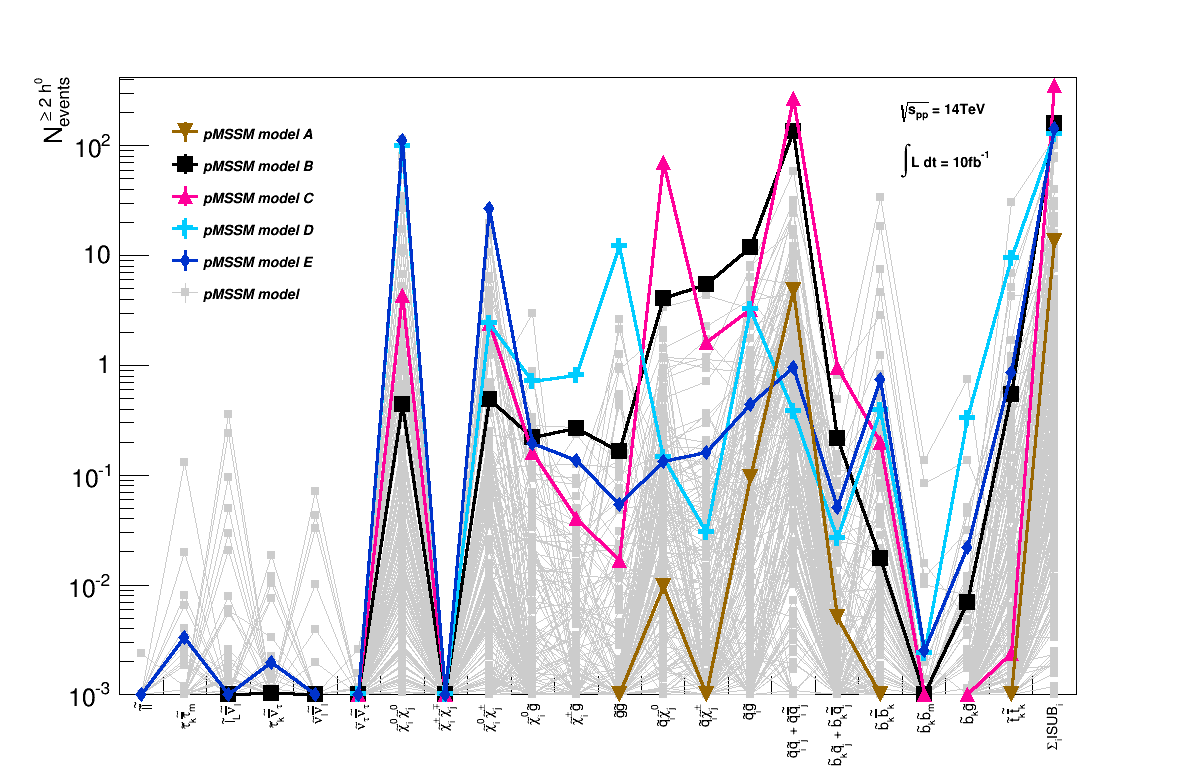}\vspace*{-8mm}
  \caption{Number of expected events containing at least two $h^0$ bosons in 
           cascades of supersymmetric origin for all candidate models. 
           The vertical axis indicates the supersymmetric final states of the 
           primary interaction process.}
  \label{fig:N_events_hh_gray}
\end{figure}

Since SUSY particles are pair produced and both particles can decay to a light 
Higgs boson, SUSY processes can be a significant source of events containing 
two $h^0$ bosons. We will show that this di-Higgs production rate can be 
significant. 
\begin{table}[htb]
    \centering
    \caption{List of most relevant SUSY processes to produce light Higgs bosons
             and the corresponding final-state topologies of interest, where 
             \etmis\ indicates the presence of missing transverse momentum in
             the final state. Numbers are given for $10$~fb$^{-1}$.}
   \begin{tabular}{|l|c|c|} 
\cline{1-3}
  processes & final state(s) & Example model \\ \cline{1-3}
$\ge 1 h^0$ production & & \\ 
processes with $\ge 100$ events & & \\ \cline{1-3}
                         $\widetilde{q}_L \widetilde{q}_L$ &  jets+$h^0$+E$^{miss}_{T}$ & $C,B$ \\     
                         $\widetilde{q}_L \widetilde{q}_R$ &  jets+$h^0$+E$^{miss}_{T}$ & $C,B$ \\         
                         $\widetilde{q}_R \widetilde{q}_R$ &  jets+$h^0$+E$^{miss}_{T}$ & $B$ \\         
                         $\widetilde{t}_2 \widetilde{\bar{t}}_2$ &  $t\overline{t}$+$h^0$+E$^{miss}_{T}$ & $(D)$ \\     
                         $\widetilde{b}_1 \widetilde{\bar{b}}_1$ &  $b$-jets+$h^0$+E$^{miss}_{T}$ & $-$ \\  
                         $\widetilde{q}_L \widetilde{\chi}^0_{2}$ &  jets+$h^0$+E$^{miss}_{T}$ & $C$ \\    
                         $\widetilde{q}_L \widetilde{\chi}^{\pm}_{1}$ & jets+(leptons)+$h^0$+E$^{miss}_{T}$ & $C$ \\   
                         $\widetilde{\chi}^0_{2} \widetilde{\chi}^{0}_{3}$ & jets+(leptons)+$h^0$+E$^{miss}_{T}$ & $E,D$ \\
                         $\widetilde{\chi}^{\pm}_{1} \widetilde{\chi}^{0}_{2}$ &jets+(leptons)+$h^0$+E$^{miss}_{T}$ & $E,D$ \\
                         $\widetilde{\chi}^{\pm}_{1} \widetilde{\chi}^{0}_{3}$ &jets+(leptons)+$h^0$+E$^{miss}_{T}$ & $E,D$ \\
                         $\widetilde{\chi}^{\pm}_{2} \widetilde{\chi}^{0}_{4}$ &jets+(leptons)+$h^0$+E$^{miss}_{T}$ & $E$ \\
                         $\widetilde{g} \widetilde{g}$ &  jets+$h^0$+E$^{miss}_{T}$, top quarks+$h^0$+E$^{miss}_{T}$ & $D$ \\     
                         $H^0$ &  2$h^0$ & $A$ \\     
                         $A^0$ &  $Z$+$h^0$ & $A$ \\ \cline{1-3}
$\ge 2 h^0$ production       & & \\ 
processes with $\ge 20$ events & & \\ \cline{1-3}
                         $\widetilde{q}_L \widetilde{q}_L$ &  jets+2$h^0$+E$^{miss}_{T}$ & $C,B$ \\     
                         $\widetilde{q}_L \widetilde{q}_R$ &  jets+2$h^0$+E$^{miss}_{T}$ & $B$ \\         
                         $\widetilde{t}_2 \widetilde{\bar{t}}_2$ &  $t\overline{t}$+2$h^0$+E$^{miss}_{T}$ & $(D)$ \\     
                         $\widetilde{b}_1 \widetilde{\bar{b}}_1$ &  $b$-jets+2$h^0$+E$^{miss}_{T}$ & $-$ \\  
                         $\widetilde{q}_L \widetilde{\chi}^0_{2}$ &  jets+2$h^0$+E$^{miss}_{T}$ & $C$ \\    
                         $\widetilde{\chi}^0_{2} \widetilde{\chi}^{0}_{3}$ &jets+(leptons)+2$h^0$+E$^{miss}_{T}$ & $E,D$ \\
                         $\widetilde{\chi}^{\pm}_{2} \widetilde{\chi}^{0}_{4}$ &jets+(leptons)+2$h^0$+E$^{miss}_{T}$ & $E$ \\
                         $H^0$ &  2$h^0$ & $A$ \\   \cline{1-3}  
$\ge 3 h^0$ production       & & \\ 
processes with $\ge 5$ events & & \\ \cline{1-3}
                         $\widetilde{\chi}^{\pm}_{2} \widetilde{\chi}^{0}_{4}$& jets+(leptons)+3$h^0$+E$^{miss}_{T}$ & $E$ \\
                         $\widetilde{q}_L \widetilde{q}_L$ &  jets+3$h^0$+E$^{miss}_{T}$ & $B$ \\     
\cline{1-3}
         \end{tabular}
\label{table:processes}
\end{table}
\clearpage
Di-Higgs production is of utmost importance in the SM to measure the
triple-Higgs coupling. As such, the measurement of di-Higgs production is a 
central research objective for the high luminosity phase of the LHC.  
The SM di-Higgs production has an expected next-to-next-to-leading order (NNLO)
cross section of roughly $40$~fb~\cite{deFlorian:2013jea}, leading to about 
$400$ events for $10$fb$^{-1}$. These events are very difficult to detect due to
the enormous SM background rate. \\[3mm]
In the MSSM another important source of di-Higgs events is the production of 
heavy Higgs bosons (see figure~\ref{fig:N_events_2hplus_HeavyHiggs} in the 
appendix). Model $A$ predicts an enourmous rate of $\,>2000$ di-Higgs 
events, visible as a di-Higgs resonance. Heavy Higgs production is discussed 
separately in the next subsection.
Di-Higgs events from processes involving SUSY particles are different due to 
the presence of large missing transverse momentum. The background from SM 
processes can be reduced by a large factor with cuts on this quantity.
Figure~\ref{fig:N_events_hh_gray} shows the di-Higgs production rate per SUSY 
process normalized to $10$fb$^{-1}$. Model $C$ predicts the largest SUSY
production rate for di-Higgs events with about $350$~events for $10$~fb$^{-1}$. 
This rate can also be compared with $10$ and $4.2$ events expected from the SM
$\,t\overline{t}h^0h^0$ or $Zh^0h^0$ processes, which have a cross section of 
$1.0$~fb at leading order~\cite{Baglio:2012np,Moretti:2004wa} and $0.42$~fb
at NNLO~\cite{Baglio:2012np}, respectively. SUSY processes can 
therefore significantly enhance di-Higgs signatures in SM di-Higgs searches. 
Any deviation from the SM expectations in these searches needs therefore to be 
interpreted carefully, since deviations could be the result of SUSY decays. 
\\[3mm]
The SUSY di-Higgs production is dominated by squark processes, followed by the 
direct production of heavy neutralinos/charginos. The most important SUSY 
di-Higgs production processes are summarized in Table~\ref{table:processes}.

\subsubsection{Tri-Higgs production via SUSY processes}

\begin{figure}[H]
  \centering
  %\subfloat[$\geq$ 3 h$^{0}$]{\label{fig:N_events_hhh_gray}
  \includegraphics[angle=270,width=0.9\textwidth]{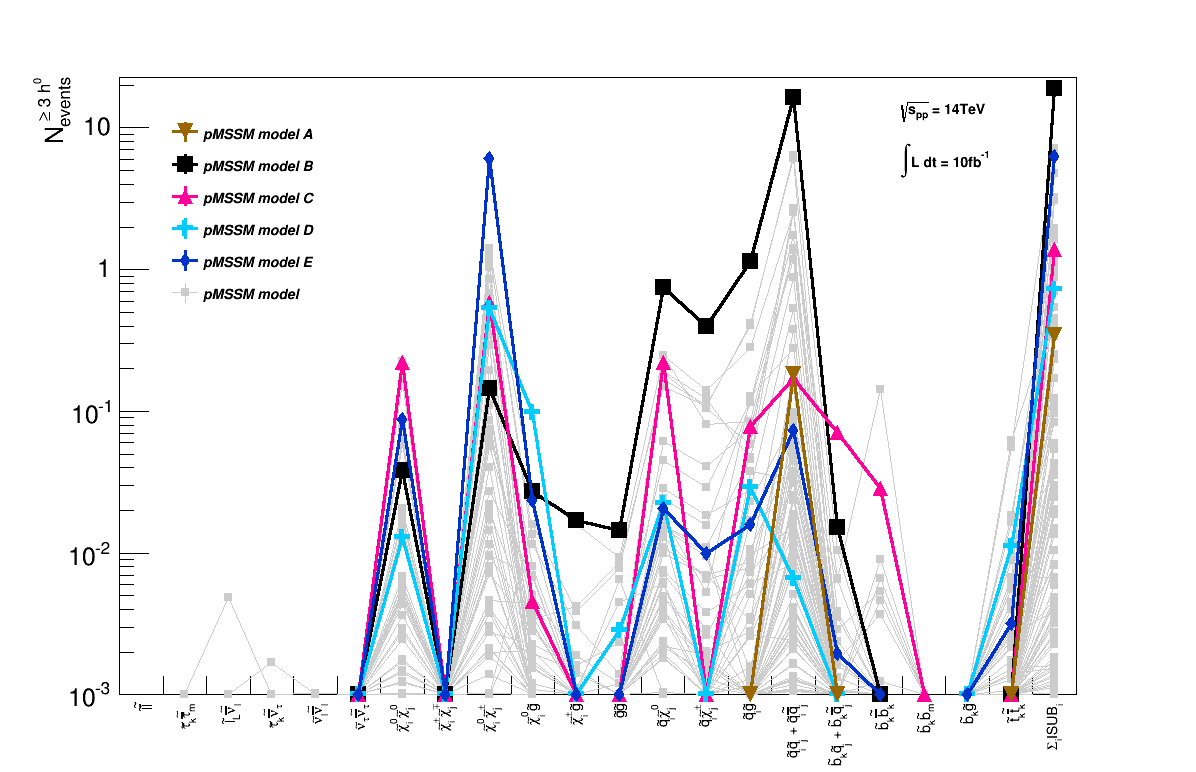}\vspace*{-8mm}
  \caption{Number of expected events containing at least three $h^0$ bosons in 
           cascades of supersymmetric origin for all candidate models. 
           The vertical axis indicates the supersymmetric final states of the 
           primary interaction process.}
    %\caption{Number of expected events containing at least two (see figure~\ref{fig:N_events_hh_gray}) or three (see figure~\ref{fig:N_events_hhh_gray}) h$^{0}$s in cascades of supersymmetric origin for the sets of $\sigma = 0.1$ and $\sigma = 0.25$ candidate models. The X-axis indicates the supersymmetric final states of the responsible first interaction process.}
  \label{fig:N_events_hhh_gray}
  %\label{fig:N_events_xh_gray}
\end{figure}

Due to the multi-step decays of heavy neutralinos there is the possibility that
one heavy neutralino can decay to two $h^0$ bosons. The corresponding  branching
fractions were discussed in \S\,4.1 and $\widetilde{\chi}^0_4$ was found to be 
the dominant source. This makes it possible to produce three Higgs bosons in one
event. Figure~\ref{fig:N_events_hhh_gray} shows the number of tri-Higgs events 
per SUSY process normalized to $10$fb$^{-1}$. Up to $20$ tri-Higgs events can be 
produced, predominantly via squark production. The dominant tri-Higgs 
production processes are summarized in Table~\ref{table:processes}.
The SM tri-Higgs production cross section is $0.044$~fb~\cite{Plehn:2005nk} 
leading to an expectation of only $0.4$ events for 10~fb$^{-1}$. Tri-Higgs 
production might become important for large luminosities or, after a LHC 
discovery, for determining e.g.~the neutralino mixing matrix. 

\subsection{The lightest Higgs boson from heavy-Higgs production processes}

\begin{figure}[H]
  \centering
  \includegraphics[width=0.9\textwidth]{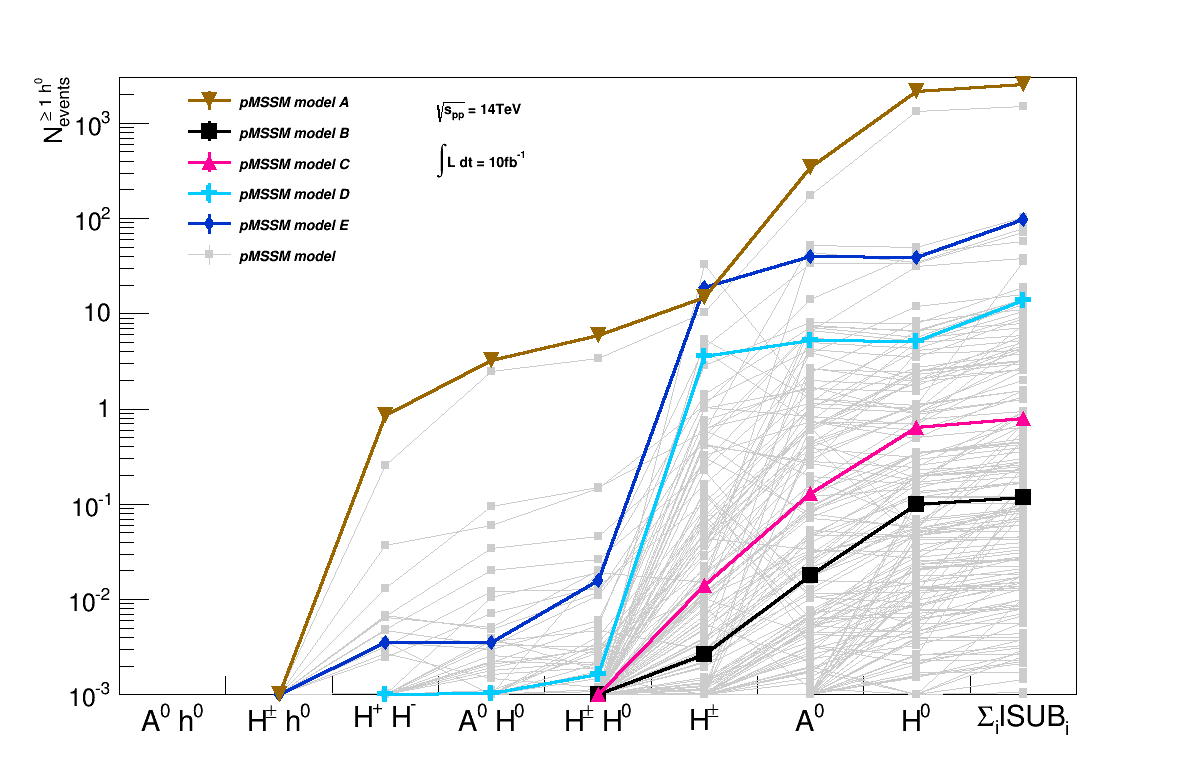}
  \caption{Number of expected events containing at least one $h^0$ boson in 
           cascades of heavy-Higgs origin for all candidate models. 
           The horizontal axis indicates the heavy Higgs particles involved in 
           the final states of the primary interaction process.}
  \label{fig:N_events_1h_HeavyHiggs}
\end{figure}

For the sake of completeness, simulated events with primary interaction
processes involving heavy Higgs particles are also investigated briefly. 
This investigation utilises the calculation of events with at least one 
$h^0$ boson according to equation~\ref{eqn:Nh}, but this time only the 
branching ratios of the heavy Higgses into one or more light Higgs boson(s) 
are taken into account. \\[3mm]
As can be seen in figure~\ref{fig:N_events_1h_HeavyHiggs}, for most models the 
$h^0$ event rates from heavy Higgs production processes is low. This is caused 
by the decoupling limit. Due to the mass constraint on the lightest Higgs-boson,
most models have an $A^0$ boson that is much heavier than the $Z$ boson. In this
decoupling limit all heavy Higgses are nearly mass degenerate and truly heavy.
As a result, the heavy-Higgs production cross sections are relatively small and
the $h^0$ event rates rather modest. The models with parameters that place them 
in the decoupling limit only reach a maximum of about 50 $h^0$ events for single
heavy-Higgs production. Exceptions are a couple of models including model A, 
which have a smaller value for $M_A$ and which are therefore less firmly in the 
decoupling limit. These models also have a relatively small value for 
$\tan\beta$, which results in a noticeable $H^0\to 2h^0$ branching ratio (see 
table~\ref{table:Parameters_MoI}) and substantially larger $h^0$ event rates
beyond 1000 events. Also heavy Higgs production can have an effect on the Standard Model
Di-Higgs production rate as discussed in \cite{Bhattacherjee:2014bca}.
\\[3mm]
It must be kept in mind, though, that the heavy Higgs particles are not 
strictly speaking supersymmetric particles and are therefore not expected to 
lead to events with a large missing transverse momentum in the detector due to 
the LSP.

\subsection{Kinematic distributions for Higgs events from SUSY}

%\begin{figure}[htp!]
%  \centering
%  \includegraphics[width=1.0\textwidth]{figures_marienew/HiggsBeta/SM_h0Production_HiggsBeta_CutsFromPaper_14TeV_100kEv_same_100plus.png}
%  \caption{Qualitative distribution (in $\%$) of the boost in terms of $\beta$~=~v$_{h^{0}}$/c of h$^{0}$ for the main h$^{0}$ production processes within t%he SM in hadron collisions (red, blue) and SUSY production for pMSSM models (grey). The distributions are each normalised to unity. Only pMSSM models that predict more than 100 h$^{0}$ produced via SUSY processes in 10~fb$^{-1}$ are presented.}
%  \label{fig:HiggsBeta100}
%\end{figure}

\begin{figure}[htp!]
  \centering
  \includegraphics[width=1.0\textwidth]{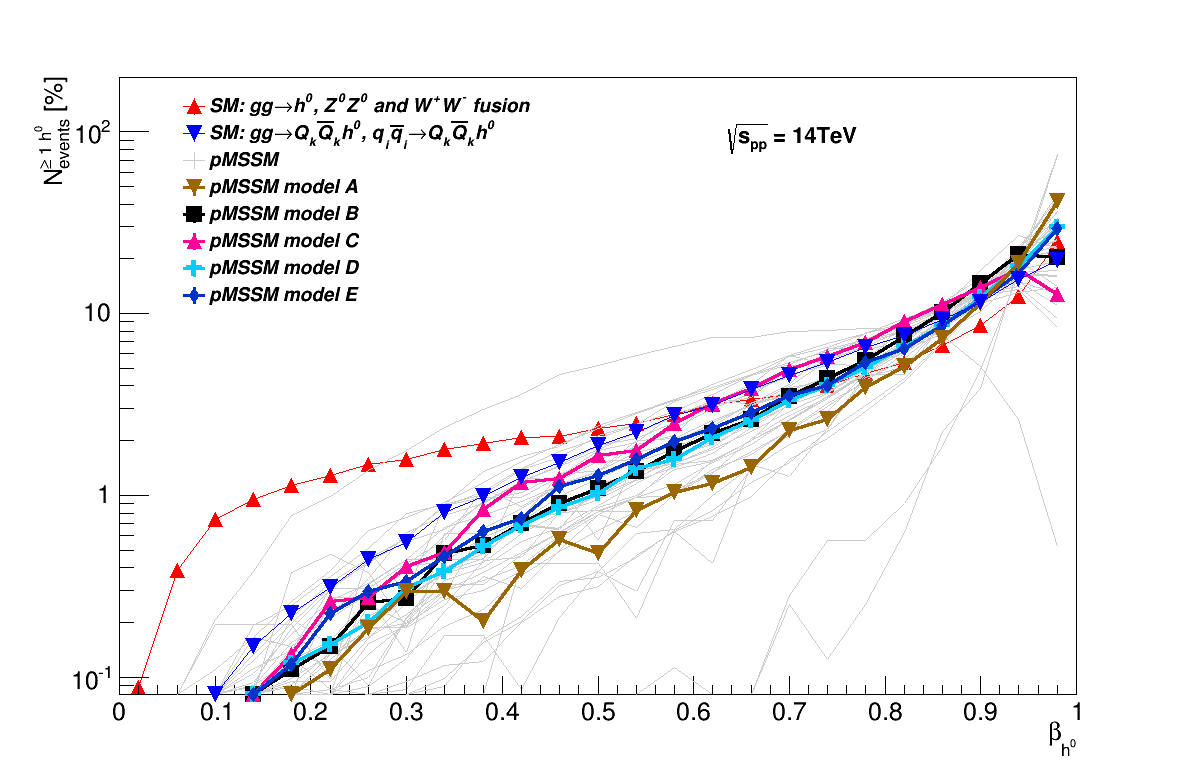}
  \caption{Qualitative distribution (in $\%$) of the $h^0$-boson boost in terms 
           of $\,\beta_{h^0}=v_{h^0}/c$ for the main $h^0$ hadroproduction 
           processes in the SM (red, blue) and in pMSSM models via SUSY 
           processes (grey). The distributions are each normalised to unity. 
           Only those pMSSM models are presented that predict more than 100 
           $h^0$ bosons produced via SUSY processes for 10~fb$^{-1}$.}
  \label{fig:HiggsBeta100}
\end{figure}

\paragraph{\boldmath Boost of the $h^0$ boson.}
When a supersymmetric particle decays into a $h^0$ boson, the mass difference 
between mother and daughter (initial and final) state can lead to a boost.
In hadronic $pp$ collisions the main contribution to $h^0$ production by SM 
processes is expected to be from gluon-gluon fusion, and to a lesser extent from
$WW/ZZ$ fusion. A second relevant contribution is expected from associated
$t\overline{t}h^0$ production, which is expected to lead to $h^0$ bosons that 
are more boosted in view of the larger (top-quark) mass scale in the process.
Both processes are shown in Figure~\ref{fig:HiggsBeta100} in order to compare 
the $h^0$-boson boost ($\beta_{h^0}$) distributions originating from SUSY and SM 
processes. Due to the larger mass scale of the SUSY processes the $h^0$ 
bosons are on average more boosted, even more than in $t\overline{t} h^0$ 
production. In extreme cases a heavy SUSY particle with mass $>1$~TeV decays to
a $h^0$ boson and a SUSY particle with a mass of ${\cal O}(100$~GeV), leading to
a very large boost. As an opposite extreme, we find one case where a squark with
$m\approx 1.5$~TeV decays to a $\widetilde{\chi}^0_3$ with $m=1.17$~TeV, which 
subsequently decays to a $h^0$ and a $\widetilde{\chi}^0_1$ with $m=1.04$~TeV. 
In such compressed scenarios the $h^0$ boost is even lower than expected from 
SM processes.
\begin{figure}[H]
  \centering
  \includegraphics[width=0.9\textwidth]{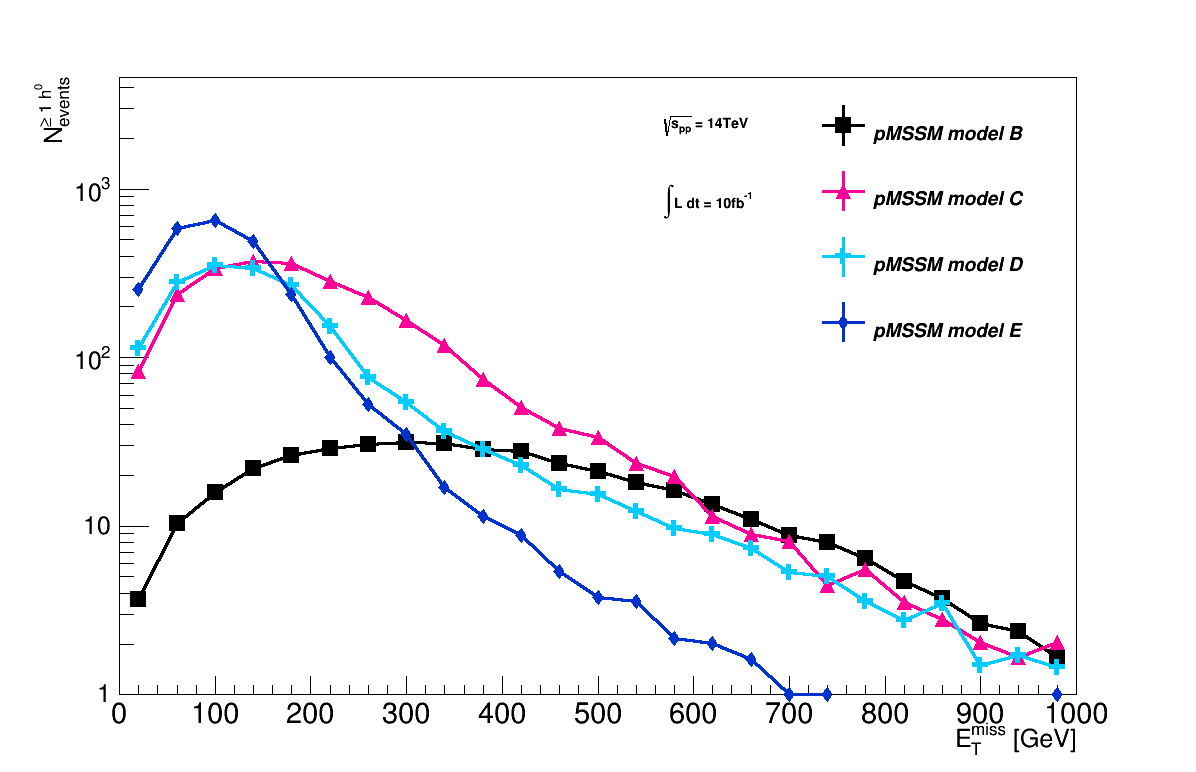}
  \caption{Number of expected events containing at least one $h^0$ boson as a 
           function of missing transverse energy \etmis\ in cascades of 
           supersymmetric origin (i.e.~without the
           production of heavy Higgs bosons).}
  \label{fig:N_events_1h_MET}
\end{figure}

\paragraph{Missing transverse momentum.}
Figure~\ref{fig:N_events_1h_MET} shows the missing transverse momentum 
distributions from SUSY processes for the selected pMSSM models. The generated 
events are normalized to an integrated luminosity of $10$fb$^{-1}$. 
All models have on average large missing transverse momentum up to several 
100~GeV, permitting the introduction of selection cuts of 100\,-200 GeV in order
to reduce backgrounds from SM processes. The production of heavy Higgs bosons 
is not considered. Model $A$ has low missing transverse momentum since the 
$h^0$ boson originates from a heavy $H^0$ boson, which does not decay to the
 LSP.

\paragraph{\boldmath Final states with $h^0$ bosons.}
After the detector response is simulated with DELPHES, the final states are 
determined. Selection cuts are applied, requiring the leptons, i.e.~electrons 
or muons, to have a transverse momentum of at least 20~GeV. For the jets this 
lower limit is chosen to be 50~GeV. B-jets and hadronic tau decays are counted 
as jets. Both leptons and jets are only considered if they are located within 
the pseudorapidity%
  \footnote{$\eta = - \ln \left[\tan(\frac{\theta}{2})\right]$ in terms of the 
            polar angle $\theta$ w.r.t.~the beam axis} 
region of $|\eta^{e, \mu, jet}| <$ 2.5. In addition to the overlay removal that is
automatically performed in DELPHES, a stricter overlap removal of 
$\Delta R_{a,b} >0.6$ is applied%
  \footnote{$\Delta R_{a,b} = \sqrt{(\Delta\eta)_{a,b}^{2}+(\Delta\phi)_{a,b}^{2}}\,$
            in terms of the pseudorapidity difference $\Delta\eta$ and the 
            difference in azimuthal angle $\Delta\phi$ between the objects 
            (leptons/jets) $a$ and $b$.}.
The generated events are again normalised to an integrated luminosity of 
10~fb$^{-1}$.\\[3mm]
\begin{figure}[H]
  \centering
  \includegraphics[angle=270,width=0.9\textwidth]{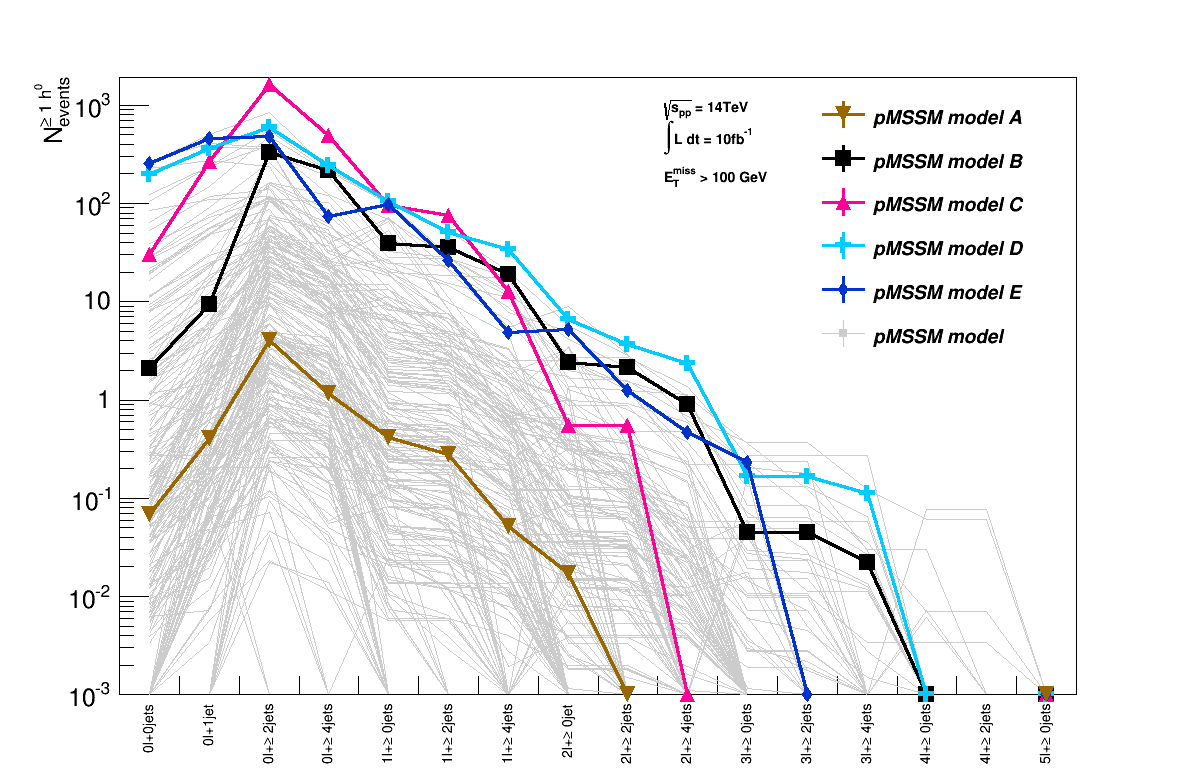}\vspace*{-8mm}
  \caption{Number of expected events containing at least one $h^0$ boson in 
           cascades of supersymmetric origin with \etmis~$>$~100~GeV. The event
           rates are split up according to specific combinations of lepton and 
           jet multiplicities.}
  \label{fig:LepJetMult_MET50_geq1h0_gray}
\end{figure}

\begin{figure}[H]
  \centering
  \includegraphics[angle=270,width=0.9\textwidth]{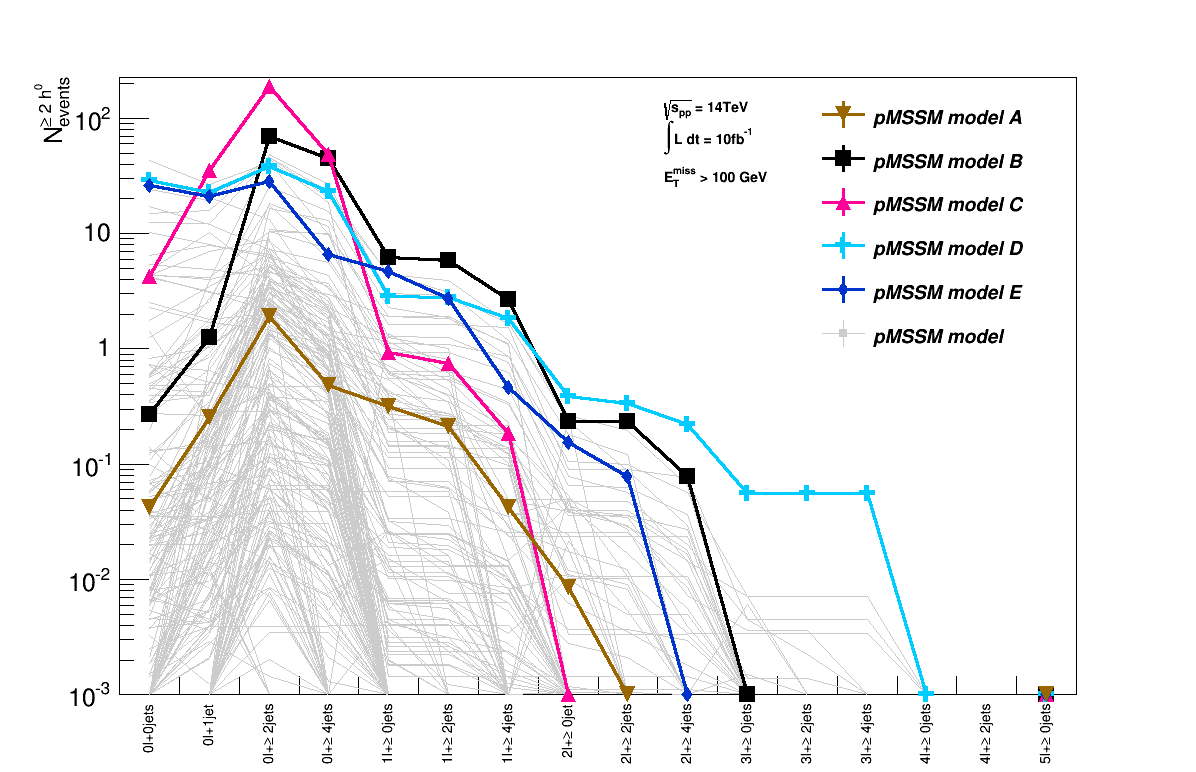}\vspace*{-8mm}
  \caption{Number of expected events containing at least two $h^0$ bosons in 
           cascades of supersymmetric origin with \etmis~$>$~100~GeV. The event
           rates are split up according to specific combinations of lepton and 
           jet multiplicities.}
  \label{fig:LepJetMult_MET50_geq2h0_gray}
\end{figure}

\begin{figure}[H]
  \centering
  \includegraphics[width=1.0\textwidth]{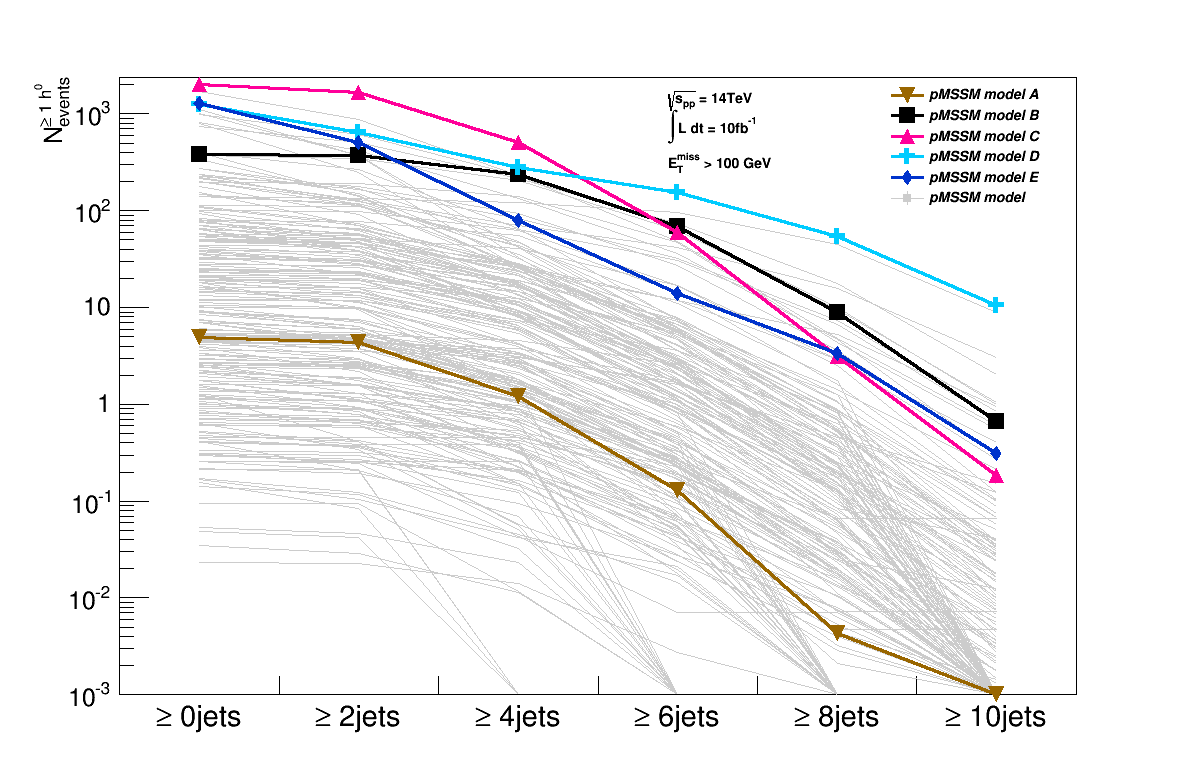}\\
  \includegraphics[width=1.0\textwidth]{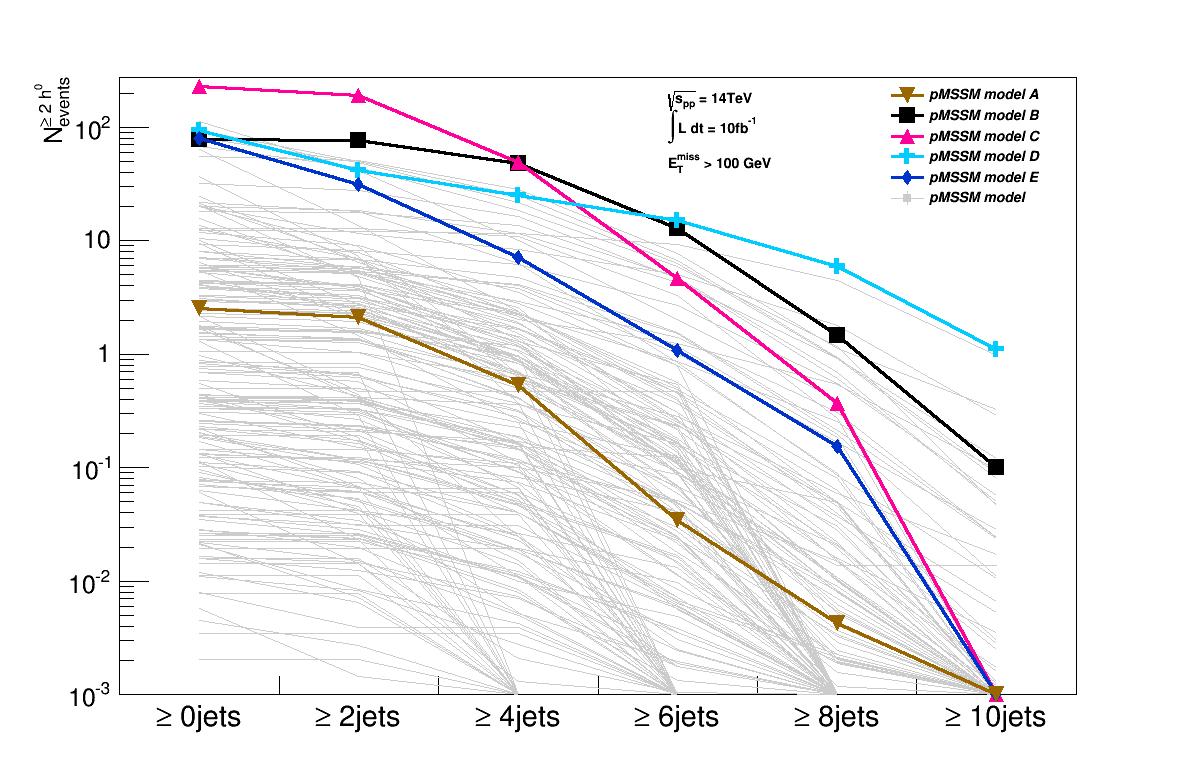}\\[4mm]
  \caption{Number of expected events containing at least one  $h^0$ boson 
           (upper figure) or two  $h^0$ bosons (lower figure) in cascades of 
           supersymmetric origin with \etmis~$>$~100~GeV. The event rates are 
           split up according to specific jet multiplicities.}
  \label{fig:JetMult_MET50_gray}
\end{figure}
\clearpage
The lepton and jet multiplicities for events requiring a missing 
transverse momentum of at least 100~GeV are shown in 
Figures~\ref{fig:LepJetMult_MET50_geq1h0_gray}, 
\ref{fig:LepJetMult_MET50_geq2h0_gray} and \ref{fig:JetMult_MET50_gray}. 
The most populated channels for single Higgs production are the channels that
contain $\,\ge 1-4$ jets, with close to 1000 events, and 
mono-higgs\footnote{The 0-lepton, 0-jet channel, which is dominated by 
                    neutralino-pair production.}
production, with up to 200 events for 10~fb$^{-1}$. In some cases very high jet 
multiplicities can occur, as can be seen in Figure~\ref{fig:JetMult_MET50_gray}.
Channels with one lepton lead to $\,\sim 100$ events and channels with two 
leptons to less than $10$ events. Higher lepton multiplicities are not 
important for $h^0$ production. Di-Higgs and tri-Higgs production is dominantly
found in channels with $\ge 2$ jets. Another notable feature is that the 
production of neutralino pairs can lead to events with two Higgs bosons,
missing transverse momentum and nothing else, i.e.~no leptons and no jets.

\section{Conclusion}
We have systematically investigated the possibilities to produce a $125$~GeV 
Higgs boson ($h^0$) via SUSY processes within the phenomenological MSSM (pMSSM).
We find the following interesting features:

\begin{itemize}
\item
Given global constraints on the pMSSM, it is possible to produce Higgs events 
with a large rate in the upcoming LHC data at the increased centre-of-mass 
energy. We have found valid pMSSM models that could produce more than $3000$ 
Higgs, $300$ di-Higgs and/or $20$ tri-Higgs events already with an integrated 
luminosity of 10~fb$^{-1}$. 
\item
A relation is observed between large Higgs-production rates via squark decays 
to heavy neutralinos and inherent difficulties to exclude such models in 
conventional (non-Higgs) LHC searches. This is caused by the fact that Higgs
production requires a less compressed neutralino mass spectrum, which can bring
the heavy neutralinos closer to the lowest-lying squark states, thereby 
reducing the available amount of energy for additional jets.
\item
In some models Higgs production is significant for almost all important SUSY 
production processes, which can have large repercussions on SM Higgs studies and
SUSY searches. 
\item
Higgs production via SUSY processes might significantly enhance the event rates
for SM Higgs and di-Higgs searches, especially in final states with 
missing transverse momentum. The allowed SUSY production rates can be reduced 
by upcoming (negative) SUSY searches at higher LHC energies, especially if new
dedicated searches for events with $h^0$ bosons and missing transverse momentum
are performed.
\item
Higgs production processes can likewise be of importance for a SUSY discovery 
via ``Higgs tagging''. We found that the different SUSY production channels
can lead to a large variety of interesting event topologies and kinematics. 
Of special interest are multi-jet channels with up to three Higgs bosons, ``mono-Higgs'' channels with up to two Higgs bosons, 
one-lepton channels with one Higgs boson and Higgs production in association with top or b-quarks, all with a sizeable amount 
of missing transverse momentum.  The list is completed by searches for heavy Higgs bosons decaying directly or via
neutralinos/charginos into $h^0$ bosons. 
\end{itemize}

\section*{Appendix: Selected illustrative figures}\vspace*{-3mm}

\begin{figure}[H]
  \centering
  \includegraphics[angle=270,width=0.38\textwidth]{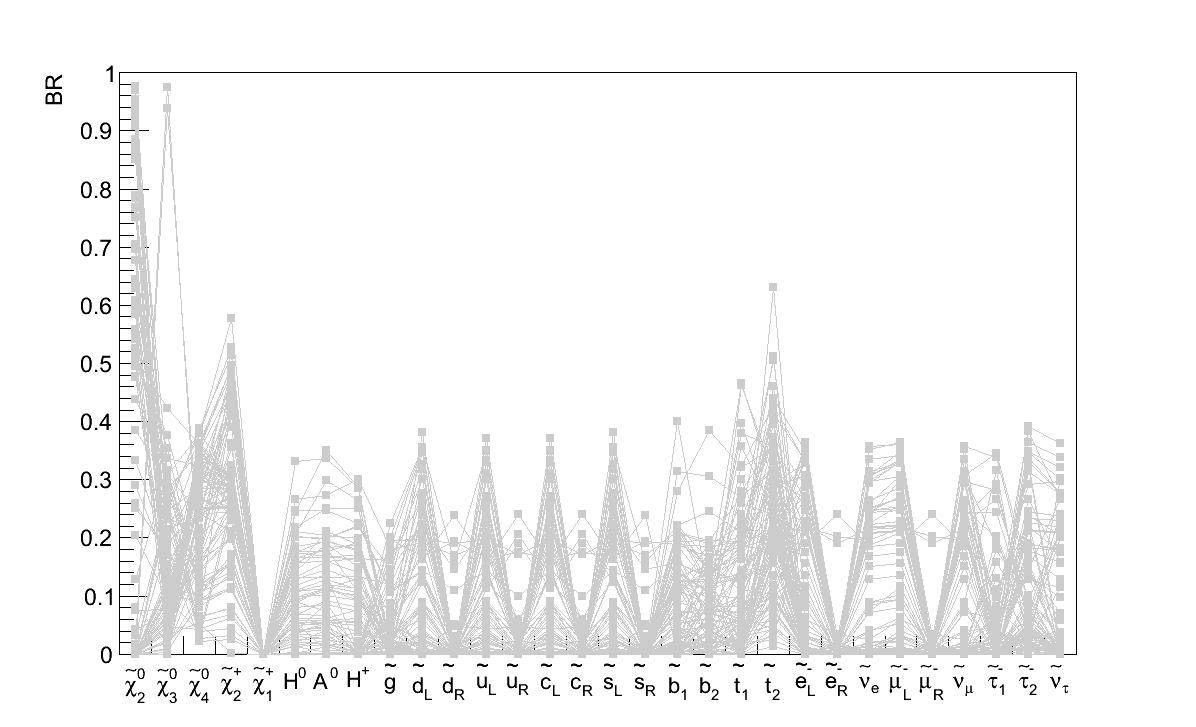}
  \includegraphics[angle=270,width=0.38\textwidth]{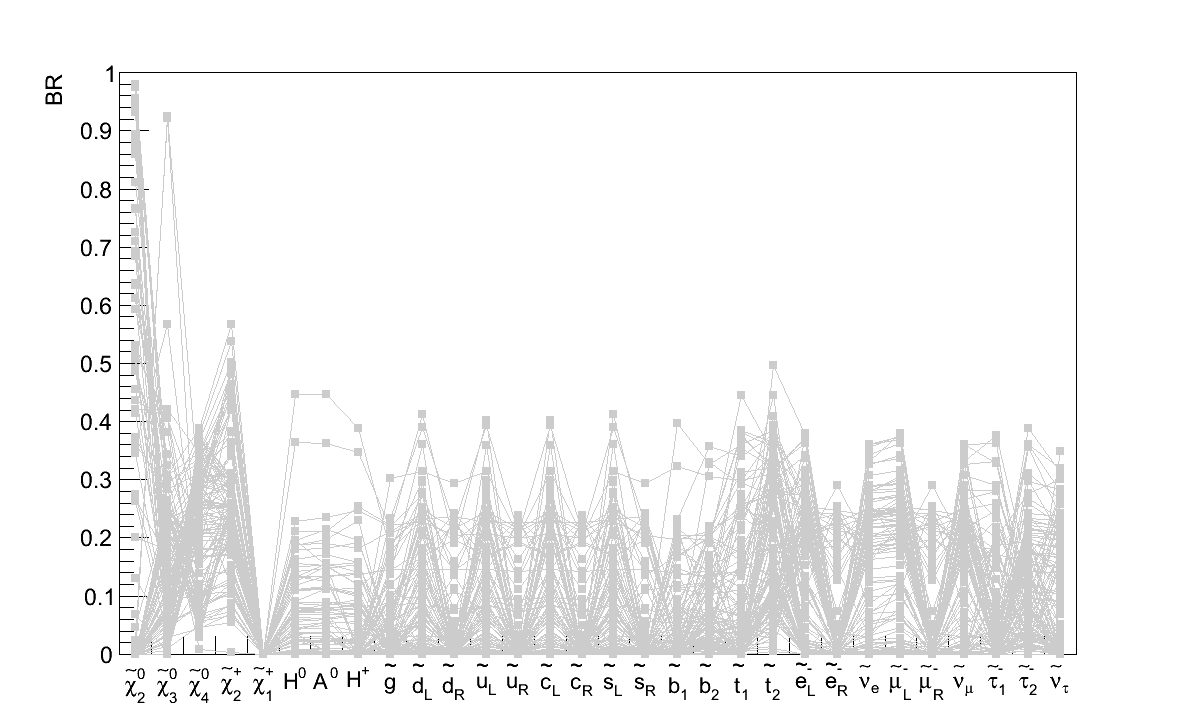}
  \includegraphics[angle=270,width=0.38\textwidth]{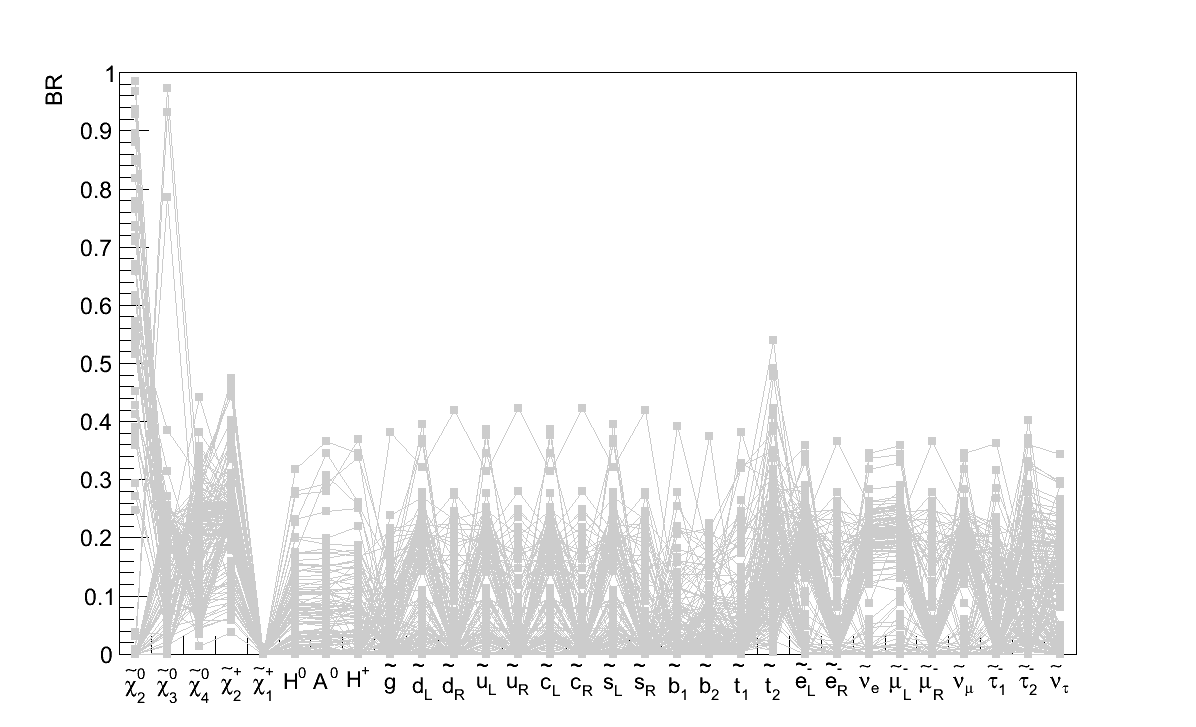}
  \includegraphics[angle=270,width=0.38\textwidth]{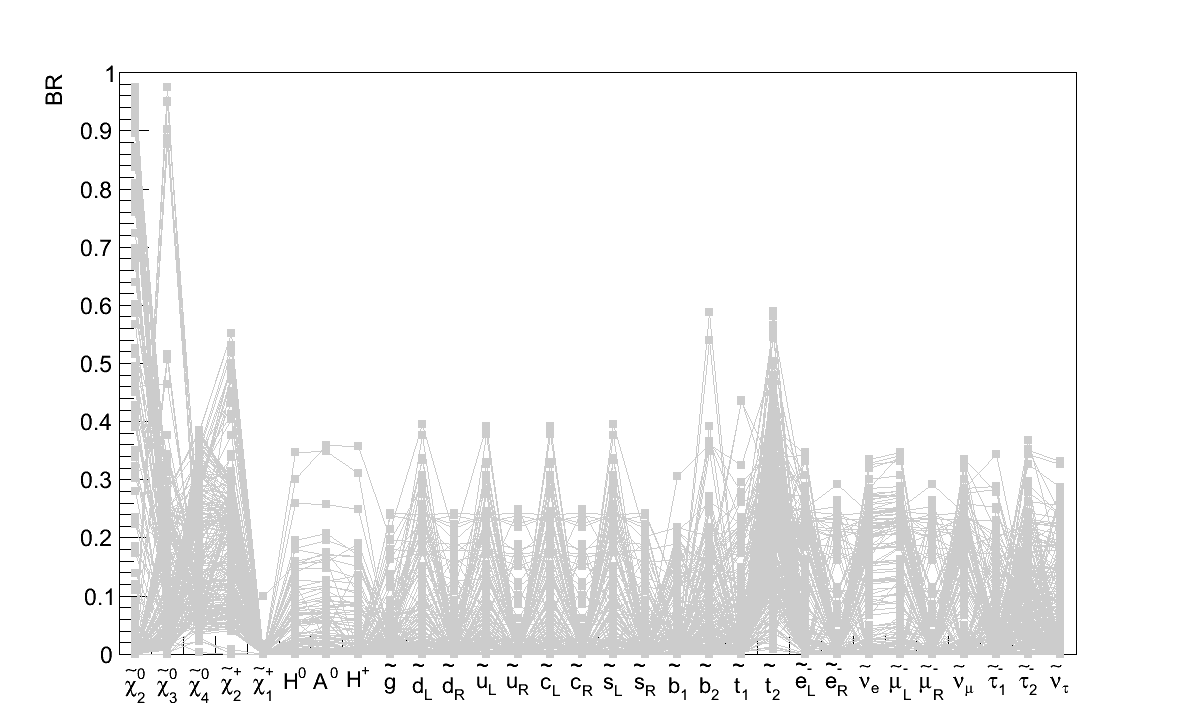}
  \caption{Branching ratios into $h^0$ for all MSSM particles. The plots show 
           set B with the width of the Gaussian particle filter set to 
           10\% (upper left), 25\% (upper right) and 40\% (lower left) of the
           full extent of the parameter space in each dimension. The lower 
           right figure shows for comparison set A, which has a different Higgs 
           branching ratio requirement. }
  \label{fig:width_comparison}
\end{figure}

\begin{figure}[H]
  \centering
  \includegraphics[width=0.7\textwidth]{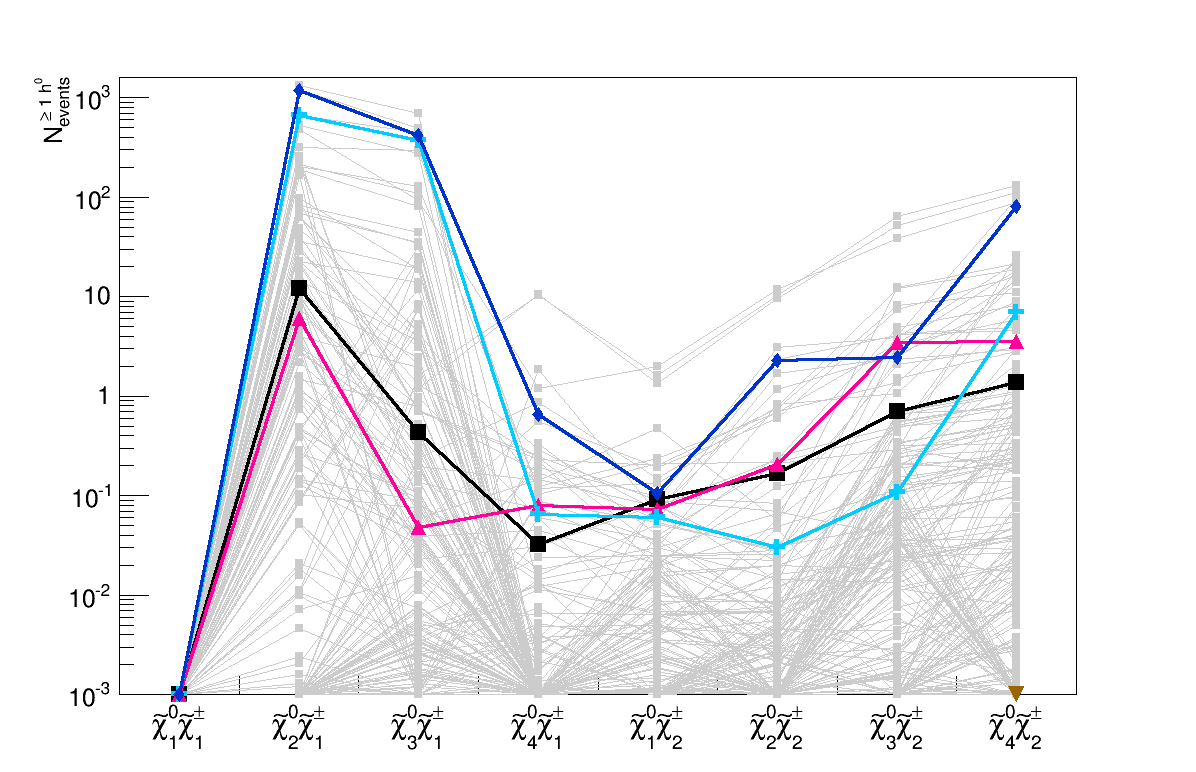}
  \includegraphics[width=0.7\textwidth]{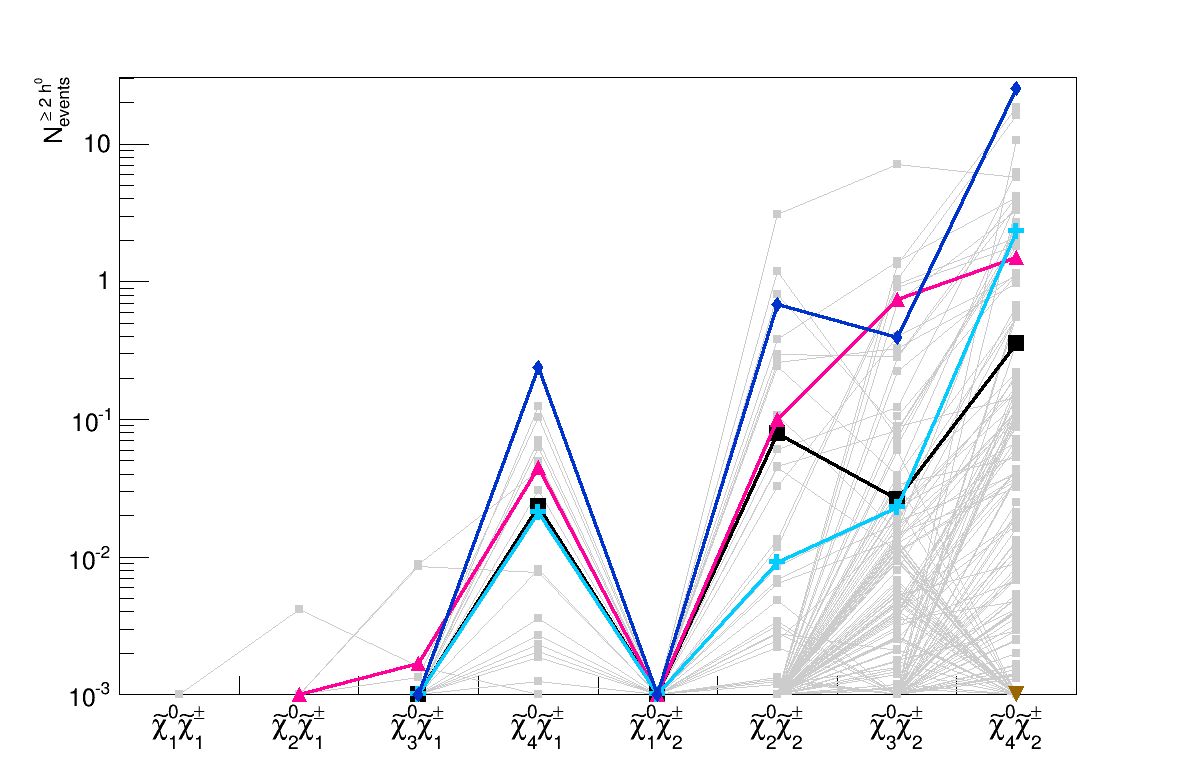}
  \includegraphics[width=0.7\textwidth]{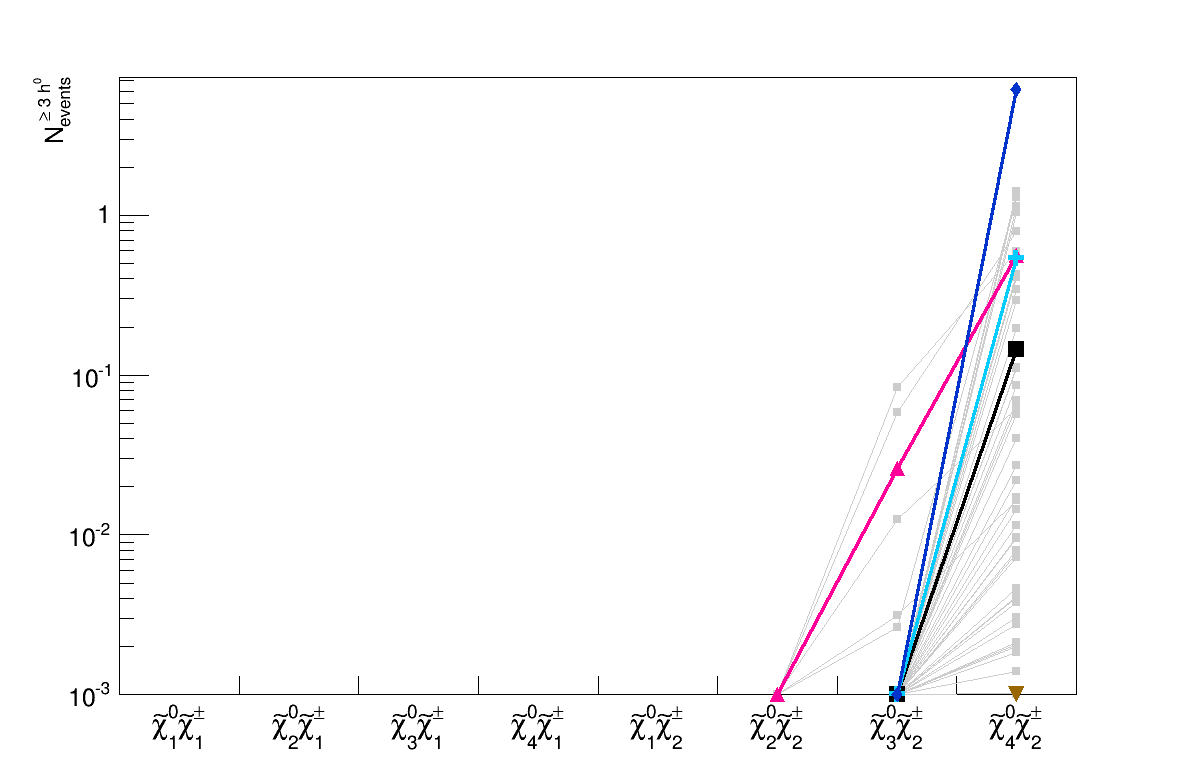}
  \caption{Number of expected events containing at least one (upper), 
           two (middle) or three (lower figure) $h^0$ boson(s) produced in 
           cascades of supersymmetric origin. The horizontal axis indicates the
           supersymmetric neutralino\,--\,chargino final states of the primary 
           interaction process.}
  \label{fig:Summary_charginos}
\end{figure}

\begin{figure}[H]
  \centering
  \includegraphics[width=0.7\textwidth]{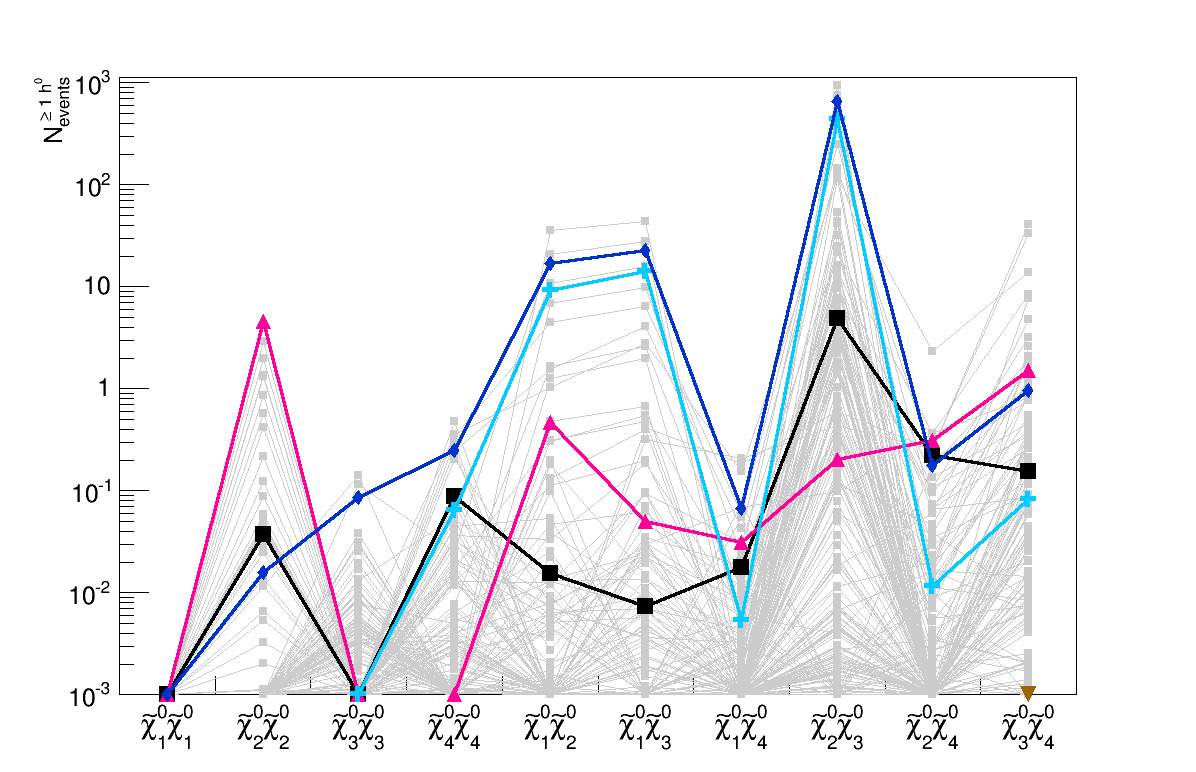}
  \includegraphics[width=0.7\textwidth]{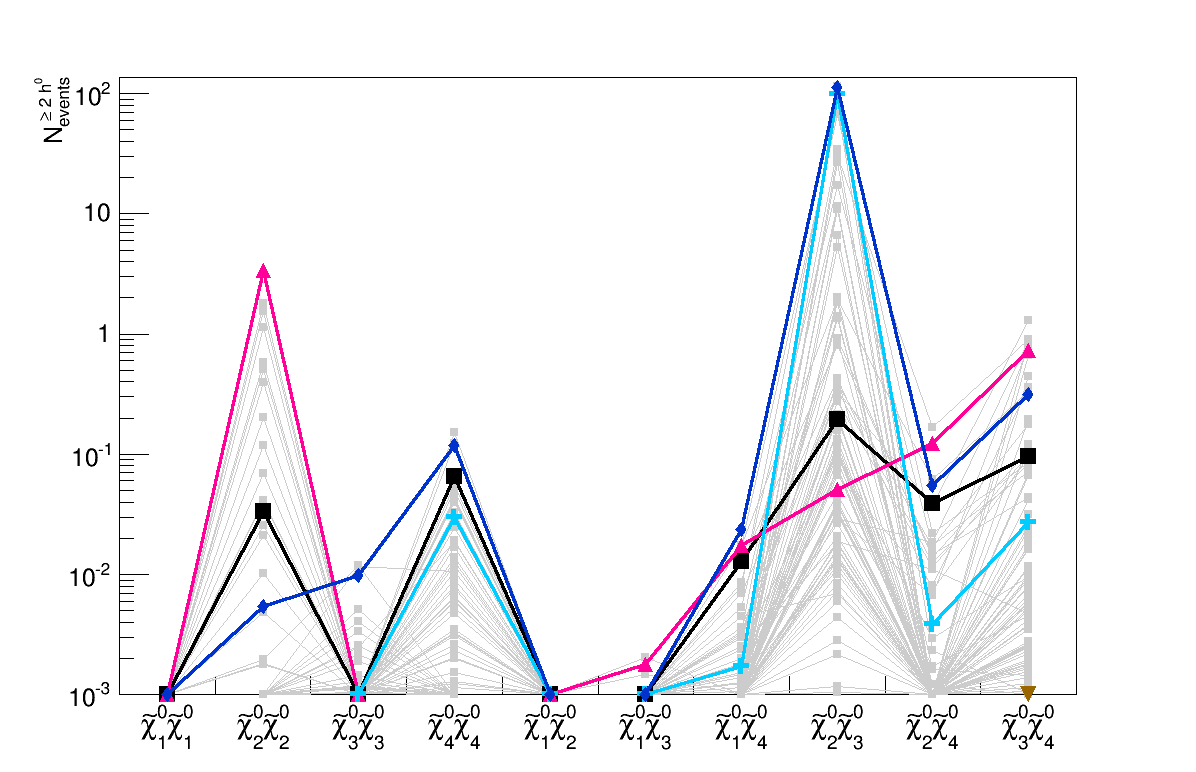}
  \includegraphics[width=0.7\textwidth]{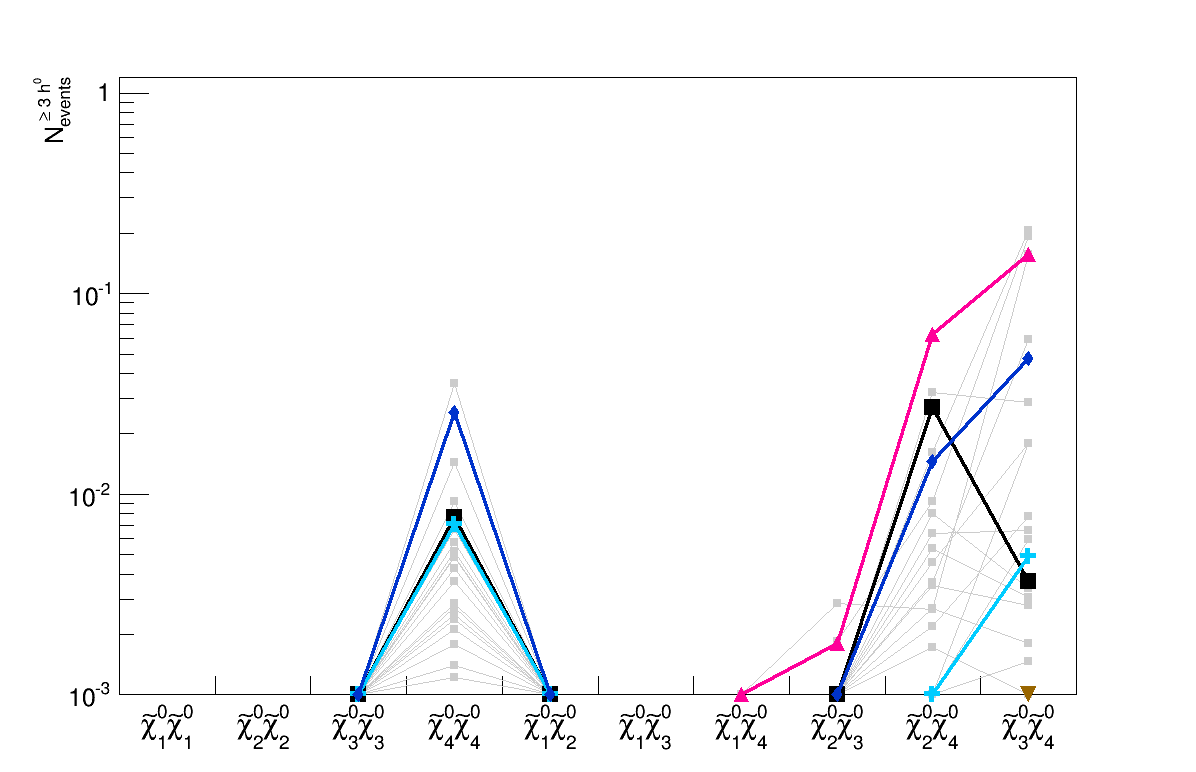}
  \caption{Number of expected events containing at least one (upper), two (middle) or three (lower figure) $h^0$ boson(s) produced in cascades 
of supersymmetric origin. The horizontal axis indicates the 
           supersymmetric neutralino\,--\,neutralino final states of the primary 
           interaction process.}
    \label{fig:Summary_neutralinos}
\end{figure}

\begin{figure}[H]
  \centering
  \includegraphics[width=0.7\textwidth]{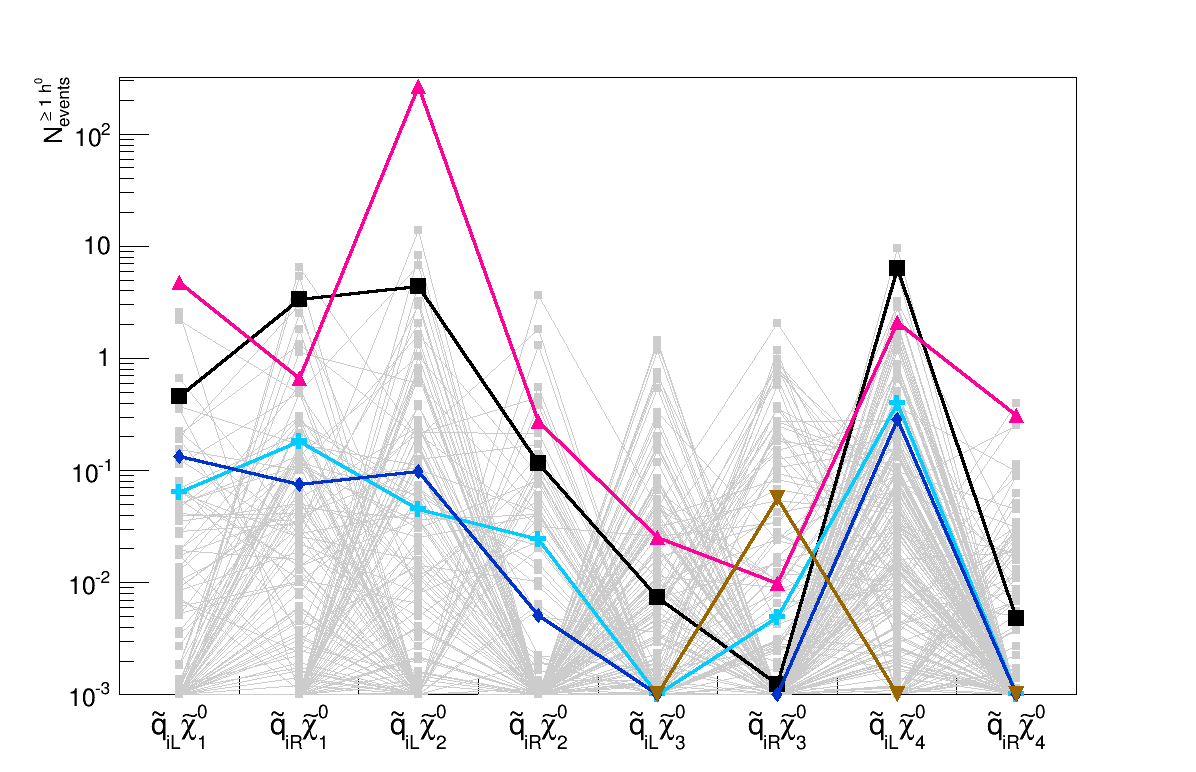}
  \includegraphics[width=0.7\textwidth]{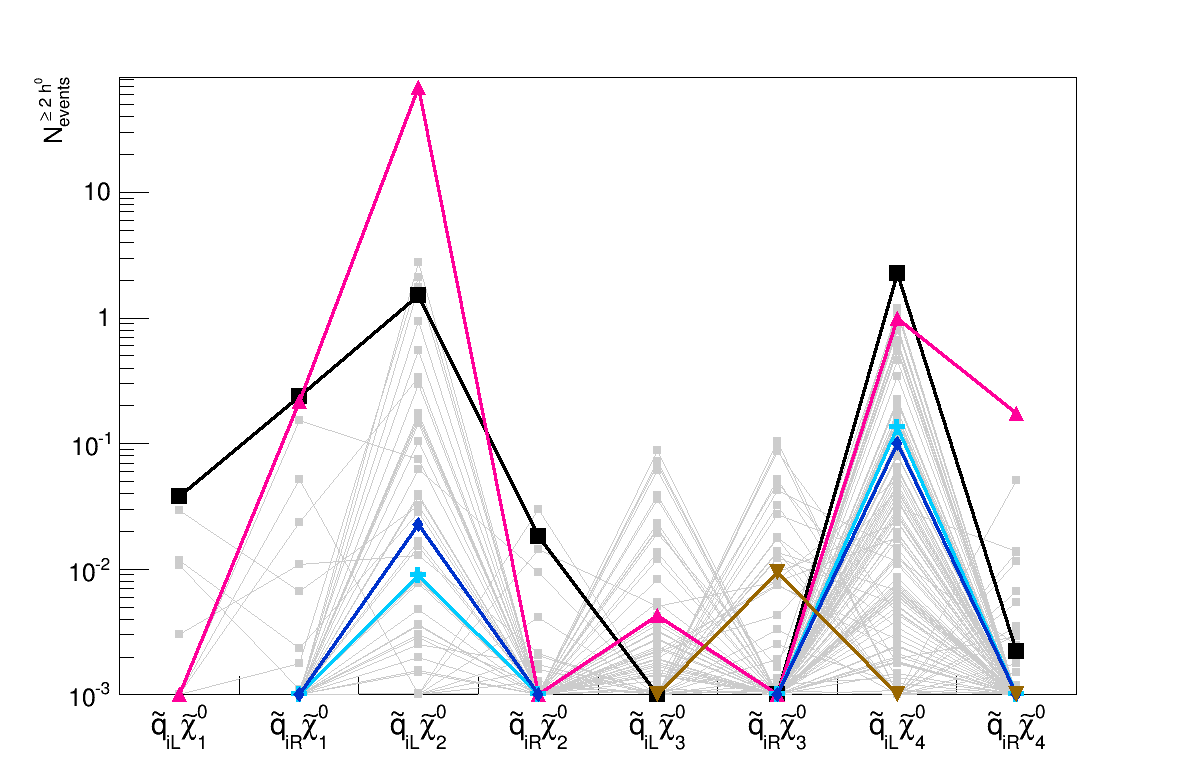}
  \includegraphics[width=0.7\textwidth]{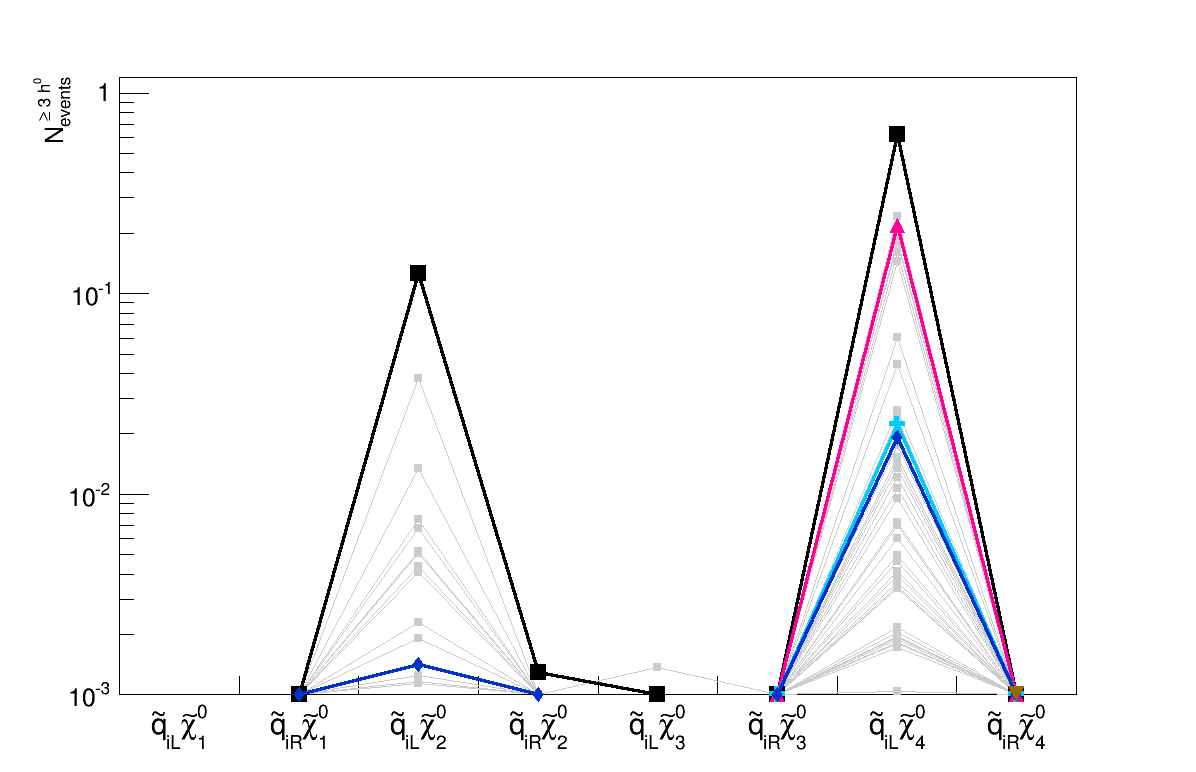}
  \caption{Number of expected events containing at least one (upper), 
           two (middle) or three (lower figure) $h^0$ boson(s) produced in 
           cascades of supersymmetric origin. The horizontal axis indicates the
           supersymmetric squark\,--\,neutralino final states of the primary
           interaction process.}
    \label{fig:Summary_squarkneutralino}
\end{figure}

\begin{figure}[H]
  \centering
  \includegraphics[width=0.7\textwidth]{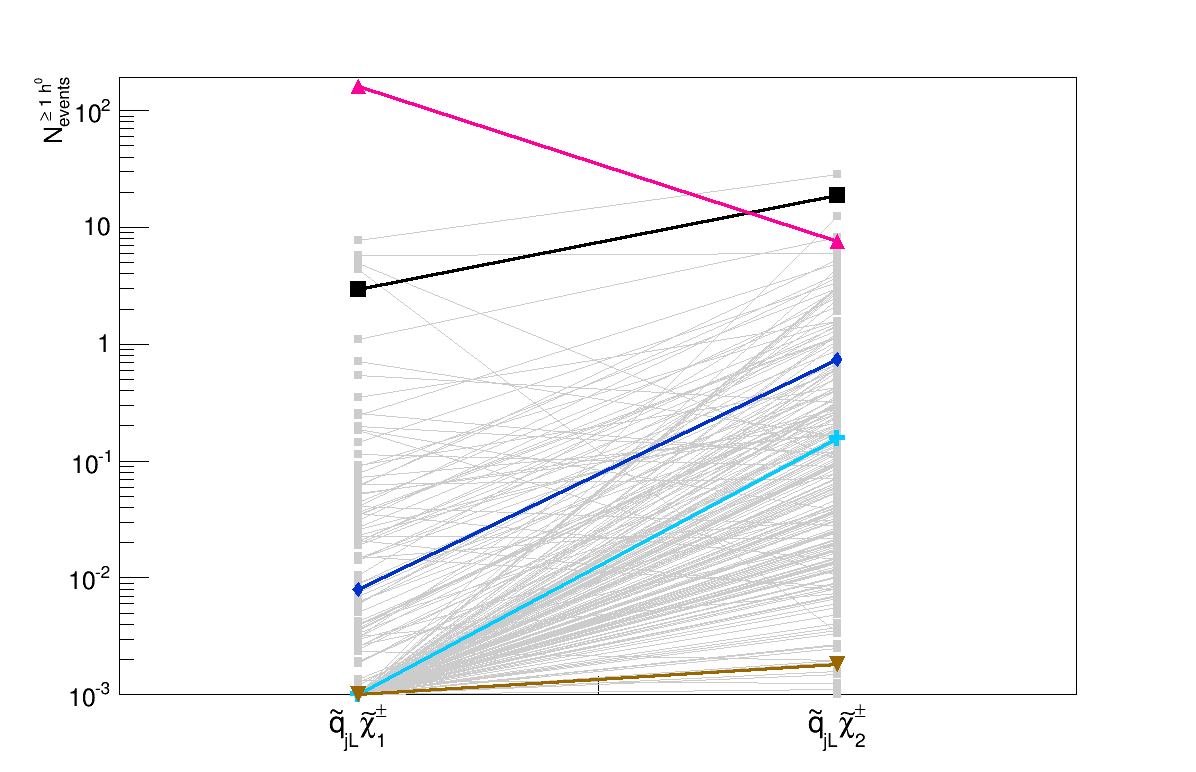}
  \includegraphics[width=0.7\textwidth]{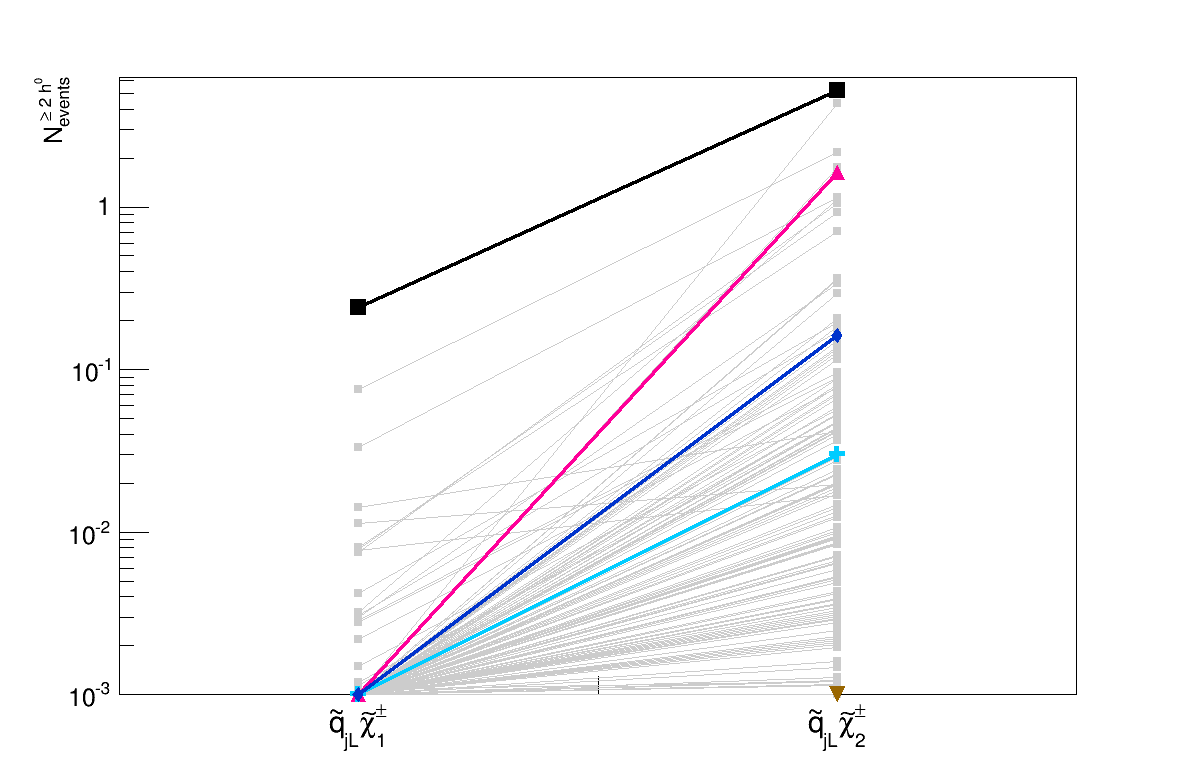}
  \includegraphics[width=0.7\textwidth]{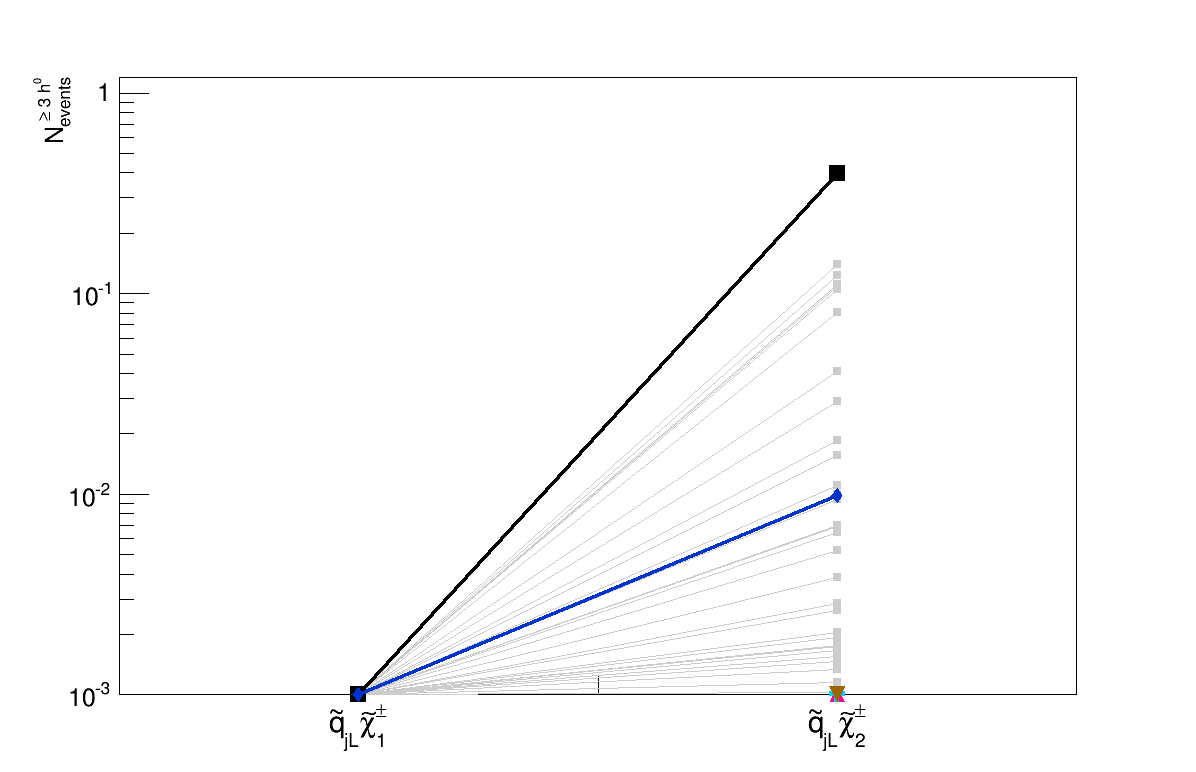}
  \caption{Number of expected events containing at least one (upper), 
           two (middle) or three (lower figure) $h^0$ boson(s) produced in 
           cascades of supersymmetric origin. The horizontal axis indicates the
           supersymmetric left-handed-squark\,--\,chargino final states of the  
           primary interaction process.}
    \label{fig:Summary_squarkchargino}
\end{figure}

\begin{figure}[H]
  \centering
  \includegraphics[width=0.7\textwidth]{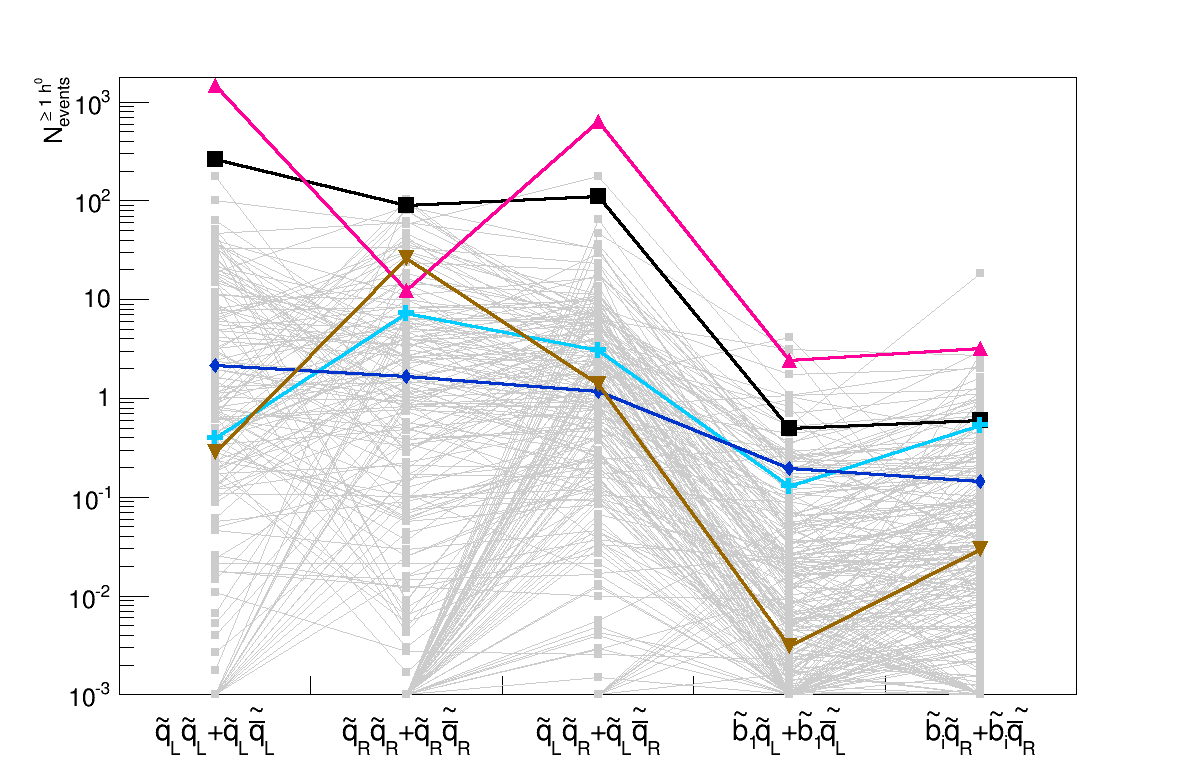}
  \includegraphics[width=0.7\textwidth]{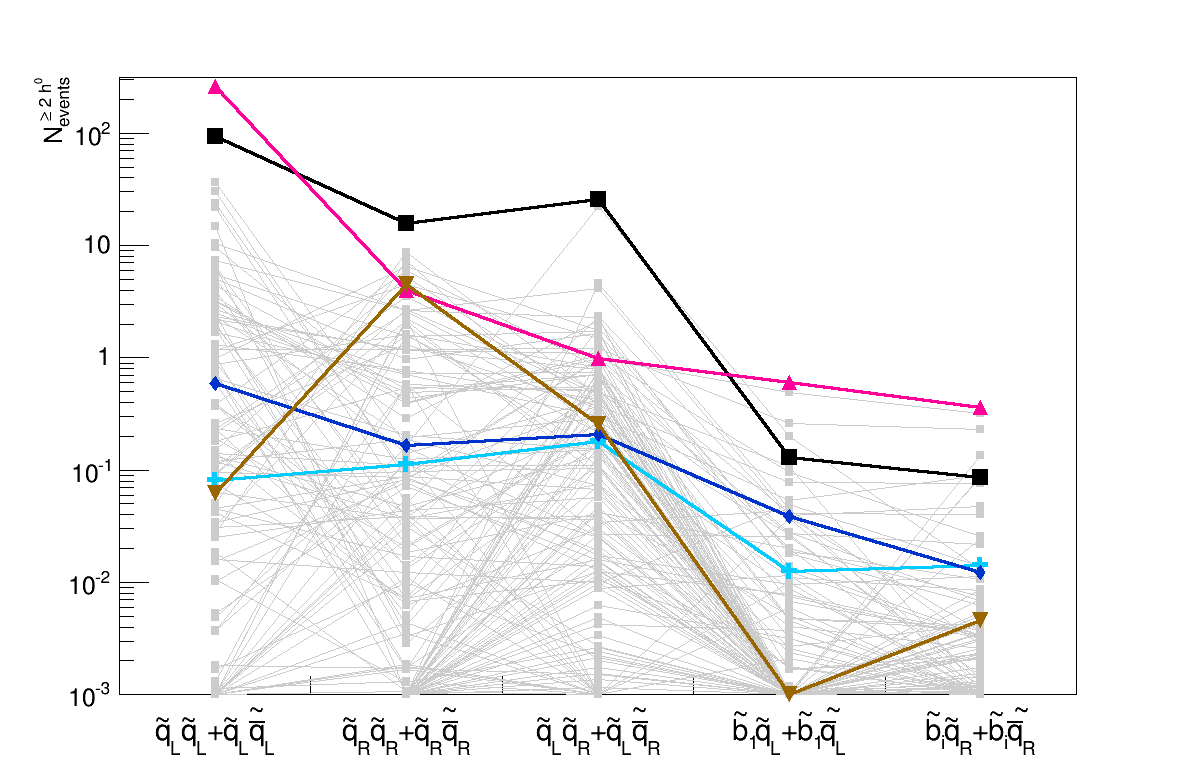}
  \includegraphics[width=0.7\textwidth]{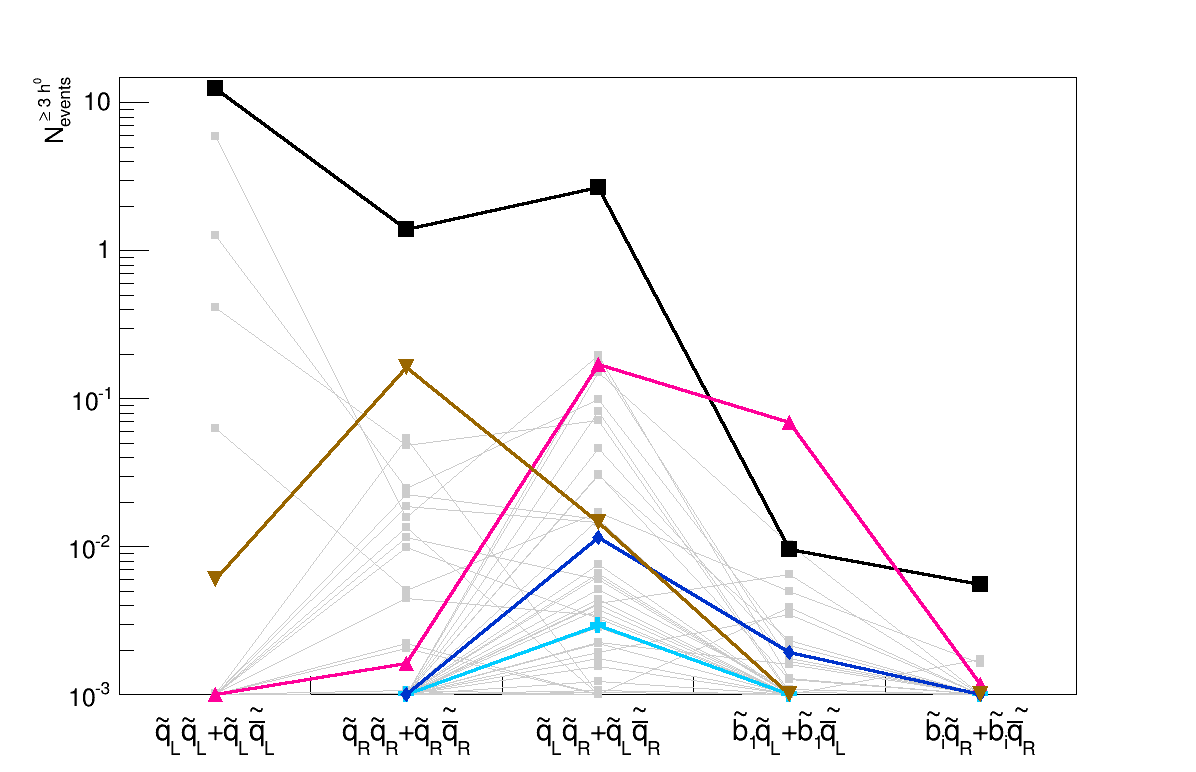}
  \caption{Number of expected events containing at least one (upper), 
           two (middle) or three (lower figure) $h^0$ boson(s) produced in 
           cascades of supersymmetric origin. The horizontal axis indicates the
           supersymmetric squark-(anti)squark final states of the primary 
           interaction process, where at least one of the squarks is of the
           1st/2nd generation.}
    \label{fig:Summary_squarksquark}
\end{figure}

\begin{figure}[H]
  \centering
  \includegraphics[width=0.7\textwidth]{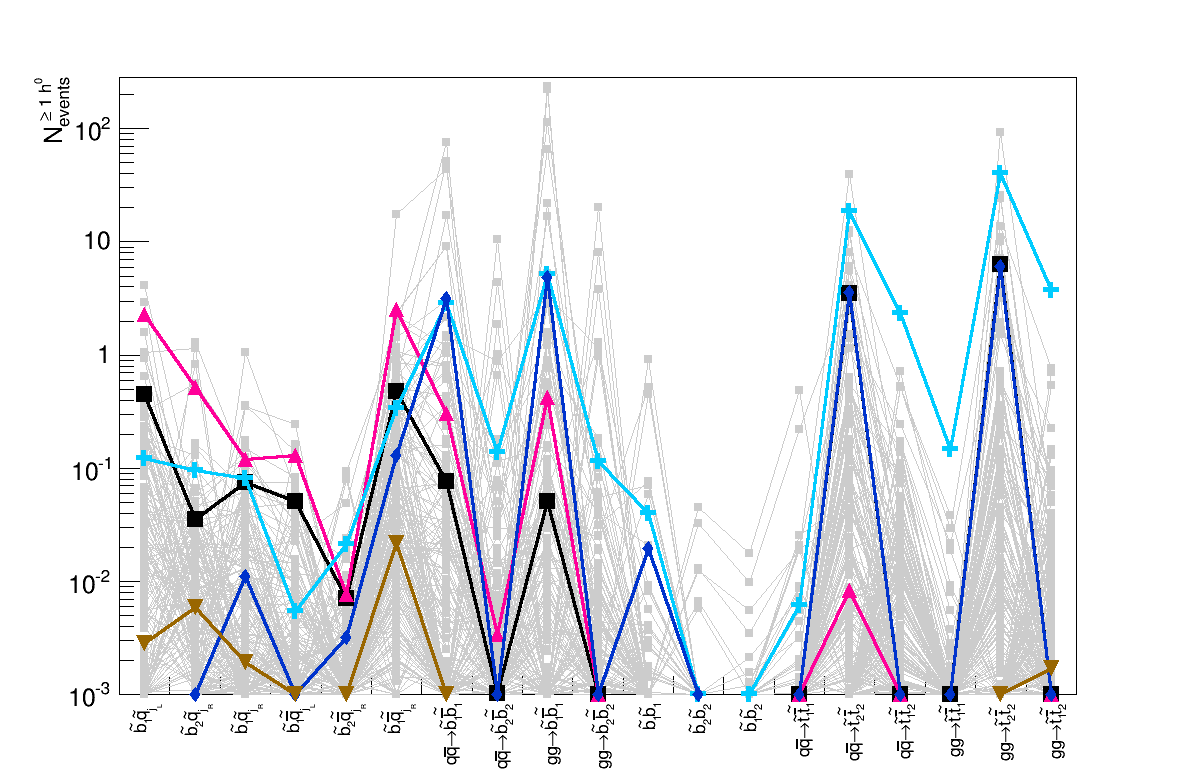}
  \includegraphics[width=0.7\textwidth]{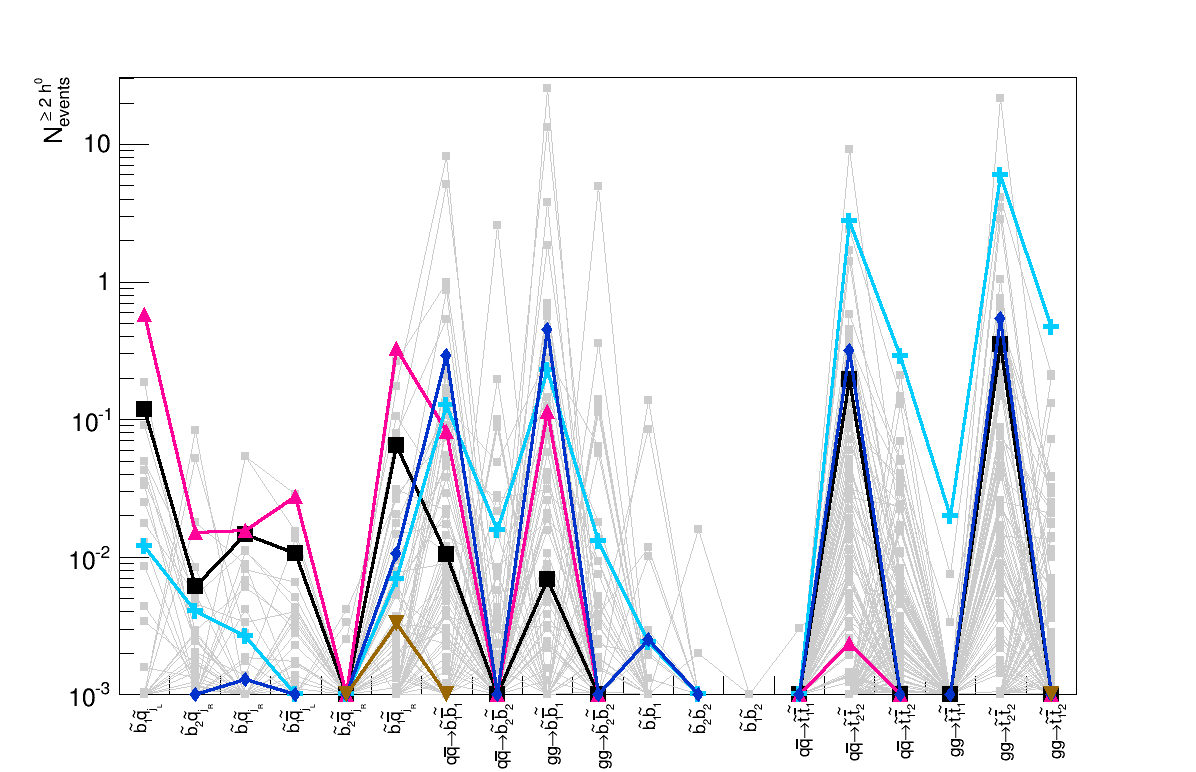}
  \includegraphics[width=0.7\textwidth]{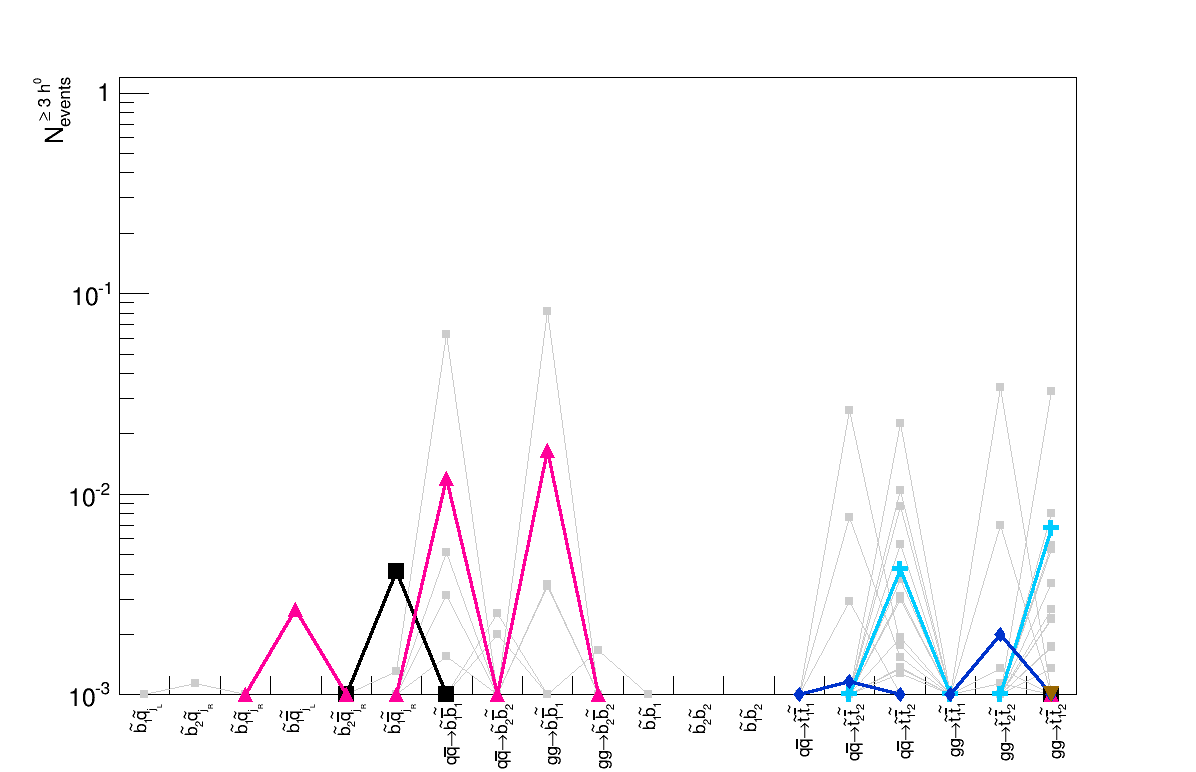}
  \caption{Number of expected events containing at least one (upper), 
           two (middle) or three (lower figure) $h^0$ boson(s) produced in 
           cascades of supersymmetric origin. The horizontal axis indicates the
           supersymmetric squark-(anti)squark final states of the primary 
           interaction process, where at least one of the squarks is of the
           3rd generation.}
    \label{fig:Summary_heavysquarks}
\end{figure}

\begin{figure}[H]
  \centering
  \includegraphics[width=1.0\textwidth]{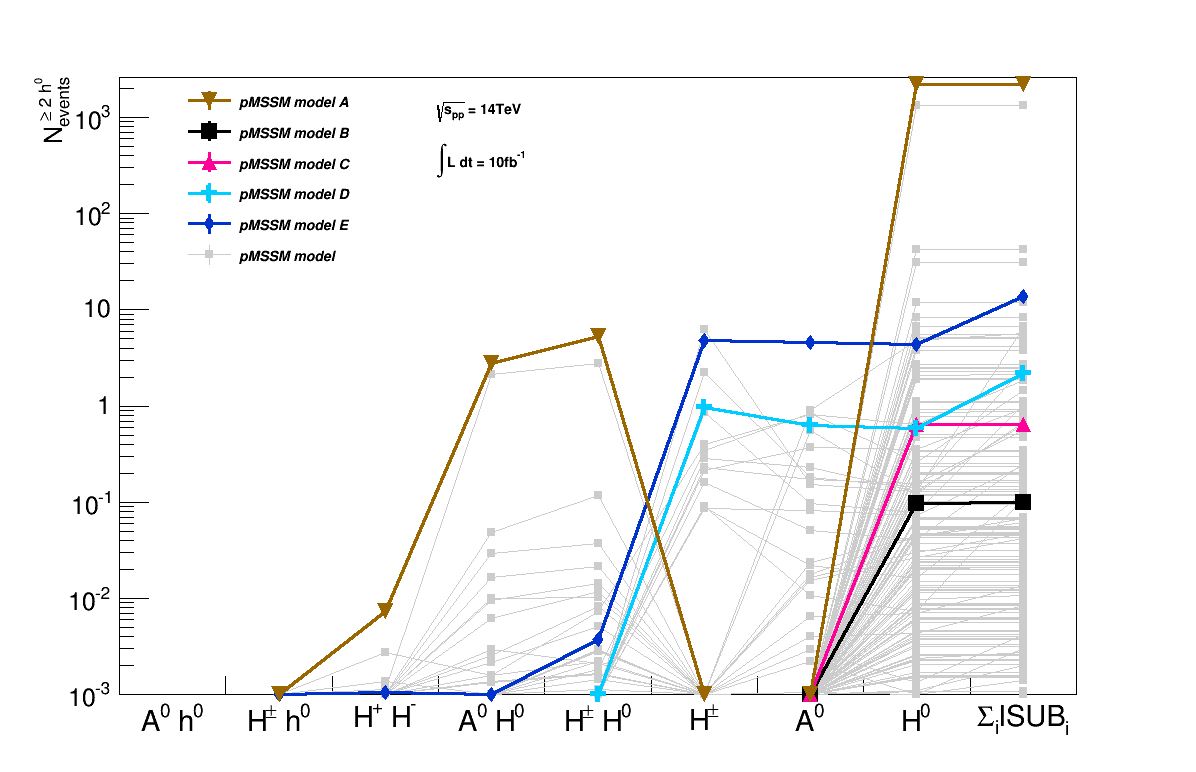}
  \includegraphics[width=1.0\textwidth]{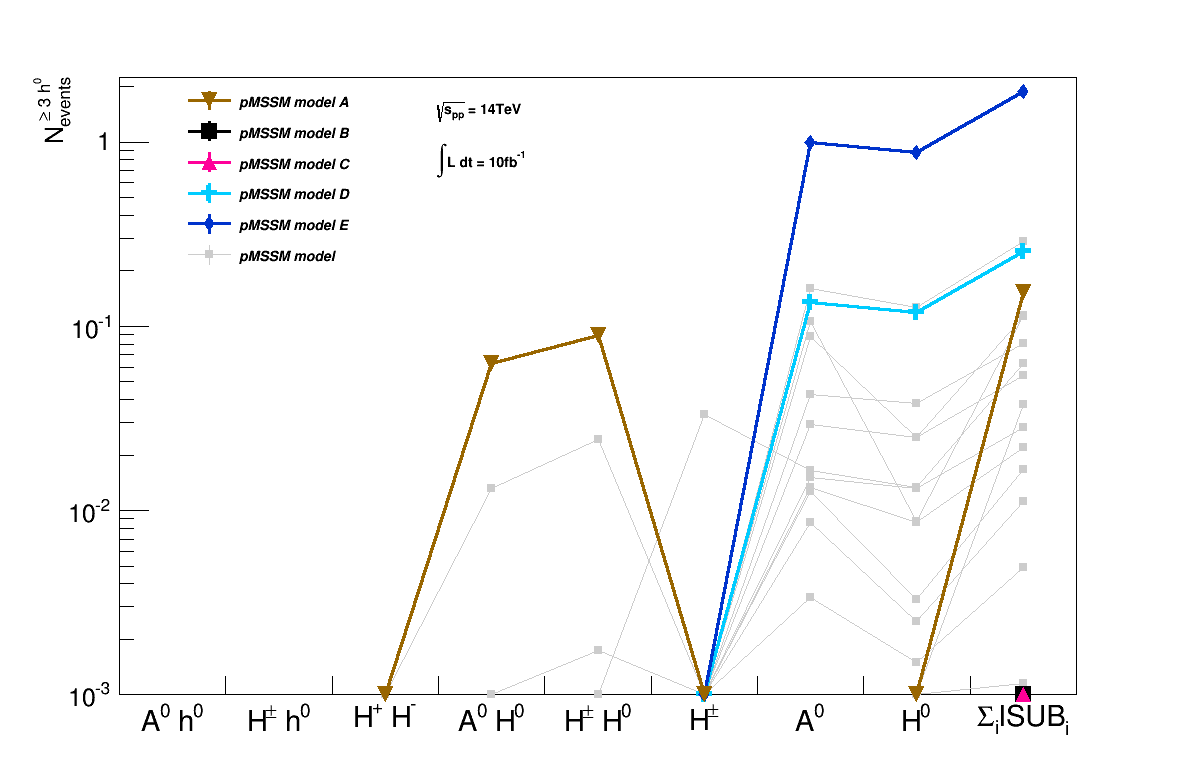}
  \caption{Number of expected events containing at least two (upper) or three 
           (lower) $h^0$ bosons in cascades of heavy-Higgs ($H^0$, $A^0$ and 
           $H^{\pm}$) origin. The horizontal axis indicates the heavy Higgs 
           particles involved in the final states of the primary interaction 
           process.}
  \label{fig:N_events_2hplus_HeavyHiggs} 
\end{figure}

\bibliographystyle{JHEP}
\bibliography{bibreport}

\appendix

\end{document}